\documentclass[showpacs,superscriptaddress,twocolumn,english,notitlepage,longbibliography,nofootinbib]{revtex4-1}
\usepackage[utf8]{inputenc}
\usepackage{color}
\usepackage{array}
\usepackage{verbatim}
\usepackage{multirow}
\usepackage{amsmath}
\usepackage{amssymb}
\usepackage{dsfont}
\usepackage{graphicx}
\usepackage{esint}
\usepackage[unicode=true,
 bookmarks=true,bookmarksnumbered=true,bookmarksopen=false,
 breaklinks=true,pdfborder={0 0 0},backref=false,colorlinks=true,linkcolor=blue]
 {hyperref}
\usepackage{enumitem}
\usepackage{titlesec}
\usepackage{cleveref}
\usepackage{longtable}
\usepackage{bbm}
\usepackage[normalem]{ulem}
\usepackage{wasysym}

\newcommand{\PreserveBackslash}[1]{\let\temp=\\#1\let\\=\temp}
\newcolumntype{C}[1]{>{\PreserveBackslash\centering}p{#1}}
\newcolumntype{R}[1]{>{\PreserveBackslash\raggedleft}p{#1}}
\newcolumntype{L}[1]{>{\PreserveBackslash\raggedright}p{#1}}

\usepackage{cleveref}
\crefname{equation}{Eq.}{Eqs.}
\crefname{figure}{Fig.}{Figs.}
\crefname{table}{Table}{Tables}
\crefname{section}{Section}{Sections}

\renewcommand{\paragraph}[1]{\vspace{0.2cm}{\bf \textit{#1}}}
\def\ie{{\it i.e.},\ }
\def\eg{{\it e.g.},\ }

\usepackage{cleveref}
\crefname{equation}{Eq.}{Eqs.}
\crefname{figure}{Fig.}{Figs.}
\crefname{table}{Table}{Tables}
\crefname{section}{Section}{Sections}
\crefname{appendix}{App.}{Apps.}

\allowdisplaybreaks 
\newcommand{\mbf}{\mathbf}
\newcommand{\mbb}{\mathbb}
\newcommand{\mcl}{\mathcal}

\newcommand{\mrm}{\mathrm}
\newcommand{\mfk}{\mathfrak}
\newcommand{\td}{\widetilde}

\def\beq#1\eeq{\begin{equation}#1\end{equation}}
\def\beqs#1\eeqs{\begin{align}#1\end{align}}

\def\pare#1{\left( #1 \right)}

\def\bra#1{\langle #1 |}
\def\ket#1{| #1 \rangle}

\def\inn#1{\langle #1 \rangle}

\def\nono{\nonumber}

\def\dg{\dagger}
\def\up{\uparrow}
\def\down{\downarrow}
\def\pr{\prime}
\def\prpr{{\prime\prime}}
\def\Tr{\mrm{Tr}}

\def\kk{\mathbf{k}}
\def\qq{\mathbf{q}}
\def\pp{\mathbf{p}}
\def\pp{\mathbf{p}}
\def\GG{\mathbf{G}}
\def\QQ{\mathbf{Q}}

\def\rr{\mathbf{r}}

\def\bb{\mathbf{b}}

\def\CC{\mathcal{P}}

\def\SZD#1{}

\makeatother

\begin{document}

\title{Twisted Bilayer Graphene V: Exact Analytic Many-Body Excitations in Twisted Bilayer Graphene Coulomb Hamiltonians: Charge Gap, Goldstone Modes and Absence of Cooper Pairing}

\author{B. Andrei Bernevig \thanks{\href{bernevig@princeton.edu}{bernevig@princeton.edu}}}
\author{Biao Lian}
\author{Aditya Cowsik}
\author{Fang Xie}
\affiliation{Department of Physics, Princeton University, Princeton, New Jersey
08544, USA}
\author{Nicolas Regnault}
\affiliation{Department of Physics, Princeton University, Princeton, New Jersey 08544, USA}
\affiliation{Laboratoire de Physique de l'Ecole normale superieure, ENS, Universit\'e PSL, CNRS,
Sorbonne Universit\'e, Universit\'e Paris-Diderot, Sorbonne Paris Cit\'e, Paris, France}
\author{Zhi-Da Song}
\affiliation{Department of Physics, Princeton University, Princeton, New Jersey
08544, USA}

\begin{abstract}
We find exact analytic expressions for the energies and wavefunctions of the charged and neutral excitations above the exact ground states (at rational filling per unit cell) of projected Coulomb Hamiltonians in twisted bilayer graphene. Our exact expressions are valid for any form of the Coulomb interaction and any form of $AA$ and $AB/BA$ tunneling. The single charge excitation energy is  a convolution of the Coulomb potential with a  quantum  geometric tensor of the TBG bands. The neutral excitations are (high-symmetry group) magnons, and their dispersion is analytically calculated in terms of the form factors of the active bands in TBG. The two-charge excitation energy and wavefunctions are also obtained, and a sufficient condition on the graphene eigenstates for obtaining a Cooper-pair from Coulomb interactions is obtained. For the actual TBG bands at the first magic angle, we can analytically show that the Cooper pair binding energy is zero in all such projected Coulomb models, implying that either phonons and/or non-zero kinetic energy are needed for superconductivity. Since [\href{https://arxiv.org/abs/2009.09413}{Vafek et al., arXiv:2009.09413 (2020)}] showed that the kinetic energy bounds on the superexchange energy are less $10^{-3}$ in Coulomb units, the phonon mechanism becomes then very likely. If nonetheless the superconductivity is due to kinetic terms which render the bands non-flat, one prediction of our theory is that the highest $T_c$ would \emph{not} occur at the highest DOS.
\end{abstract}

\date{\today}
\maketitle

\section{Introduction}

The rich physics of the experimentally observed insulating states in magic angle twisted bilayer graphene (TBG) at integer number of electrons per unit cell and the superconducting phase with finite doping above the insulating states has attracted considerable interest \cite{bistritzer_moire_2011,cao_correlated_2018,cao_unconventional_2018, lu2019superconductors, yankowitz2019tuning, sharpe_emergent_2019, saito_independent_2020, stepanov_interplay_2020, liu2020tuning, arora_2020, serlin_QAH_2019, cao_strange_2020, polshyn_linear_2019,  xie2019spectroscopic, choi_imaging_2019, kerelsky_2019_stm, jiang_charge_2019,  wong_cascade_2020, zondiner_cascade_2020,  nuckolls_chern_2020, choi2020tracing, saito2020,das2020symmetry, wu_chern_2020,park2020flavour, saito2020isospin,rozen2020entropic, lu2020fingerprints, burg_correlated_2019,shen_correlated_2020, cao_tunable_2020, liu_spin-polarized_2019, chen_evidence_2019, chen_signatures_2019, chen_tunable_2020, burg2020evidence, tarnopolsky_origin_2019, zou2018, fu2018magicangle, liu2019pseudo, Efimkin2018TBG, kang_symmetry_2018, song_all_2019,po_faithful_2019,ahn_failure_2019,adrien2019, hejazi_multiple_2019, lian2020, hejazi_landau_2019, padhi2020transport, xu2018topological,  koshino_maximally_2018, ochi_possible_2018, xux2018, guinea2018, venderbos2018, you2019,  wu_collective_2020, Lian2019TBG,Wu2018TBG-BCS, isobe2018unconventional,liu2018chiral, bultinck2020, zhang2019nearly, liu2019quantum,  wux2018b, thomson2018triangular,  dodaro2018phases, gonzalez2019kohn, yuan2018model,kang_strong_2019,bultinck_ground_2020,seo_ferro_2019, hejazi2020hybrid, khalaf_charged_2020,po_origin_2018,xie_superfluid_2020,julku_superfluid_2020, hu2019_superfluid, kang_nonabelian_2020, soejima2020efficient, pixley2019, knig2020spin, christos2020superconductivity,lewandowski2020pairing, xie_HF_2020,liu2020theories, cea_band_2020,zhang_HF_2020,liu2020nematic, daliao_VBO_2019,daliao2020correlation, classen2019competing, kennes2018strong, eugenio2020dmrg, huang2020deconstructing, huang2019antiferromagnetically,guo2018pairing, ledwith2020, repellin_EDDMRG_2020,abouelkomsan2020,repellin_FCI_2020, vafek2020hidden, fernandes_nematic_2020, Wilson2020TBG, wang2020chiral, ourpaper1, ourpaper2,ourpaper3,ourpaper4,ourpaper6}.
The single-particle picture predicts a gapless metallic state at electron number $\pm (3,2,1)$, and hence the insulating states have to follow from many-body interactions. The initial observations of the insulating states \cite{cao_correlated_2018,cao_unconventional_2018, lu2019superconductors,yankowitz2019tuning} were then followed by the experimental discovery by both scanning tunneling microscope \cite{nuckolls_chern_2020,choi2020tracing} and transport \cite{serlin_QAH_2019, sharpe_emergent_2019, saito2020,das2020symmetry, wu_chern_2020,park2020flavour} that these states might exhibit Chern numbers, even when the TBG substrate is not aligned with hBN, which would indicate a many-body origin of the Chern insulator.

These remarkable experimental advances have been followed by extensive theoretical efforts aimed at their explanation \cite{xu2018topological,  koshino_maximally_2018, ochi_possible_2018, xux2018, guinea2018, venderbos2018, you2019,  wu_collective_2020, Lian2019TBG,Wu2018TBG-BCS, isobe2018unconventional,liu2018chiral, bultinck2020, zhang2019nearly, liu2019quantum,  wux2018b, thomson2018triangular,  dodaro2018phases, gonzalez2019kohn, yuan2018model,kang_strong_2019,bultinck_ground_2020, seo_ferro_2019, hejazi2020hybrid, khalaf_charged_2020,po_origin_2018,xie_superfluid_2020,julku_superfluid_2020, hu2019_superfluid, kang_nonabelian_2020, soejima2020efficient, pixley2019, knig2020spin, christos2020superconductivity, lewandowski2020pairing, xie_HF_2020,liu2020theories, cea_band_2020,zhang_HF_2020,liu2020nematic, daliao_VBO_2019,daliao2020correlation, classen2019competing, kennes2018strong, eugenio2020dmrg, huang2020deconstructing, huang2019antiferromagnetically,guo2018pairing, ledwith2020, repellin_EDDMRG_2020,abouelkomsan2020,repellin_FCI_2020, vafek2020hidden, fernandes_nematic_2020}.  Using a strong-coupling approach where the interaction is projected into a  Wannier basis, Kang and Vafek \cite{kang_strong_2019} constructed a special Coulomb Hamiltonian, of an enhanced symmetry, where the ground state (of Chern number $0$) at $\pm 2$ electrons per unit cell can be exactly obtained (with rather weak assumptions). In Ref.~\cite{ourpaper4} we have showed that the type of Kang-Vafek type Hamiltonians \cite{kang_strong_2019} (hereby called positive semi-definite Hamiltonians - PSDH)  are actually \emph{generic} in projected Hamiltonians, and that the presence of extra symmetries \cite{kang_strong_2019,zou2018,bultinck2020} renders some Slater determinant states to be exact eigenstates of PSDH. We found at zero filling, these states are the ground states of PSDH. At nonzero integer filling, these states are the ground states of the PSDH under weak assumptions (first considered by Kang and Vafek \cite{kang_strong_2019}). 
With a unitary  particle-hole (PH) symmetry first derived in Ref.~\cite{song_all_2019}, the  PSDH  projected to the active bands has enhanced U(4) (in all the parameter space) and U(4)$\times$U(4) (in a certain, first chiral limit) symmetries first mentioned in Refs.~\cite{kang_strong_2019, bultinck_ground_2020,seo_ferro_2019}. 
We showed \cite{ourpaper3, ourpaper4} that these symmetries are valid for PSDHs of TBG irrespective of the number of projected bands. We also found that, for two projected bands in the first chiral limit (a second chiral limit, of U(4)$\times$U(4) defined in Ref.~\cite{ourpaper3} was also found), ground states of different Chern numbers are exactly degenerate \cite{ourpaper4}. 
These ground states are all variants of U(4) ferromagnets (FM) in valley/spin. When kinetic energy is added or away from the chiral limit, the lowest/highest Chern number becomes the ground state in low/high magnetic field, which explains/is consistent with experimental findings \cite{nuckolls_chern_2020,choi2020tracing,serlin_QAH_2019, sharpe_emergent_2019, saito2020,das2020symmetry, wu_chern_2020,park2020flavour}.

In this paper we show that the Kang-Vafek type of PSDH also allow, remarkably, for an exact expression of the charge $\pm 1$ excitation (relevant for transport gaps) energy and eigenstate, neutral excitation (relevant for the Goldstone and thermal transport), and charge $\pm 2$ excitation (relevant for possible Cooper pair binding energy).  We show that the charge excitation dispersion is fully governed by a generalized ``quantum geometric tensor" of the projected bands, convoluted with the Coulomb interaction. The smallest  charge $1$ excitation gap is at the $\Gamma_M$ point. The neutral, and charge $\pm 2$ excitation, on top of every FM ground state can also be obtained as a single-particle diagonalization problem, despite the state having a thermodynamic number of particles. The neutral excitation has an exact zero mode, which we identify with the FM U(4)-spin wave, and whose low-momentum dispersion (velocity) can be computed exactly. 
The charge $\pm2$ excitations allows for a simple check of the Richardson criterion \cite{richardson,richardson1964,richardson_numerical_1966,richardson1977} of superconductivity: We check if states appear below the non-interacting 2-particle continuum. We find a sufficient criterion for the appearance/lack of Cooper binding energy in these type of PSDH Hamiltonian systems based on the eigenvalues of the generalized ``quantum geometric tensor". We analytically show that, generically, the projected Coulomb Hamiltonians cannot exhibit Cooper pairing binding energy. As such, this implies that either phonons or non-zero kinetic energy are needed for superconductivity. Since the Ref.~\cite{vafek2020hidden} showed that the kinetic energy bounds on the superexchange energy are less $10^{-3}$ in Coulomb units, the phonon mechanism becomes becomes likely. If however, experimentally, the kinetic energy is stronger, a Coulomb mechanism for superconductivity is still possible. Since we proved that flat bands cannot Cooper pair under Coulomb, a prediction of a Coulomb with non-flat bands mechanism for superconductivity would be that the highest superconducting temperature does \emph{not} happen at the point of highest density of states DOS. This is in agreement with recent experimental data \cite{park2020flavour}.

\section{The positive semi-definite Hamiltonian and its ground states}
We generically consider the TBG system with a Coulomb interaction Hamiltonian projected to the active $8$ lowest bands (2 per spin-valley flavor) obtained by diagonalizing the single particle Bistritzer-MacDonald (BM) \cite{bistritzer_moire_2011} TBG Hamiltonian (see \cref{p5:singpartRevappendix} for a brief review, and more detail in Refs. \cite{ourpaper1, ourpaper2}). The projected single-particle Hamiltonian reads

\begin{equation}
H_0=\sum_{n=\pm 1}\sum_{\mathbf{k} \eta s} \epsilon_{n,\eta}(\mathbf{k}) c^\dag_{\mathbf{k},n,\eta, s} c_{\mathbf{k},n,\eta, s}\ ,\label{p5:eq:HHMT}
\end{equation} where we define
$\eta=\pm$ for graphene valleys $K$ and $K'$, $s=\uparrow,\downarrow$ for electron spin, and $n=\pm1$ for the lowest conduction/valence bands in each spin-valley flavor. $c^\dag_{\mathbf{k},n,\eta, s}$ is the electron creation operator of energy band $n$, with the origin of $\mathbf{k}$ chosen at $\Gamma$ point of the moir\'e Brillouin zone (MBZ).

The density-density Coulomb interaction, when projected into the active bands of \cref{p5:eq:HHMT}, always takes the form of a positive  semidefinite Hamiltonian (PSDH) (see proof in Ref.~\cite{ourpaper3}, see also brief review in \cref{p5:interactionRevappendix}):
\begin{equation}
H_{I}=\frac{1}{2\Omega_{\text{tot}}}\sum_{\mathbf{G}} \sum_{\mathbf{q} \in {\rm MBZ}} O_{\mathbf{q,G}} O_{\mathbf{-q,-G}}, \label{p5:eq:nonnegative-HintMT}
\end{equation}
where $\Omega_{\text{tot}}$ is the sample area, and $\GG$ runs over all vectors in the (triangular) moir\'e reciprocal lattice $\mathcal{Q}_0$. This Hamiltonian is of a same positive semidefinite form as that Kang and Vafek \cite{kang_strong_2019} obtained by projecting the Coulomb interaction into the Wannier basis of the active bands.
In this work we will omit the kinetic energy.
Due Ref. \cite{ourpaper4}, the energy splitting between the degenerate ground states of \cref{p5:eq:nonnegative-HintMT} is smaller than 0.1meV per electron. As shown in the rest of this work, the characteristic energy of charged and neutral excitations is about 10meV. Thus, it is safe to neglect the kinetic energy for most of the excitations. But some of the U(4) Goldstone modes might be opened a small gap due to the kinetic energy.
We leave this effect of kinetic energy to future studies.

The $O_{\mathbf{q,G}}$ operator takes the form:
\begin{align}
O_{\mathbf{q,G}}=& \sum_{\mathbf{k},m,n,\eta,s} \sqrt{V(\mathbf{G}+\mathbf{q})} M_{m,n}^{\left(\eta\right)}\left(\mathbf{k},\mathbf{q}+\mathbf{G}\right) \nonumber \\ &\times  \left(\rho_{\mathbf{k,q},m,n,s}^\eta-\frac{1}{2}\delta_{\mathbf{q,0}}\delta_{m,n}\right), \label{p5:Oqdef1MT}
\end{align}
where  $V(\qq)$ is the Fourier transform of the Coulomb interaction,  $\rho_{\mathbf{k,q},m,n,s}^\eta=c^\dag_{\mathbf{k+q},m,\eta,s}c_{\mathbf{k},n,\eta,s}$ is the density operator in band basis, and the $-\frac{1}{2}\delta_{\mathbf{q,0}}\delta_{m,n}$ factor is a chemical potential added to respect many-body charge conjugation symmetry (see \cref{p5:gaugefixingRevappendix} and Ref. \cite{ourpaper3}). For theoretical derivations we shall keep $V(\qq)$ general except that we assume $V(\qq)\ge0$ and only depends on $q=|\qq|$; although for numerical calculations we will take $V(\mathbf{q})=2\pi e^2\xi\tanh(q\xi/2)/\epsilon q$ for dielectric constant $\epsilon (\sim6)$ and screening length $\xi(\sim10$nm) (see App. \ref{p5:interactionRevappendix}). In particular, $O_{-\qq,-\GG}=O_{\qq,\GG}^\dagger$, and thus $H_I$ in Eq. (\ref{p5:eq:nonnegative-HintMT}) is a PSDH. An important quantity in Eq. (\ref{p5:Oqdef1MT}) for our many-body Hamiltonian are the \emph{form factors}, or  the overlap matrices, of a set of bands $m,n$ (App. \ref{p5:gaugefixingRevappendix})
\begin{equation}
M_{m,n}^{\left(\eta\right)}\left(\mathbf{k},\mathbf{q}+\mathbf{G}\right)=\sum_{\alpha \QQ} u_{\mathbf{Q}-\mathbf{G},\alpha;m\eta}^{*}\left(\mathbf{k}+\mathbf{q}\right)u_{\mathbf{Q},\alpha;n\eta}\left(\mathbf{k}\right) , \label{p5:eq:M-defMT}
\end{equation}
where $u_{\mathbf{Q}\alpha;n\eta}$ is the Bloch wavefunction of band $n$ and valley $\eta$ (here $\alpha=A,B$ denotes the microscopic graphene sublattices, and $\mathbf{Q}$ are sites of a honeycomb momentum lattice with definition in App. \ref{p5:singpartRevappendix}, see also \cite{ourpaper1} for details). 
A nonzero Berry phase of the projected bands renders the spectra of the PSDH \cref{p5:eq:nonnegative-HintMT} not analytically solvable: the $O_{\mathbf{q,G}}$'s at different $\qq, \GG$ generically do not commute (unless in the stabilizer code limit discussed in Refs.~\cite{ourpaper3,ourpaper4}), and hence the PSDH is not solvable.  The properties of the PSDH Eq. (\ref{p5:eq:nonnegative-HintMT})  depend on the quantitative and qualitative (symmetries) properties of the form factors in \cref{p5:eq:M-defMT}, which are detailed in Refs. \cite{ourpaper1,ourpaper2,ourpaper3} and briefly reviewed in \cref{p5:gaugefixingRevappendix}. 
First, in Ref. \cite{ourpaper1} we showed that $M_{m,n}^{\left(\eta\right)}\left(\mathbf{k},\mathbf{q}+\mathbf{G}\right)$ falls off exponentially with $|\GG|$, and can be neglected for $|\GG|>\sqrt{3}k_\theta$, where $k_\theta=2|\mathbf{K}|\sin(\theta/2)$ is the distance between the $K$ points of two graphene sheets. Furthermore, we showed in Refs. \cite{ourpaper2,ourpaper3} that by gauge-fixing the $C_{2z}$, $T$, and unitary particle-hole symmetry $P$ \cite{song_all_2019}, the form factors can be rewritten into a matrix form in the $n,\eta$ basis as (see \cref{p5:eq:M-para})
\begin{equation}\label{p5:eq-Mmn}
M_{mn}^{(\eta)}\left(\mathbf{k},\mathbf{q}+\mathbf{G}\right)=\sum_{j=0}^3 (M_j)_{m,\eta;n,\eta} \alpha_j(\mathbf{k,q+G})\ ,
\end{equation}
where $M_0=\zeta^0\tau^0$, $M_1=\zeta^x\tau^z$, $M_2=i\zeta^y\tau^0$, and $M_3=\zeta^z\tau^z$, and $\alpha_j(\mathbf{k,q+G})$ are real scalar functions satisfying (\cref{p5:eq:alpha-cond1,p5:eq:alpha-cond2} in \cref{p5:sec:exact-GS}).

A further simplification \cite{bultinck_ground_2020} happens in a region of the parameter space where the $AA$ interlayer coupling $w_0=0$ \cite{bultinck_ground_2020}, which is called the (first) chiral limit \cite{tarnopolsky_origin_2019} (a similar simplification occurs in a second chiral limit \cite{ourpaper3}).  
In this limit there is another chiral symmetry $C$ anticommuting with the single-particle Hamiltonian, which further imposes the constraints  $\alpha_1 (\kk,\qq+\GG)=\alpha_3 (\kk,\qq+\GG)=0$ (see Ref. \cite{ourpaper4} and \cref{p5:sec:exact-GS}). 
The first chiral limit also allows for the presence of a \emph{Chern band basis} in which bands of Chern number $e_Y=\pm1$ are created by the operators
\beq 
d^\dagger_{\kk,e_Y,\eta,s} = \frac{1}{\sqrt2} ( c^\dg_{\kk,+,\eta,s} + i e_Y c^\dg_{\kk,-,\eta,s})\ ,
 \label{p5:eq:Chern-bandMT}
\eeq  
In Ref. \cite{ourpaper2} we detail the gauge-fixing for this basis. The Chern basis is also discussed in Refs.~\cite{ourpaper2,bultinck_ground_2020,hejazi2020hybrid}. The form factors under the Chern basis take the simple diagonal form
\begin{equation}\label{p5:eq:chiral-MqG1MT}
M_{e_Y}^{(\eta)}(\kk,\qq+\GG)=\alpha_0(\mathbf{k,q+G})+ie_Y\alpha_2(\mathbf{k,q+G}).
\end{equation} 

The symmetries of the projected Hamiltonian in the nonchiral ($w_0, w_1 \ne 0$) and two chiral $w_0=0$ or $w_1=0$ limits are important. 
We will use the matrices $\zeta^a, \tau^a, s^a$ with $a=0,x,y,z$ as identity and $x,y,z$ Pauli matrices in (particle-hole related) band, valley and spin-space respectively. 
In Ref. \cite{ourpaper3} (short review in \cref{p5:projectedinteractionRevappendix}), we have showed that the PSDH has a U(4) symmetry  in the nonchiral limit (with single-particle representations of generators $s^{ab}=\{\zeta^y \tau^y s^a, \zeta^y \tau^x s^a,\zeta^0 \tau^0 s^a, \zeta^0 \tau^z s^a\}$ with $a,b=0,x,y,z$ in the energy band basis $c^\dg_{\kk,+,\eta,s}$), and a U(4)$\times$U(4) symmetry in the two chiral-flat limits (with single-particle representations of generators $s^{ab}_\pm=(1\pm e_Y)\tau^a s^b/2$ in the Chern band basis $d^\dagger_{\kk,e_Y,\eta,s}$, \cite{ourpaper2} \cref{p5:chernbasisRevappendix}), mirroring the results obtained by Refs. \cite{kang_strong_2019, seo_ferro_2019, kang_symmetry_2018, bultinck_ground_2020} for projection into the two active bands. We note that in Ref.~\cite{ourpaper3} we showed these symmetries hold for any number of PH symmetric projected bands. 
In \cref{p5:projectedinteractionRevappendix,p5:enhancedU4U4symmetryRevappendix} we provide a summary of these detailed results. 
Adding the kinetic term in the first chiral limit breaks the U(4)$\times$U(4) symmetry of the projected interaction to a U(4) subset ( with generators $\widetilde{s}^{ab}=\zeta^0 \tau^a s^b$ in the energy band basis, $(a,b=0,x,y,z)$). The symmetries we found in the first chiral and nonchiral limits agrees with that in Ref.~\cite{bultinck_ground_2020}, and the relation between our U(4) symmetry generators and those of Kang and Vafek \cite{kang_strong_2019} are given in Ref. \cite{ourpaper3}. We will restrict our study within the nonchiral-flat limit and first chiral-flat limit in this paper. Thus, without ambiguity, we will simply call the first chiral limit the ``chiral limit".

With these symmetries, in the nonchiral-flat limit (where the projected kinetic Hamiltonian $H_0=0$), one can write down exact eigenstates of the PSDH \cref{p5:eq:nonnegative-HintMT}, which we have analyzed in full detail in Ref. \cite{ourpaper4} and review in \cref{p5:exactgroundstatenonchirallimitRevappendix}. 
In the nonchiral limit, $O_{\mathbf{q,G}}$ is diagonal in $\eta$ and $s$, and filling both $n=\pm$ bands of any valley/spin gives a Chern number $0$ eigenstate for all even fillings $\nu=0,\pm2,\pm4$ (along with any U(4) rotation) \cite{ourpaper4}: 
\begin{equation}\label{p5:eq:U(4)-GSMT}
|\Psi_\nu\rangle=\prod_{\mathbf{k}} \left(\prod_{j=1}^{(\nu+4)/2}c^\dag_{\mathbf{k},+,\eta_{j},s_{j}} c^\dag_{\mathbf{k},-,\eta_{j},s_{j}}\right)|0\rangle  ,
\end{equation} 
where $\{\eta_j,s_j\}$ are distinct valley-spin flavors which are fully occupied. They form the $[(2N_M)^{(\nu+4)/2}]_4$ irreducible representation (irrep) of the nonchiral-flat limit U(4) symmetry group, where $[\lambda^p]_4$ is short for the Young tableau notation $[\lambda,\lambda,\cdots]_4$ with $0\le p\le4$ identical rows of length $\lambda$ (see Ref.~\cite{ourpaper4} for a brief review). With $M_{m,n}^{\left(\eta\right)}\left(\mathbf{k},\mathbf{q}+\mathbf{G}\right)$ in Eq. (\ref{p5:eq:M-para}), we have that the state $|\Psi_\nu\rangle$ is an eigenstate of $O_{\mathbf{q,G}}$ satisfying $ O_{\mathbf{q,G}}|\Psi_\nu\rangle= \delta_{\qq0} N_M A_\GG |\Psi_\nu\rangle$, where $A_\GG$ is given by  (\cref{p5:eq:U(4)U(4)-OqG-action})
\begin{equation} \label{p5:neweq:AG}
A_\mathbf{G} = \nu\frac{\sqrt{V(\mathbf{G})}}{N_M}\sum_\kk \alpha_0(\mathbf{k,G}),
\end{equation}
where $N_M$ is the total number of moir\'e unit cells. 
For $\nu=0$, the state \cref{p5:eq:U(4)-GSMT} is always a ground state as it is annihilated by $O_{\qq,\GG}$ \cite{ourpaper4}.  

In the first chiral-flat limit (where $H_0=0$ and $w_0=0$),  the projected Hamiltonian \cref{p5:eq:nonnegative-HintMT} has as eigenstates, the \emph{filled band} wavefunctions \cite{ourpaper4} (see \cref{p5:exactgroundstatechirallimitRevappendix} for brief review):
\begin{equation}\label{p5:eq:U(4)U(4)-GSMT}
|\Psi_{\nu}^{\nu_+,\nu_-}\rangle =\prod_{\mathbf{k}} \left(\prod_{j_1=1}^{\nu_+}d^\dag_{\mathbf{k},+1,\eta_{j_1},s_{j_1}} \prod_{j_2=1}^{\nu_-}d^\dag_{\mathbf{k},-1,\eta_{j_2},s_{j_2}}\right)|0\rangle
\end{equation}
where $\nu_+-\nu_-=\nu_C$ is the total Chern number of the state, and $\nu_++\nu_-=\nu+4$ ($0\le \nu_\pm\le 4$) is the total number of electrons per moir\'e unit cell in the projected bands, $\kk$ runs over the entire MBZ and the occupied spin/valley indices $\{\eta_{j_1},s_{j_1}\}$ and $\{\eta_{j_2},s_{j_2}\}$ can be arbitrarily chosen. 
Moreover, these eigenstates of  \cref{p5:eq:nonnegative-HintMT} are also eigenstates of $O_{\mathbf{q,G}}$ in \cref{p5:Oqdef1MT}, satisfying $O_{\mathbf{q,G}}|\Psi_\nu^{\nu_+,\nu_-}\rangle= \delta_{\qq0} N_M A_\GG |\Psi_\nu^{\nu_+,\nu_-}\rangle$, where $A_\GG$ is still given by Eq. (\ref{p5:neweq:AG}). They form the $\left([N_M^{\nu_+}]_4,[N_M^{\nu_-}]_4\right)$ irrep of  U(4)$\times$U(4) (Young tableaux notation, see Ref.~\cite{ourpaper4}). 
For a fixed integer filling factor $\nu$, we found that the states with different Chern numbers $\nu_C$ are all degenerate in the chiral-flat limit \cite{ourpaper4}. 
In particular, at charge neutrality $\nu=0$, the U(4)$\times$U(4) multiplet of  $|\Psi_0^{\nu_+,\nu_-}\rangle$ with Chern number $\nu_C=\nu_+-\nu_-=0,\pm 2,\pm 4$ are exact degenerate ground states.  At nonzero fillings $\nu$, we cannot guarantee that the $\nu\neq 0$ eigenstates are the ground states.

\SZD{To do: add an appendix to explain the U(4) representation. Add a figure of Yang tableaux and use it to explain the excitations. }

In Ref. \cite{ourpaper4} we found that under a weak condition, the  eigenstates \cref{p5:eq:U(4)-GSMT,p5:eq:U(4)U(4)-GSMT} become the ground states of $H_I$ for  all integer fillings $-4\le\nu\le 4$ ($\nu$ even in Eq. \ref{p5:eq:U(4)-GSMT}). 
If the $\qq=\mathbf{0}$ component of the form factor $M_{m,n}^{\left(\eta\right)}\left(\mathbf{k},\mathbf{G}\right)$ is independent of $\kk$ for all $\mathbf{G}$'s, \ie
\begin{equation}\label{p5:eqn-condition-at-nuMT}
\text{Flat Metric Condition:}\;\; M_{m,n}^{\left(\eta\right)}\left(\mathbf{k},\mathbf{G}\right)=\xi(\mathbf{G})\delta_{m,n},
\end{equation} 
then all the states in \cref{p5:eq:U(4)-GSMT,p5:eq:U(4)U(4)-GSMT} become ground states of $H_I$ by an operator shift (\cref{p5:eq:shifted-HI}) \cite{kang_strong_2019,ourpaper3,ourpaper4} (see \cref{p5:exactgroundstatenonchirallimitRevappendix,p5:exactgroundstatechirallimitRevappendix}). We noted in Ref.~\cite{ourpaper1} that this flat metric condition is always true for $\mathbf{G}=\mathbf{0}$, for which $ M_{m,n}^{\left(\eta\right)}\left(\mathbf{k},0\right) = \delta_{mn}$ from wavefunction normalization. 
In Ref. \cite{ourpaper1} we have shown that, around the first magic angle,  $M_{m,n}^{\left(\eta\right)}\left(\mathbf{k},\mathbf{G}\right) \approx 0$ for $|\mathbf{G}| >\sqrt{3}k_\theta$ for $i=1,2$. 
Hence, the condition \cref{p5:eqn-condition-at-nuMT} is valid for all $\mathbf{G}$ with the exception of the 6 smallest nonzero $\mathbf{G}$ satisfying $|\mathbf{G}|=\sqrt{3}k_\theta$. 
Hence, the condition is largely valid, and our numerical analysis \cite{ourpaper1} confirms its validity for $\kk$ in a large part of the MBZ. 
The idea to impose a similar condition as Eq.~(\ref{p5:eqn-condition-at-nuMT}) first used by Kang and Vafek \cite{kang_strong_2019} to find the $\nu=\pm2$ ground state for their PSDH. 
Due to a slightly different U(4) symmetry, our U(4) FM states are different, but overlap with the Kang and Vafek ones in the chiral limit, as discussed in detail in Refs. \cite{ourpaper3, ourpaper4}.

We note that for $\nu\neq0$, the states in Eqs.~(\ref{p5:eq:U(4)-GSMT}) and~(\ref{p5:eq:U(4)U(4)-GSMT}) still remain the exact ground states if the flat metric condition Eq.~(\ref{p5:eqn-condition-at-nuMT}) is not violated too much \cite{ourpaper4,ourpaper6}. This is because they correspond to gapped insulator eigenstates \cite{ourpaper4,ourpaper6} when condition Eq.~(\ref{p5:eqn-condition-at-nuMT}) is satisfied, and the flat metric condition Eq.~(\ref{p5:eqn-condition-at-nuMT}) has to be largely broken to bring down another state into the ground state.  From now on, we ``call" \cref{p5:eq:U(4)-GSMT,p5:eq:U(4)U(4)-GSMT}  ground states of the system. 

Remarkably, as we will show in the rest of our paper below, one can analytically find a large series of excitations above the ground states  \cref{p5:eq:U(4)-GSMT,p5:eq:U(4)U(4)-GSMT}. 

Our excitations will be build out of acting with the band creation and annihilation operators on the ground states in \cref{p5:eq:U(4)-GSMT,p5:eq:U(4)U(4)-GSMT}. 
We first need to compute the commutators in the non-chiral Hamiltonian (see \cref{p5:ChargeCommutationRelationsAppendix} in particular \ref{p5:ChargeCommutationRelationsNonChiralAppendix})
{\small
\begin{align}
&[O_{\mathbf{q,G}}, c_{\mathbf{k},n,\eta,s}^\dag]=\sum_{m}\sqrt{V(\mathbf{G}+\mathbf{q})} M_{m,n}^{\left(\eta\right)}\left(\mathbf{k},\mathbf{q}+\mathbf{G}\right) c_{\mathbf{k+q},m,\eta,s}^\dag, \nonumber \\ & [O_{\mathbf{q,G}}, c_{\mathbf{k},n,\eta,s}]= -\sum_{m}\sqrt{V(\mathbf{G}+\mathbf{q})} M_{m,n}^{\left(\eta \right)* }\left(\mathbf{k},-\mathbf{q}-\mathbf{G}\right) c_{\mathbf{k-q},m,\eta,s}, 
\end{align}
}
where we have used the property $M_{m,n}^{\left(\eta \right)* }\left(\mathbf{k},-\mathbf{q}-\mathbf{G}\right) = M_{n,m}^{\left(\eta\right)}\left(\mathbf{k}-\mathbf{q},\mathbf{q}+\mathbf{G}\right) $ \cite{ourpaper3}. 
In the chiral limit, the same operators read in the Chern basis (see \cref{p5:ChargeCommutationRelationsChiralAppendix})
{\small
\begin{align}
&[O_{\mathbf{q,G}}, d_{\mathbf{k},e_Y,\eta,s}^\dag]= \sqrt{V(\mathbf{G}+\mathbf{q})} M_{e_Y}(\kk,\qq+\GG) d^\dagger_{\kk + \qq,e_Y,\eta,s}, \nonumber \\ &[O_{\mathbf{q,G}}, d_{\mathbf{k},e_Y,\eta,s}]=-\sqrt{V(\mathbf{G}+\mathbf{q})}M^*_{e_Y}(\kk,-\qq-\GG)  d_{\kk - \qq,e_Y,\eta,s}.
\end{align}}%
From these equations, we can obtain the commutators of $O_{\mathbf{-q,-G}}O_{\mathbf{q,G}}$ with the band electron creation operators in the non-chiral case as

\begin{align} \label{p5:neweq:OqGcommutator}
&\ [O_{\mathbf{-q,-G}}O_{\mathbf{q,G}}, c_{\mathbf{k},n,\eta,s}^\dag]=\sum_{m}P_{mn}^{\left(\eta\right)}\left(\mathbf{k},\mathbf{q}+\mathbf{G}\right)c_{\mathbf{k},m,\eta,s}^\dagger \nono\\ 
& + \sqrt{V(\mathbf{G}+\mathbf{q})} \sum_{m} \Big( M_{m,n}^{\left(\eta\right)}\left(\mathbf{k},\mathbf{q}+\mathbf{G}\right) c_{\mathbf{k+q},m,\eta,s}^\dag O_{-\mathbf{q,-G}} \nono \\
&\qquad + \sum_{m} M_{m,n}^{\left(\eta\right)}\left(\mathbf{k},-\mathbf{q}-\mathbf{G}\right) c_{\mathbf{k-q},m,\eta,s}^\dag O_{\mathbf{q,G}} \Big)
\end{align}
and in the first chiral limit in Chern basis as
\begin{align}\label{p5:chirallimitOqGcommutatorsMT}
& [O_{\mathbf{-q,-G}}O_{\mathbf{q,G}}, d_{\mathbf{k},e_Y,\eta,s}^\dag]=P\left(\mathbf{k},\mathbf{q}+\mathbf{G}\right)d_{\mathbf{k},e_Y,\eta,s}^\dag \nono \\
&+\sqrt{V(\mathbf{G}+\mathbf{q})}  \Big(M_{e_Y}\left(\mathbf{k},\mathbf{q}+\mathbf{G}\right) d_{\mathbf{k+q},e_Y,\eta,s}^\dag O_{-\mathbf{q,-G}} \nono  \\&+   M_{e_Y}\left(\mathbf{k},-\mathbf{q}-\mathbf{G}\right) d_{\mathbf{k-q},e_Y,\eta,s}^\dag O_{\mathbf{q,G}} \Big),
\end{align}
respectively.
Similar relations for $[O_{\mathbf{-q,-G}}O_{\mathbf{q,G}}, c_{\mathbf{k},n,\eta,s}]$ and $[O_{\mathbf{-q,-G}}O_{\mathbf{q,G}}, d_{\mathbf{k},e_Y,\eta,s}]$, where $M^{(\eta)}(\kk,\qq+\GG) \to M^{(\eta)*}(\kk,-\qq-\GG)$, are derived in \cref{p5:ChargeCommutationRelationsAppendix}.
The matrix factor $P$ is the convolution of the Coulomb potential and the form factor matrices. 
In the non-chiral case, $P$ is a matrix given by 
\begin{equation} \label{p5:neweq:P-matrix}
P_{mn}^{\left(\eta\right)}\left(\mathbf{k},\mathbf{q}+\mathbf{G}\right)  =
V(\mathbf{G}+\mathbf{q}) (M^{\left(\eta\right)\dagger }M^{\left(\eta\right)})_{mn} \left(\mathbf{k},\mathbf{q}+\mathbf{G}\right).
\end{equation}
In the first chiral limit, it is a number independent on $e_Y$:
\begin{align} \label{p5:PinchirallimitMT}
& P\left(\mathbf{k},\mathbf{q}+\mathbf{G}\right) ) = V(\mathbf{G}+\mathbf{q})|M_{e_Y}\left(\mathbf{k},\mathbf{q}+\mathbf{G}\right)|^2  \nonumber  \\ 
&\quad= V(\mathbf{G}+\mathbf{q})(\alpha^2_0(\mathbf{k,q+G})+\alpha^2_2(\mathbf{k,q+G})  ),
\end{align} 
where $ \alpha_0(\mathbf{k,q+G})$ and $\alpha_2(\mathbf{k,q+G})$ are the decomposition of the form factors in \cref{p5:eq:chiral-MqG1MT}. 
The above commutators and the existence of exact eigenstates \cref{p5:eq:U(4)U(4)-GSMT,p5:eq:U(4)-GSMT}, which are ground states with the flat metric condition \cref{p5:eqn-condition-at-nuMT}, allow for the computation of part of the low energy excitations with polynomial efficiency. We now show the summary of the computation for the bands of charge $+1$, $+2$ and neutral excitations. 
The charge $-1$,$-2$ excitations can be found in \cref{p5:charge-1excitationappendix,p5:charge-2excitations}, respectively. 

\section{Charge \texorpdfstring{$\pm1$}{+-1} excitations}

\SZD{To do: explain the reason we can do $\pm1$ excitations is that the single particle Hilbert spaces is spanned by $[(2N_M)^{(\nu+4)/2},1]_4$ for U(4) and is spanned by $([N_M^{\nu_+},1]_4,[N_M^{\nu_-},0]_4) \oplus ([N_M^{\nu_+},0]_4,[N_M^{\nu_-},1]_4)$ for U(4)$\times$U(4), which have very small dimensions (one or two) at each momentum.}

\subsection{Method to obtain the \texorpdfstring{$\pm1$}{+-1} excitation spectrum} \label{p5:sec:method-charge1}

\begin{figure}[t]
\centering
\includegraphics[width=\linewidth]{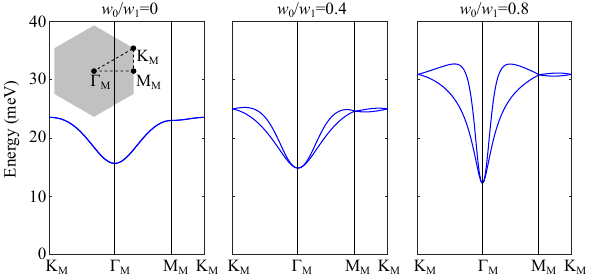}
\caption{Exact charge $\pm1$ excitations given by the simplified excitation matrix (\cref{p5:neweq:R-simple}) for three different $w_0/w_1$ at the twist angle $\theta=1.05^\circ$.
Here we change $w_0$ while keeping $w_1=110\mrm{meV}$ fixed.
Other parameters are given in \cref{p5:newapp:Ham}.
These excitations are exact at the charge neutrality point ($\nu=0$) for generic state and are exact at finite integer fillings if the flat metric condition is  satisfied. 
The charge +1 and $-1$ excitations are degenerate.
The exact charge $\pm1$ excitations obtained using the full excitation matrix (\cref{p5:eq:excitation-RmnMT}) without assuming the flat metric condition are given in \cref{p5:fig:E1appendix,p5:fig:E1appendix20} in \cref{newapp:plot1}. The charge gap in those cases shrinks considerably.
}
\label{p5:fig:E1main}
\end{figure}

To find  the charge one excitations (adding an electron into the system), we sum the commutators in \cref{p5:neweq:OqGcommutator} over $\qq, \GG$, and use the fact that the ground states in \cref{p5:eq:U(4)-GSMT,p5:eq:U(4)U(4)-GSMT} satisfy $(O_{\mathbf{q,G}}-A_{\GG} N_M\delta_{\mathbf{q},0})|\Psi \rangle =0$ for coefficient $A_\GG$ in Eq. (\ref{p5:neweq:AG}) in their corresponding limits. 
For any state $|\Psi\rangle$ in \cref{p5:eq:U(4)-GSMT,p5:eq:U(4)U(4)-GSMT}, we find:
\begin{equation}\label{p5:eq:excitation-HI}
\left[H_I-\mu N, c_{\mathbf{k},n,\eta,s}^\dag \right] |\Psi\rangle =\frac{1}{2\Omega_{\text{tot}}} \sum_{m} R_{mn}^\eta(\mathbf{k}) c_{\mathbf{k},m,\eta,s}^\dag|\Psi\rangle\ ,
\end{equation}
where $N$ is the electron number operator, and the matrix
{\small
\begin{align}\label{p5:eq:excitation-RmnMT}
&R_{mn}^\eta(\mathbf{k}) = \sum_{\GG\qq m'}V(\mathbf{G}+\mathbf{q}) M_{m'm}^{\left(\eta\right)*}\left(\mathbf{k},\mathbf{q}+\mathbf{G}\right) M_{m'n}^{\left(\eta\right)}\left(\mathbf{k},\mathbf{q}+\mathbf{G}\right) \nonumber \\
&\qquad + \sum_{\GG}2N_M A_{-\GG} \sqrt{V(\mathbf{G})} M_{m,n}^{\left(\eta\right)}\left(\mathbf{k},\mathbf{G}\right)-\mu\delta_{mn}.
\end{align} }%
We hence see that, if $|\Psi\rangle$  is  one of the $|\Psi_{\nu}^{\nu_+,\nu_-}\rangle$  Eq.~(\ref{p5:eq:U(4)U(4)-GSMT}) or one of the $|\Psi_\nu\rangle$ Eq.~(\ref{p5:eq:U(4)-GSMT}) eigenstates of $H_I$, then $c_{\mathbf{k},m,\eta,s}^\dag|\Psi\rangle\ $ can be recombined as eigenstates of $H_I$ with eigenvalues obtained by diagonalizing the $2\times2$ matrix $R_{mn}^\eta(\mathbf{k})$. 

In the nonchiral case, the eigenstates $|\Psi_\nu\rangle$ we found in Ref. \cite{ourpaper4} (and re-written in Eq. (\ref{p5:eq:U(4)-GSMT})) have both active bands $n=\pm1$ in each valley $\eta$ and spin $s$ either fully occupied or fully empty. 

In this case, we can consider two charge $+1$ states $c_{\mathbf{k},n,\eta,s}^\dag|\Psi\rangle$ ($n=\pm$) at a fixed $\kk$ in a fully empty valley $\eta$ and spin $s$. 
These two states then form a closed subspace with a $2\times2$ subspace Hamiltonian $R^\eta(\mathbf{k})$ defined by Eq. (\ref{p5:eq:excitation-RmnMT}). Diagonalizing the matrix $R^\eta(\mathbf{k})$ then gives the excitation eigenstates and excitation energies.
Furthermore, at $\nu=0$, the state $|\Psi_{\nu=0}\rangle$ in Eq. (\ref{p5:eq:U(4)-GSMT}) is the \emph{ground state} of the interaction Hamiltonian $H_I$ regardless of the flat metric condition Eq.~(\ref{p5:eqn-condition-at-nuMT}), and hence $c_{\mathbf{k},n,\eta,s}^\dag|\Psi_{\nu=0} \rangle$ always gives the charge excitation above the ground state.  

If we further assume the flat band condition Eq.~(\ref{p5:eqn-condition-at-nuMT}) (or its violation is small enough), all eigenstates $|\Psi_\nu\rangle$ become exact ground states and the second row of \cref{p5:eq:excitation-RmnMT} vanishes (see \cref{p5:charge1excitationBandNonChiralappendix}). 
Since the U(4) irrep of the ground state $|\Psi_\nu\rangle$ is $[(2N_M)^{(\nu+4)/2}]_4$, the U(4) irrep of the charge 1 excited state is given by $[(2N_M)^{(\nu+4)/2},1]_4$. 
A similar equation for the charge $-1$ excitations is derived in in \cref{p5:charge-1excitationappendix}, where we denote the excitation matrix as $\td{R}$.

As explained in \cref{p5:sec:exact-GS}, when the flat metric condition is satisfied, the second term in $R^\eta$ (\cref{p5:eq:excitation-RmnMT}) can be canceled by the chemical potential term (the third term), and thus we obtain a simplified expression for $R^\eta$ {\it independent} of $\nu$:
{\small
\begin{align}\label{p5:neweq:R-simple}
R_{mn}^{\eta}(\mathbf{k})
=& \sum_{\GG\qq m'}V(\mathbf{G}+\mathbf{q}) M_{m'm}^{\left(\eta\right)*}\left(\mathbf{k},\mathbf{q}+\mathbf{G}\right)  M_{m'n}^{\left(\eta\right)}\left(\mathbf{k},\mathbf{q}+\mathbf{G}\right).
\end{align}}%
It is worth noting that \cref{p5:neweq:R-simple} is {\it exact} for $\nu=0$ even without the flat metric condition Eq.~(\ref{p5:eqn-condition-at-nuMT}), because the coefficient $A_\GG$ (\cref{p5:neweq:AG}) in the second term of $R^\eta$ (\cref{p5:eq:excitation-RmnMT}) and the chemical potential in the third term of $R^\eta$ vanish at $\nu=0$. 

The simplified matrix $\td{R}^\eta$ for charge $-1$ excitation with the flat metric condition Eq.~(\ref{p5:eqn-condition-at-nuMT}) is the complex conjugation of $R^\eta$, \ie $\td{R}^\eta_{mn}(\kk)={R}^{\eta*}_{mn}(\kk)$. This shows that, the charge $-1$ excitations are degenerate with the charge $+1$ excitations if either $\nu=0$ or the flat metric condition Eq.~(\ref{p5:eqn-condition-at-nuMT}) is satisfied.

The charge $+1$ excitation dispersion determined by \cref{p5:neweq:R-simple} (which does not depend on $\nu$) is plotted in \cref{p5:fig:E1main}.

The parameters used in the calculation to obtain the spectrum are given in \cref{p5:newapp:Ham}.
We find that, with the flat metric condition imposed, the charge $\pm1$ excitation (\cref{p5:fig:E1main}) is gapped, and the minimum is at the $\Gamma$ point, with a large dispersion velocity. 
The exact charge $\pm1$ excitations at different fillings obtained using the full $R^\eta$ matrix (\cref{p5:eq:excitation-RmnMT}) of realistic parameters (which break the flat metric condition Eq.~(\ref{p5:eqn-condition-at-nuMT})) are given in \cref{p5:fig:E1appendix,p5:fig:E1appendix20} in \cref{newapp:plot1}.
 
The degeneracy of the excitation spectrum depends on the filling $\nu$ of the ground state. 
In the nonchiral-flat U(4) limit, $R^{\eta}$ does not depend on spin, and $R^+$, $R^-$ have the same eigenvalues because they are related by the symmetry $C_{2z}P$, where $P$ is a single-body unitary PH symmetry (\cref{p5:gaugefixingRevappendix}) \cite{song_all_2019,ourpaper3,ourpaper2}.
Thus charge $+1$ excitations in different valley-spin flavors have the same energy. 
For the state $\ket{\Psi_\nu}$ (\cref{p5:eq:U(4)-GSMT}), the $+1$ excitations in the empty $(4-\nu)/2$ spin-valley flavors are degenerate.
Correspondingly, $-1$ excitations in the occupied $(4+\nu)/2$ spin-valley flavors are also degenerate.

In the (first) chiral-flat limit, and with the flat metric condition Eq.~(\ref{p5:eqn-condition-at-nuMT}) (or at $\nu=0$ without (\ref{p5:eqn-condition-at-nuMT})), the expression for the charged excitations in the Chern basis $d^\dag_{\kk,e_Y,\eta,s}|\Psi_{\nu}^{\nu_+,\nu_-}\rangle$ becomes diagonal and independent of $e_Y$ (see \cref{p5:charge1excitationChiralLimitappendix} for the chiral-flat limit without the flat metric condition Eq.~(\ref{p5:eqn-condition-at-nuMT})):
\begin{align}\label{p5:eq:excitation-HIMT}
\left[H_I-\mu N, d_{\mathbf{k},e_Y,\eta,s}^\dag \right] |\Psi\rangle =& \frac{1}{2\Omega_{\text{tot}}}R_0(\kk)  d_{\mathbf{k},e_Y,\eta,s}^\dag|\Psi\rangle, \nonumber \\ 
 R_0(\kk) =&   \sum_{\GG,\qq} P(\kk, \qq+\GG),
\end{align}
provided that the Chern band $e_Y (=\pm1)$ in valley $\eta$ and spin $s$ is fully empty and $P(\kk, \qq+\GG)$ given in \cref{p5:PinchirallimitMT}.
We obtain
\begin{equation} \label{p5:R0InTheChiralLimitWithFlatConditionMT}
R_0(\kk)=  
\sum_{\GG,\qq} V(\GG+\qq) [\alpha_0 (\kk,\qq+\GG)^2 +\alpha_2 (\kk,\qq+\GG)^2]  .
\end{equation}
The spectrum at the magic angle is shown in Fig. \ref{p5:fig:E1main}. 
The U(4)$\times$U(4) irrep of the charge +1 excited states with $e_Y=1$ and $e_Y=-1$ are given by $\left([N_M^{\nu_+},1]_4,[N_M^{\nu_-}]_4\right)$ and $\left([N_M^{\nu_+}]_4,[N_M^{\nu_-},1]_4\right)$, respectively. The charge $-1$ excitation details can be found in \cref{p5:charge-1excitationappendix}.

Since in the chiral-flat limit the scattering matrix $R_0(\kk)$ is identity in the $e_Y$ space, the excitation has degeneracy in addition to the valley-spin degeneracies. 
For a state in \cref{p5:eq:U(4)U(4)-GSMT} with filling $\nu$, the charge $+1$ and $-1$ excitations have degeneracies $4-\nu$ and $4+\nu$, respectively.

\subsection{Bounds on the charge \texorpdfstring{$\pm1$}{+-1} excitation gap}
\label{szdsec:gapbound}

In this subsection, we will focus on the charge neutrality point ($\nu=0$), where the second and third terms in $R^\eta(\mathbf{k})$ (\cref{p5:eq:excitation-RmnMT}) vanish, and nonzero integer fillings $\nu=\pm1,\pm2,\pm3$ with the flat metric condition Eq.~(\ref{p5:eqn-condition-at-nuMT}) such that the second and third terms in $R^\eta(\mathbf{k})$ cancel each other.
In these cases $R^\eta(\mathbf{k})$ is a positive semidefinite matrix and hence has non-negative eigenvalues.  
We are able to obtain some analytical bounds for the gap of the $\pm1$ excitation.
Detail calculations are given in \cref{p5:charge1excitationSpectrumPropertiesappendix}. Since charge $\pm1$ excitations in this case are degenerate, our conclusion below for charge +1 excitations also apply to charge -1 excitations.

We rewrite the $R^\eta(\mathbf{k})$ matrix as  $R_{mn}^\eta(\mathbf{k}) =     (M^{\left(\eta\right)\dagger}(\kk) V  M^{\left(\eta\right)}(\kk))_{mn}$, where now $  M^{\left(\eta\right)}(\kk)$ with given $\eta$ and $\kk$ is a matrix of the dimension $2N_M \cdot N_{\GG} \times 2$ (with $2$ because we are projecting into the two active TBG bands).
$N_M$ is number of moir\'e unit cells, $N_\GG$ is the number of plane waves (MBZs) taken into consideration. 
By separating the $\{\qq, \GG\}=0$ contribution, and using Weyl's inequalities we find in \cref{p5:charge1excitationSpectrumPropertiesappendix} that the energies of the excited states are  $\ge \frac12 V(\qq=0)/\Omega_{\text{tot}}$.
The bound $\frac12 V(\qq=0)/\Omega_{\text{tot}}$ is small but nonzero for large but finite $\Omega_{\rm tot}$.
This shows that the states $c_{\mathbf{k},n,\eta,s}^\dag|\Psi\rangle$ are not exactly degenerate to the ground state $|\Psi\rangle$ (note that we did not prove these are the unique ground states). 

The excited states of the PSDH appears to give rise to finite gap charge 1 excitations.  The largest gap happens in the atomic limit or a material, where $\langle{ u_m(\kk+\qq) } | u_n(\kk) \rangle = \delta_{mn} $, for which $R_{mn} =\delta_{mn} \sum_{\qq,\GG} V(\qq + \GG) = \delta_{mn} \Omega_{\rm tot} V(\bf{r}=0)$. 
Hence the gap is $\frac12 V(\rr=0)$.
Away from the atomic limit, the gap is reduced, but will generically remain finite.  We now give an argument for this. 
Since we know that TBG is far away from an atomic limit - the bands being topological, we expect a reduction in this gap. We perform a different decomposition of the matrix $R_{mn}^\eta$: we separate it into $\GG=\mathbf{0}$ and $\GG\ne0$ sums (see \cref{p5:charge1excitationGapappendix}). 
The $\GG\ne0$ part, besides being negligible for $|\GG| \ge \sqrt{3}k_\theta$ \cite{ourpaper1}, is also positive semidefinite, and the eigenvalues of $R_{mn}^\eta$ are bounded by (and close to) the $\GG=0$ part:
\begin{equation}
R_{mn}^\eta(\mathbf{k}) \ge   \sum_\qq V(\qq) M_{m',m}^{\left(\eta\right)*}\left(\mathbf{k},\mathbf{q}\right) M_{m',n}^{\left(\eta\right)}\left(\mathbf{k},\mathbf{q}\right),
\end{equation}
where $\qq$ is summed over the MBZ, and the inequality means that the eigenvalues of the left hand side are equal to or larger than the eigenvalues of the right hand side. 
We then re-write the right hand side as  $\sum_\qq V(\qq) (\delta_{mn} - 	\mathfrak{G}_\eta^{mn}(\kk,\qq))$, where we call the {\it positive semi-definite} matrix $\mathfrak{G}^{mn}(\kk,\qq)$ the generalized ``quantum geometric'', whose trace is the generalized Fubini-Study metric. 
For small momentum transfer $\qq$, we can show that $\mfk{G}^{mn}(\kk,\qq) = \sum_{ij} q_{i} q_{j} \mathfrak{G}^{mn}_{ij}(\mathbf{k}) + \mcl{O}(q^3)$.
where $\mathfrak{G}^{mn}_{ij}(\mathbf{k})$ is the conventional quantum geometric tensor (and the Fubini-Study metric) \cite{xie_superfluid_2020,titus2013metric} defined by

\begin{align}
\mathfrak{G}^{mn}_{ij}(\mathbf{k}) &=  \sum_{a,b=1}^N \partial_{k_i} u^*_{a,m}(\mathbf{k}) \bigg(\delta_{a,b} - \sum_{l\in \mcl{B}} u_{a,l}(\mathbf{k}) u^*_{b,l}(\mathbf{k}) \bigg) \nono\\
& \times \partial_{k_j}u_{b,n}(\mathbf{k}),
\end{align}
in which $m,n\in\mcl{B}$ are energy band indices and $i,j$ are spatial direction indices of the orthonormal vectors $u_m(\mathbf{k})$ in a $N$ dimensional Hilbert space, with $\mathbf{k}$ being the momentum (or other parameter). The $\mfk{G}^{mn}(\kk,\qq)$ tensor quantifies the distance between two eigenstates in momentum space. 

Generically, we expect \cite{xie_superfluid_2020} that the  inner product between two functions at $\kk$ and $\kk +\bf{q}$ to fall off as $\bf{q}$ increases, leaving a finite term in $R_{mn}^\eta(\mathbf{k})$, the electron gap, at every $\bf{k}$.
In trivial bands in the atomic limit, the positive semi-definite matrix $\mfk{G}^{mn}(\kk,\qq)$ reaches its theoretical lower bound 0 and hence the charge 1 gap is maximal.
In topological bands, such as TBG, the quantum metric has a lower bound and hence the charge 1 gap is reduced.

\section{Charge neutral excitations}

\SZD{To do: explain the reason we can do neutral excitations with a single particle-hole pair is that the corresponding Hilbert spaces is spanned by $[(2N_M)^{(\nu+4)/2}-1,1]_4$ for U(4) and is spanned by $([N_M^{\nu_+}-1,1]_4,[N_M^{\nu_-}]_4) ~ \oplus ([N_M^{\nu_+}-1]_4,[N_M^{\nu_-},1]_4) ~ \oplus ([N_M^{\nu_+},0]_4,[N_M^{\nu_-}-1,1]_4) ~ \oplus
([N_M^{\nu_+}+1]_4,[N_M^{\nu_-}-1]_4)
$ for U(4)$\times$U(4), which have very small dimensions ($\sim N_M$) at each momentum.
}

\SZD{To do: point out that we prove that ALL the Goldstone modes are in the single particle-hole pair Hilbert space. (i) We know the number of Goldstone modes from group theory counting. (ii) We can show that the number of zero modes in the single particle-hole pair sector equals to the number of Goldstone modes. (iii) Andrei already proves this for U(4)$\times$U(2). I find a similar proof for U(4). }

\SZD{To do: obtain the excitation spectrum at $\nu=-1,-3$ in the nonchiral-flat limit and relate it to the phase transition in paper-VI.}

\SZD{To do: discuss the number of Goldstone modes if the symmetry is broken to U(2)$\times$U(2).}

\subsection{Method to obtain charge neutral excitations}

To obtain the charge neutral excitations, we choose the natural basis $c_{\kk+\pp, m_2, \eta_2, s_2}^\dagger c_{\kk, m_1, \eta_1, s_1}|\Psi \rangle$, where $|\Psi\rangle$ is any of the exact ground states and/or eigenstates in \cref{p5:eq:U(4)U(4)-GSMT,p5:eq:U(4)-GSMT} and $\pp$ is the momentum of the excited state. 
The scattering matrix of these basis can be solved as easily as a one-body problem, despite the fact that \cref{p5:eq:U(4)U(4)-GSMT,p5:eq:U(4)-GSMT} hold a thermodynamic number of electrons.  The details are given in \cref{p5:NeutralExcitationappendix}.
For $\ket{\Psi}$ being a state in \cref{p5:eq:U(4)-GSMT}, the scattering of $c_{\kk+\pp, m_2, \eta_2, s_2}^\dagger c_{\kk, m_1, \eta_1, s_1}|\Psi \rangle$ by the interaction is: 
\begin{align}
& \left[H_I-\mu N,c_{\kk+\pp, m_2, \eta_2, s_2}^\dagger c_{\kk, m_1, \eta_1, s_1} \right] |\Psi_\nu\rangle \nono\\
=& \frac{1}{2\Omega_{\text{tot}}}\sum_{m,m'}
    \sum_\qq S^{(\eta_2,\eta_1)}_{m,m';m_2,m_1}(\mathbf{k}+\qq,\kk;\pp)\nono\\
& \qquad \times c_{\kk+\pp+\qq, m, \eta_2, s_2}^\dagger 
    c_{\kk+\qq, m', \eta_1, s_1}|\Psi_\nu\rangle, 
\end{align}
{\small
\begin{align}\label{p5:eq:neutralexcitation-HIMT}
& S^{(\eta_2,\eta_1)}_{m,m';m_2,m_1}(\mathbf{k}+\qq,\kk;\pp) \\
=& \delta_{\qq,0}(\delta_{m,m_2} \widetilde{R}_{m'm_1}^{\eta_1}(\mathbf{k}) +\delta_{m',m_1} R_{mm_2}^{\eta_2}(\mathbf{k}+\pp) ) \nono\\
-& 2\sum_\GG V(\mathbf{G}+\mathbf{q})   M_{m,m_2}^{\left(\eta_2\right)}\left(\mathbf{k}+\pp,\mathbf{q}+\mathbf{G}\right)  M_{m',m_1}^{\left(\eta_1\right)*}\left(\mathbf{k},\mathbf{q}+ \mathbf{G}\right), \nono
\end{align}}%
where $R_{mn}^\eta(\mathbf{k})$ (\cref{p5:eq:excitation-RmnMT}) and $\widetilde{R}_{mn}^\eta(\mathbf{k})$ are the charge $\pm1$ excitation matrices. 
A valley-spin flavor in $|\Psi_\nu\rangle$ (\cref{p5:eq:U(4)-GSMT}) is either fully occupied or fully empty, thus $\{\eta_1, s_1\}$ belongs to the valley-spin flavors which are fully occupied, while  $\{\eta_2, s_2\}$ belongs to the valley-spin flavors which are not occupied. \cref{p5:eq:neutralexcitation-HIMT} shows that  the neutral excitation scattering matrix is a sum of the two single-particle energies ($\delta_{m,m_2} \widetilde{R}_{mm_1}^{\eta_1}(\mathbf{k}_1) +\delta_{m,m_1} R_{mm_2}^{\eta_2}(\mathbf{k}_2)$) plus an interaction term. 
By translation invariance, the scattering preserves the total momentum $\mathbf{p}$.
The spectrum of the charge neutral excitations at each $\pp$ is a diagonalization problem of a matrix of the dimension $4N_M\times 4N_M$, where the left and right indices are $(\kk+\qq,m,m')$ and $(\kk,m_2,m_2)$, respectively.

\begin{figure}[t]
\centering
\includegraphics[width=\linewidth]{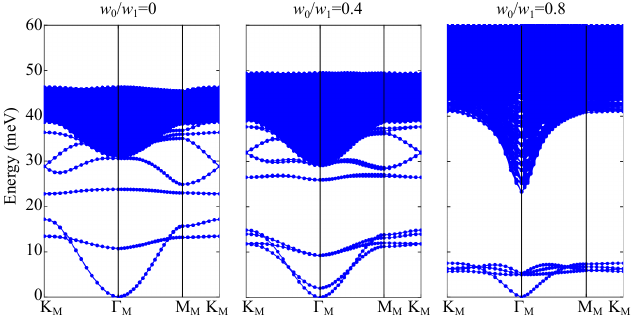}
\caption{
Exact charge neutral excitations with the flat metric condition being imposed for three different $w_0/w_1$ at the twist angle $\theta=1.05^\circ$.
Here we change $w_0$ while keeping $w_1=110\mrm{meV}$ fixed.
Other parameters are given in \cref{p5:newapp:Ham}.
These excitations are exact at the charge neutrality point ($\nu=0$) for generic states and are exact at finite integer fillings if the flat metric condition is satisfied. 
The exact charge neutral excitations at different fillings without imposing flat metric condition are given in \cref{p5:fig:E0appendix,p5:fig:E0appendix20} in \cref{newapp:plot0}. Note the softening of further Goldstone modes from finite to zero $w_0$, reflecting the symmetry enhancement of the first chiral limit. The continuum above the Goldstone modes is fundamentally made of independent particle-hole excitations}
\label{p5:fig:E0main}
\end{figure}

The excitation spectrum with the flat metric condition Eq.~(\ref{p5:eqn-condition-at-nuMT}) being imposed, \ie with the $R^\eta$ (\cref{p5:eq:excitation-RmnMT}) being replaced by the simplified \cref{p5:neweq:R-simple}, is shown in \cref{p5:fig:E0main}.
As explained in \cref{p5:sec:method-charge1}, the simplified charge $\pm1$ matrices $R$ and $\td{R}$ do not depend on the filling $\nu$.
Thus the obtained charge neutral excitation dispersion also do not depend on $\nu$.
\cref{p5:fig:E0main} is exact for $\nu=0$ even when the flat metric condition is not satisfied since \cref{p5:neweq:R-simple} is exact for $\nu=0$. 
The exact charge neutral excitations at different fillings without imposing the flat metric condition Eq.~(\ref{p5:eqn-condition-at-nuMT}) are given in \cref{p5:fig:E0appendix,p5:fig:E0appendix20} in \cref{newapp:plot0}. 

It is worth noting that, in the \cref{p5:fig:E0main,p5:fig:E0appendix,p5:fig:E0appendix20} we just plot the eigenvalues of the scattering matrix \cref{p5:eq:neutralexcitation-HIMT}, which does not assume any information of the occupied valley-spin flavors in the ground state. 
In practice, for a given ground state $\ket{\Psi}$, the spectrum branch annihilating (creating) electrons in empty (occupied) states does not exist. 

\subsection{Goldstone modes}\label{szdsec:GM}

Solving \cref{p5:eq:neutralexcitation-HIMT} provides us with the expression for the neutral excitations at momentum $\mathbf{p}$ on top of the TBG ground states, including the Goldstone mode, whose dispersion relation can be obtained in terms of the quantum geometry factors of the TBG. 
In general, the scattering matrix is not guaranteed to be positive semi-definite, and negative energy would imply instability of the ground states.
However, in a large (physical) range of parameters (\cref{p5:newapp:Ham}) of TBG at the twist angle $\theta=1.05^\circ$, we find that, as shown in \cref{p5:fig:E0main,p5:fig:E0appendix,p5:fig:E0appendix20}, the energies of charge neutral excitations of the exact ground states $|\Psi_\nu\rangle$ in Eq. (\ref{p5:eq:U(4)-GSMT}) in the nonchiral-flat limit and $|\Psi_{\nu}^{\nu_+,\nu_-}\rangle$ in Eq. (\ref{p5:eq:U(4)U(4)-GSMT}) in the chiral-flat limit are non-negative, implying these are indeed stable ground states. 
As shown in \cref{p5:fig:E1appendix,p5:fig:E0appendix} and discussed in \cref{newapp:plot1,newapp:plot0}, strong (first) chiral symmetry breaking may lead to an instability to a metallic phase.

\begin{table}[t]
\centering
\begin{tabular}{c|c|c}
\hline
Little group & Number of GMs & Ground states \\
\hline 
U(4)$\times$U(4) & 0 & $\ket{\Psi_{0}^{4,0}}$ \\
U(1)$\times$U(3)$\times$U(4) &  3  & $\ket{\Psi_{-3}^{1,0}}$, $\ket{\Psi_{-1}^{3,0}}$  \\
U(2)$\times$U(2)$\times$U(4) &  4  & $\ket{\Psi_{-2}^{2,0}}$ \\
U(1)$\times$U(3)$\times$U(1)$\times$U(3) & 6 & $\ket{\Psi_{-2}^{1,1}}$, $\ket{\Psi_0^{3,1}}$ \\
U(2)$\times$U(2)$\times$U(1)$\times$U(3) & 7 & $\ket{\Psi_{-1}^{2,1}}$ \\
U(2)$\times$U(2)$\times$U(2)$\times$U(2) & 8 & $\ket{\Psi_{0}^{2,2}}$ \\
\hline
\end{tabular}
\caption{The little groups (remaining symmetry subgroups) and the number of Goldstone modes (denoted by GMs in the table) of the ground states $\ket{\Psi_{\nu}^{\nu_+,\nu_-}}$ in the (first) chiral-flat U(4)$\times$U(4) limit.
Only $\nu\le0$ states are tabulated since the symmetry and Goldstone modes of $\nu>0$ states are same as the $\nu<0$ states since they are related by the many-body charge-conjugation operator (\cref{p5:projectedinteractionRevappendix}) \cite{ourpaper3}.  
Only states with $\nu_+\ge \nu_-$ are tabulated since $\ket{\Psi_{\nu}^{\nu_+,\nu_-}}$ and $\ket{\Psi_{\nu}^{\nu_-,\nu_+}}$ have the equivalent little groups upon interchanging of the two U(4)s, and thus have the same number of Goldstone modes.}
\label{p5:tab:Goldstone-U4xU4}
\end{table}

\begin{table}[t]
\centering
\begin{tabular}{c|c|c}
\hline
Little group & Number of GMs & Ground states \\
\hline
U(1)$\times$U(3) & 3 & $\ket{\Psi_{-2}}$, $\ket{\Psi_{2}}$ \\
U(2)$\times$U(2) & 4 & $\ket{\Psi_0}$ \\ 
\hline
\end{tabular}
\caption{The little groups (remaining symmetry subgroups) and the number of Goldstone modes (GMs) of the ground states $\ket{\Psi_{\nu}}$ in the nonchiral-flat U(4) limit.}
\label{p5:tab:Goldstone-U4}
\end{table}

In \cref{p5:tab:Goldstone-U4,p5:tab:Goldstone-U4xU4} we have tabulated the little group (defined as the remaining symmetry subgroup of the state) and the number of Goldstone modes for each ground state in \cref{p5:eq:U(4)-GSMT,p5:eq:U(4)U(4)-GSMT}.
As examples, here we only derive the little groups and number of Goldstone modes for $\ket{\Psi_{-2}^{1,1}}$ (\ref{p5:eq:U(4)U(4)-GSMT}) and $\ket{\Psi_{-2}}$ (\ref{p5:eq:U(4)-GSMT}).
The little groups and Goldstone modes for other states can be obtained by the same method.
First we consider the ground state $\ket{\Psi_{-2}^{1,1}}$ in the (first) chiral-flat U(4)$\times$U(4) limit, which has vanishing total Chern number.
Recall that the U(4)$\times$U(4) irrep of $\ket{\Psi_{-2}^{1,1}}$ is $([N_M^1]_4,[N_M^1]_4)$. 
In each of the $e_Y=\pm1$ sectors, only one U(4) spin-valley flavor is occupied.
Hence the little group of the state $\ket{\Psi_{-2}^{1,1}}$ in each $e_Y$ sector is U(1)$\times$U(3), where the U(1) is the phase rotation in the occupied flavor and the U(3) is the unitary rotations within the 3 empty flavors. 
Thus, the total little group of the state $\ket{\Psi_{-2}^{1,1}}$ is U(1)$\times$U(3)$\times$U(1)$\times$U(3), which has the rank (number of independent generators) 20. Since the Hamiltonian has a symmetry group U(4)$\times$U(4) which has rank $32$, we find the number of broken symmetry generators to be 32-20=12. On the other hand, since all the Goldstone modes we derived are quadratic (similar to the SU(2) ferromagnets, see Eq. (\ref{p5:eq:GMquadratic})), it is known that \cite{nielsen1976} the number of Goldstone modes is equal to $1/2$ of the number of broken generators, namely, $12/2=6$. This is because a quadratic Goldstone mode is always a complex boson, which is equivalent to two real boson degrees of freedom corresponding to 2 broken generators.  

Next, we consider the ground state $\ket{\Psi_{-2}}$ in the nonchiral-flat U(4) limit. 
Since the U(4) irrep of $\ket{\Psi_{-2}}$ is $[2N_M]_4$, only one U(4) spin-valley flavor is occupied.
Thus the little group of $\ket{\Psi_{-2}}$ is U(1)$\times$U(3), where the U(1) is within the occupied flavor and the U(3) is within the 3 empty flavors. 
Hence the number of broken generators is $16-10=6$, where 16 and 10 are the ranks of U(4)$\times$U(4) and U(1)$\times$U(3), and the number of (quadratic) Goldstone modes is 6/2=3.

In the above paragraph we have shown that state $\ket{\Psi_{-2}^{1,1}}$ in the chiral-flat limit has three more Goldstone modes than $\ket{\Psi_{-2}}$ in the nonchiral-flat limit, although their wavefunctions are identical. 
This is because, if we slightly go away from the (first) chiral-flat limit towards the nonchiral-flat limit, \ie take the parameter $0<w_0\ll w_1$, some branches of the Goldstone modes will be gapped by a finite $w_0$, as shown in \cref{p5:fig:E0main,p5:fig:E0appendix,p5:fig:E0appendix20}.

The number of Goldstone modes can also be obtained by examining the scattering matrix in Eq. (\ref{p5:eq:neutralexcitation-HIMT}). 
Here we take $\ket{\Psi_{-2}^{1,1}}$ as an example. 
As discussed in \cref{p5:sec:goldstonestiffness}, in the first chiral limit, the state $d^\dag_{\mathbf{k+p},e_{Y2},\eta_2,s_2}  d_{\mathbf{k},e_{Y1},\eta_1,s_1} |\Psi_{-2}^{1,1}\rangle$ will be scattered to $d^\dag_{\mathbf{k'+p},e_{Y2},\eta_2,s_2}  d_{\mathbf{k'},e_{Y1},\eta_1,s_1}|\Psi_{-2}^{1,1}\rangle$ through the scattering matrix $S_{e_{Y2},e_{Y1}}(\kk',\kk;\pp)$, which does not depend on $\eta_1,s_1,\eta_2,s_2$, and $S_{e_{Y2},e_{Y1}}(\kk',\kk;\mathbf{0})$ has an exact zero state for $e_{Y2}=e_{Y1}$ (\cref{p5:Goldstoneappendix}).
Now we count the number of Goldstone modes on top of $\ket{\Psi_{-2}^{1,1}}$ using this property of scattering matrix. 

Suppose the occupied flavors in $\ket{\Psi_{-2}^{1,1}}$ are $\{e_Y,\eta,s\}=\{+1,+,\up\},\{-1,+,\up\}$.
Then, for the state $d^\dag_{\mathbf{k+p},e_{Y1},\eta_2,s_2}  d_{\mathbf{k},e_{Y1},\eta_1,s_1} |\Psi_{-2}^{1,1}\rangle$ to be non-vanishing, $\{e_{Y_1},\eta_1,s_1\}$ can only take the values in the two $e_Y$-valley-spin flavors $\{+1,+,\up\},\{-1,+,\up\}$, and $\{\eta_2,s_2\}$ can only take values in the other three valley-spin flavors in each $e_Y$ sector. 
There are in total 6 non-vanishing channels.
Since each channel has an zero mode given by the zero of $S_{e_{Y1},e_{Y1}}(\kk',\kk;\mathbf{0})$, there are 6 Goldstone modes, consistent with the group theory analysis in \cref{p5:tab:Goldstone-U4xU4}.

\subsection{Exact Goldstone mode and its stiffness in the (first) chiral-flat \texorpdfstring{U(4)$\times$U(4)}{U4xU4} limit}\label{p5:sec:goldstonestiffness}

In the first chiral limit, we are able to obtain the Goldstone modes analytically.
We pick the basis as $d^\dag_{\mathbf{k+p},e_{Y2},\eta_2,s_2}  d_{\mathbf{k},e_{Y1},\eta_1,s_1} |\Psi_{\nu}^{\nu_+,\nu_-}\rangle$, where the valley-spin flavor $\{\eta_1,s_1\}$ with Chern band basis $e_{Y1}$ is fully occupied and the valley-spin flavor  $\{\eta_2,s_2\}$  and Chern band basis $e_{Y2}$ is fully empty.  
The PSDH scatters the basis to 
{
\begin{equation}
\sum_{\kk'} S_{e_{Y2},e_{Y1}}(\kk',\kk;\pp)  d^\dag_{\mathbf{k'+p},e_{Y2},\eta_2,s_2}  d_{\mathbf{k'},e_{Y1},\eta_1,s_1}|\Psi_{\nu}^{\nu_+,\nu_-}\rangle,
\end{equation}}%
where the scattering matrix $S$ does not depend on $\eta_1$, $\eta_2$, $s_1$, $s_2$. 
The simple commutators between $O_{-\qq,-\GG} O_{\qq,\GG}$ and fermion creation and annihilation operators in the chiral limit (\cref{p5:chirallimitOqGcommutatorsMT}) lead to a simple scattering matrix. 
We here focus on the $e_{Y1}=e_{Y2}$, $\pp=0$ channel.
For generic $\nu=0$ states and the $\nu=\pm1,\pm2,\pm3$ states with flat metric condition (\cref{p5:eqn-condition-at-nuMT}), we have
{\small
\begin{align}\label{p5:eq:neutralexcitationflatband-HIMT}
& S _{e_{Y};e_{Y}}(\kk+\qq,\kk;0) =\nono\\
& 2\delta_{\qq,0}\sum_{\GG,\qq'} V(\GG+\qq') 
    [\alpha_0 (\kk,\qq'+\GG)^2 +\alpha_2 (\kk,\qq'+\GG)^2 ]  \nono\\
&  -2\sum_\GG V(\mathbf{G}+\mathbf{q}) [\alpha_0(\mathbf{k,q+G})^2 + \alpha_2(\mathbf{k,q+G})^2 ].
\end{align}}%
The general expression of $S_{e_{Y2};e_{Y1}}(\mathbf{k}+\qq,\kk;\pp)$ for all channels without imposing the flat metric condition is given in \cref{p5:NeutralExcitationChiralappendix}.
We first show the presence of an exact zero eigenstate of \cref{p5:eq:neutralexcitationflatband-HIMT} by remarking that the scattering matrix $S _{e_{Y};e_{Y}}(\kk+\qq,\kk;\mathbf{0})$ satisfies (irrespective of $\eta_{1,2},s_{1,2}$):
\begin{equation}
\sum_{\qq} S _{e_{Y};e_{Y}}(\kk+\qq,\kk;\mathbf{0}) =0.
\end{equation} 
This guarantees that the rank of the scattering matrix is not maximal, and that there is at least one \emph{exact}  zero energy eigenstate, with  equal  amplitude on every state in the Hilbert space: $\sum_{\kk} d^\dag_{\mathbf{k},e_{Y},\eta_2,s_2}  d_{\mathbf{k},e_{Y},\eta_1,s_1}| \Psi_{\nu}^{\nu_+,\nu_-}\rangle$. 
More details are given in \cref{p5:Goldstoneappendix,p5:zeroenergyneutralappendix}. 
The U(4)$\times$ U(4) multiplet of this state is also at zero energy. 
Moreover, the scattering matrix  $S_{e_{Y};e_{Y}}(\mathbf{k+q},\kk;\mathbf{0})$ is positive semi-definite.
The details of this proof can be found in \cref{p5:zeroenergyneutralappendix}.

\begin{figure}[t]
\centering
\includegraphics[width=.9\linewidth]{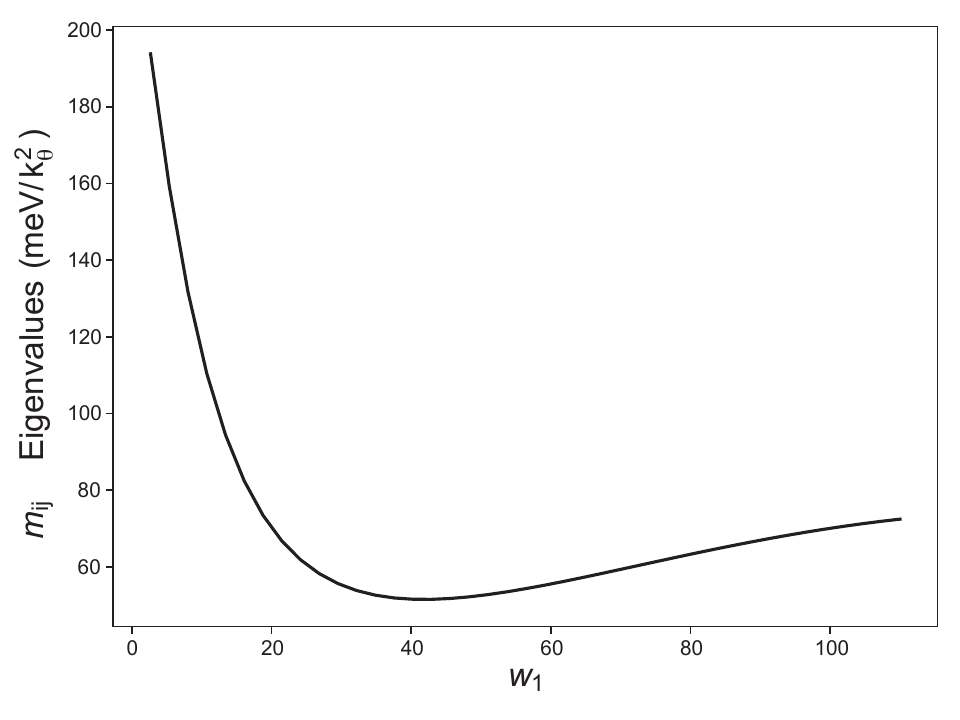}
\caption{The eigenvalues of the mass tensor of the Goldstone mode in the first chiral limit (\ref{p5:eq:mass_tensor}). Here $w_1$ is in units of meV.
} 
\label{p5:fig:mij_eigenvalues}
\end{figure}

Since the $\pp=\mathbf{0}$ state has zero energy, for small $\pp$, by continuity, there will be low-energy states in the neutral continuum. By performing a k$\cdot$p perturbation in the $\pp=0$ states in \cref{p5:zeromomentumgoldstone}, one can compute the dispersion of the low-lying states. 
Full details are given in \cref{p5:GoldstoneStiffnessappendix}. 
In the chiral limit, and imposing the flat metric condition \cref{p5:eqn-condition-at-nuMT} we find, by using $\alpha_a(\kk,\qq+\GG) = \alpha_a(-\kk,-\qq-\GG)$ for $a=0,2$ and as expected for the Goldstone of a FM, the linear term in $\pp$ vanishes and
\beq\label{p5:eq:GMquadratic}
E_{\text{Goldstone}} (\mathbf{p}) = \frac{1}{2} \sum_{ij=x,y} m_{ij} p_i p_j\ ,
\eeq
to second order in $\pp$. 
We find the Goldstone stiffness
{\scriptsize
\begin{align}
&m_{ij} = \frac{1}{2\Omega_{\text{tot}}}\sum_{\kk, \qq,\GG} V(\GG+\qq) 
    \Big[\alpha_0 (\kk,\qq+\GG) \partial_{k_i}\partial_{k_j} \alpha_0 (\kk,\qq+\GG) +\nono \\
&\alpha_2 (\kk,\qq+\GG) \partial_{k_i}\partial_{k_j} \alpha_2 (\kk,\qq+\GG) + 2\partial_{k_i} \alpha_0 (\kk,\qq+\GG) \partial_{k_j} \alpha_0 (\kk,\qq+\GG)\nono\\
&+2\partial_{k_i} \alpha_2 (\kk,\qq+\GG) \partial_{k_j} \alpha_2 (\kk,\qq+\GG)\Big]. \label{p5:eq:mass_tensor}
\end{align}}%
Since $C_{3z}$ symmetry is unbroken in the ground states in Eq. (\ref{p5:eq:U(4)U(4)-GSMT}), an isotropic mass tensor $m_{ij}\propto \delta_{ij}$ is expected. The eigenvalues of $m_{ij}$ with different values of $w_1$ are plotted in \cref{p5:fig:mij_eigenvalues}.

\section{Charge \texorpdfstring{$\pm2$}{+-2} excitations}

\SZD{To do: explain the reason we can do $\pm2$ excitations is that the corresponding Hilbert spaces is spanned by $[(2N_M)^{(\nu+4)/2},2]_4$ for U(4) and is spanned by 
$([N_M^{\nu_+},2]_4,[N_M^{\nu_-}]_4) ~ \oplus
([N_M^{\nu_+},1,1]_4,[N_M^{\nu_-}]_4) ~ \oplus
([N_M^{\nu_+},1]_4,[N_M^{\nu_-},1]_4) ~ \oplus 
([N_M^{\nu_+}]_4,[N_M^{\nu_-},2]_4) ~ \oplus
([N_M^{\nu_+}]_4,[N_M^{\nu_-},1,1]_4)
$ 
for U(4)$\times$U(4), which have very small dimensions ($\sim N_M$) at each momentum.
}

\subsection{Method to obtain \texorpdfstring{$\pm2$}{+-2} excitations}
We now derive the charge $\pm 2$ excitations. 
We choose a basis for the charge +2 excitations as $c_{\kk+\pp, m_2, \eta_2, s_2}^\dagger c^\dagger_{-\kk, m_1, \eta_1, s_1}|\Psi \rangle$ where $|\Psi \rangle$ is any of  the exact ground states and or eigenstates in \cref{p5:eq:U(4)-GSMT,p5:eq:U(4)U(4)-GSMT} (for which $(O_{\mathbf{q,G}}-A_{\GG} N_M\delta_{\mathbf{q,0}})|\Psi \rangle =0$) and $\pp$ is the momentum of the excited state.
Hence $\{\eta_1,s_1\}$, $\{\eta_2, s_2\}$ belong to the valley-spin flavors which are not occupied. 
The details of the commutators of the Hamiltonian and the basis are given in \cref{p5:charge2appendix}. 
We find
\begin{align}
& \left[H_I-\mu N,c_{\kk+\pp, m_2, \eta_2, s_2}^\dagger c^\dagger_{-\kk, m_1, \eta_1, s_1} \right] |\Psi\rangle \nono\\ 
=& \frac{1}{2\Omega_{\text{tot}}} 
    \sum_{m,m',\qq} T^{(\eta_2,\eta_1)} _{m,m';m_2,m_1}(\kk+\qq,\kk;\pp) \nono\\
& \times c_{\kk+\qq+\pp, m, \eta_2, s_2}^\dagger c^\dagger_{-\kk-\qq, m', \eta_1, s_1}|\Psi\rangle, 
\end{align}
{\small
\begin{align}\label{p5:eq:charge2excitation-HIMT}
& T^{(\eta_2,\eta_1)}_{m,m';m_2,m_1}(\kk+\qq,\kk;\pp) \\
=& \delta_{\qq,0}(\delta_{m,m_2} R_{mm_1}^{\eta_1}(-\mathbf{k}) +\delta_{m,m_1} R_{mm_2}^{\eta_2}(\mathbf{k+p}) ) \nono\\
+&2\sum_\GG V(\mathbf{G}+\mathbf{q})   M_{mm_2}^{\left(\eta_2\right)}\left(\kk+\pp,\mathbf{q}+\mathbf{G}\right)  M_{m'm_1}^{\left(\eta_1\right)}\left(-\kk,-\mathbf{q}- \mathbf{G}\right),\nono
\end{align} }%
where $R_{mn}^\eta(\mathbf{k})$ are the charge $+1$ excitation matrices in \cref{p5:eq:excitation-RmnMT}. 
We see that the charge $+2$ excitation energy is a sum of the two single-particle energies plus an interaction energy. 
By the translational invariance, scattering preserves the momentum ($\pp$) of the excited state.
The spectrum of the excitations at a given $\pp$ is a diagonalization problem of a matrix of the dimension $4N_M\times 4N_M$.
The scattering matrix $\td{T}$ of the charge $-2$ excitations is derived in \cref{p5:charge-2excitations}. 
It has the same form as $T$ here except that the charge $+1$ excitation matrix $R^\eta$ is replaced by charge $-1$ excitation matrix $\td{R}^\eta$ and the $M$ matrix is replaced by the complex conjugation of $M$. 

\begin{figure}[t]
\centering
\includegraphics[width=\linewidth]{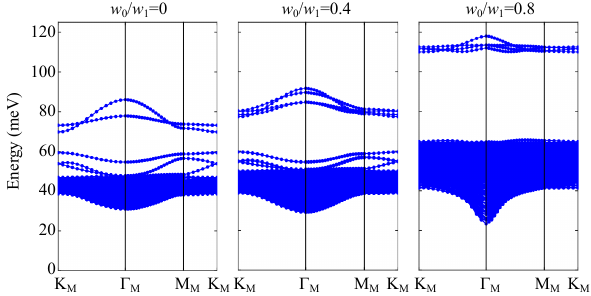}
\caption{Exact charge $\pm2$ excitations with the flat metric condition being imposed for three different $w_0/w_1$ at the twist angle $\theta=1.05^\circ$. 
Here we change $w_0$ while keeping $w_1=110\mrm{meV}$ fixed.
Other parameters are defined in \cref{p5:newapp:Ham}.
These excitations are exact at charge neutrality point ($\nu=0$) for generic states and are exact at finite integer fillings if the flat metric condition is satisfied. 
The charge +2 and $-2$ excitations are degenerate.
The exact charge $\pm2$ excitations at different fillings without imposing the flat metric condition are given in \cref{p5:fig:E2appendix,p5:fig:E2appendix20} in \cref{newapp:plot2}. Note bound states \emph{above but not below} the two-particle continuum, confirming our analytic proof and showing the lack of Cooper pairing in TBG projected Coulomb Hamiltonians}
\label{p5:fig:E2main}
\end{figure}

The spectrum of charge $\pm2$ excitations with the flat metric condition Eq.~(\ref{p5:eqn-condition-at-nuMT}) imposed is shown in \cref{p5:fig:E2main}.
By imposing the flat metric condition, we can replace the $R^{\eta}$ matrix (\cref{p5:eq:excitation-RmnMT}) in \cref{p5:eq:charge2excitation-HIMT} by the simplified \cref{p5:neweq:R-simple}. 
Since \cref{p5:neweq:R-simple} does not depend on $\nu$, the obtained charge $+2$ excitation dispersion also do not depend on $\nu$. 
\cref{p5:fig:E2main} is exact for $\nu=0$ even when the flat metric condition is not satisfied since \cref{p5:neweq:R-simple} is exact for $\nu=0$.
Due to the many-body charge-conjugation symmetry at $\nu=0$ \cite{ourpaper3}, the charge $-2$ excitations are degenerate with the charge $+2$ excitations. 
Exact charge $\pm2$ excitations without imposing the flat metric condition Eq.~(\ref{p5:eqn-condition-at-nuMT}) at different fillings are given in \cref{p5:fig:E2appendix,p5:fig:E2appendix20} in \cref{newapp:plot2}.

\subsection{Absence of Cooper pairing in the projected Coulomb Hamiltonian}\label{p5:newsec:noCooper}

The exact expression of the 2-particle excitation spectrum (\cref{p5:eq:charge2excitation-HIMT}) allows for the determination of the Cooper pair binding energy (if any).
We notice the scattering matrix \cref{p5:eq:charge2excitation-HIMT}, $T^{(\eta_2,\eta_1)}(\kk+\qq,\kk;\pp)$, differs by a sign from the neutral charge energy \cref{p5:eq:neutralexcitation-HIMT}: It is the sum of energies of two charge $+1$ excitations at momenta $\kk+\pp, -\kk$ plus an interaction matrix, while \cref{p5:eq:neutralexcitation-HIMT} is the sum of charge $+1$ and $-1$ excitations minus an interaction matrix.
This allows us to use the Richardson criterion \cite{richardson,richardson1964,richardson_numerical_1966,richardson1977} for the existence of Cooper pairing by examining the binding energy as follows:  
\begin{equation}
\Delta(N+2)  = E(N+2) + E(N)- 2E(N+1) <0
\end{equation} where $E(N)$ is the energy of the lowest state at $N$-particles. 
We now assume that the lowest state of the  charge $+2$ excitation continuum obtained by diagonalizing the matrices \cref{p5:eq:charge2excitation-HIMT} is the lowest energy state at two particles above the ground state, which is confirmed by numerical calculations for a range of parameters \cite{ourpaper6}. 
We note that \cref{p5:eq:excitation-RmnMT} is the charge $+1$ excitation. 
The lowest energy of the non-interacting $2$-particle spectrum is $2 \min (R)$, where $\min (R)$ is the smallest eigenvalue of $R^\eta(\kk)$ over $\kk\in$ MBZ and valley flavors $\eta=\pm$. 

Hence we can write the binding energy as $\min(T)-2 \min(R)$, where $\min(T)$ represents the minimal eigenvalues of $T^{(\eta_2,\eta_1)}(\kk+\qq,\kk;\pp)$ over momenta $\pp\in$ MBZ and different valley flavors $\eta_2,\eta_1$.  
For later convenience, we denote the sum of the first two terms of \cref{p5:eq:charge2excitation-HIMT} as
\begin{align}
 & T^{(\eta_2,\eta_1)\pr}_{m,m';m_2,m_1}(\kk+\qq,\kk;\pp) \nono\\
=& \delta_{\qq,0}(\delta_{m,m_2} R_{mm_1}^{\eta_1}(-\mathbf{k}) 
    +\delta_{m,m_1} R_{mm_2}^{\eta_2}(\mathbf{k+p}) )
\end{align}
and the last term of \cref{p5:eq:charge2excitation-HIMT} as 
\begin{align} \label{p5:neweq:Tpp}
& T^{(\eta_2,\eta_1)\prpr}_{m,m';m_2,m_1}(\kk+\qq,\kk;\pp) 
= 2\sum_\GG V(\mathbf{G}+\mathbf{q}) \nono\\
&\quad \times M_{m m_2}^{\left(\eta_2\right)}\left(\kk+\pp,\mathbf{q}+\mathbf{G}\right)  M_{m' m_1}^{\left(-\eta_1\right)*}\left(\kk,\mathbf{q}+ \mathbf{G}\right).
\end{align}
We therefore have $T=T'+T''$ in short notations. Here we have used the time-reversal symmetry: $M^{(\eta_1)}_{m' m_1}(-\kk,-\qq-\GG)=M^{(-\eta_1)*}_{m' m_1}(\kk,\qq+\GG)$, as explained in \cref{p5:Cooperpairappendix}.
We use Weyl's inequalities to find sufficient conditions for the presence and absence of superconductivity. 
In particular, for given $\pp,\eta_2,\eta_1$, the smallest eigenvalue of $T^{(\eta_2,\eta_1)}(\kk+\qq,\kk;\pp)$ is smaller than the smallest eigenvalue of $T^{(\eta_2,\eta_1)\pr}(\kk+\qq,\kk;\pp)$ plus the largest eigenvalue of $T^{(\eta_2,\eta_1)\prpr}(\kk+\qq,\kk;\pp)$.
Hence we have $\min(T)\le \min(T')+\max(T'')=2\min(R)+\max(T'')$.

Therefore, a sufficient criterion for the \emph{presence} of Cooper pairing binding energy is that $T^{(\eta_2,\eta_1)\prpr}(\kk+\qq,\kk;\pp)$ has all eigenvalues negative:
\begin{equation} \label{p5:neweq:Cooper}
\forall \eta_1, \eta_2, \pp,\quad \text{Eig}[T^{(\eta_2,\eta_1)\pr\pr}_{mm';m_2 m_1}(\kk+\qq,\kk;\pp)] < 0.
\end{equation}
On the other hand, for given $\pp,\eta_2,\eta_1$, the smallest eigenvalue of $T^{(\eta_2,\eta_1)}(\kk+\qq,\kk;\pp)$ is larger than the the smallest eigenvalue of $T^{(\eta_2,\eta_1)\pr}(\kk+\qq,\kk;\pp)$ plus the smallest eigenvalue of $T^{(\eta_2,\eta_1)\prpr}(\kk+\qq,\kk;\pp)$.
Hence we have $\min(T)\ge \min(T')+\min(T'')=2\min(R)+\min(T'')$.
Therefore, a sufficient criterion for the \emph{absence} of Cooper pairing binding energy is that $T^{(\eta_2,\eta_1)\prpr}(\kk+\qq,\kk;\pp)$ is positive semi-definite:
\begin{equation} \label{p5:neweq:noCooper}
\forall \eta_1, \eta_2, \pp,\quad \text{Eig}[T^{(\eta_2,\eta_1)\pr\pr}_{mm'; m_2 m_1}(\kk+\qq,\kk;\pp)] \ge 0.
\end{equation}

From the charge $+2$ excitation spectra in \cref{p5:fig:E2main,p5:fig:E2appendix,p5:fig:E2appendix20} we can see that the the spectrum of $T$ consists of two parts: the two-particle continuum, which is given by the sums of two charge $+1$ excitations, and a set of charge $+2$ collective modes above the the two-particle continuum.
Thus it seems that $T''$ are always non-negative positive.

In \cref{p5:Cooperpairappendix} we proved that, for the projected Coulomb Hamiltonian with the time-reversal symmetry, the matrix $T^{(\eta,-\eta)\prpr}(\kk+\qq,\kk;\pp)$, which corresponds to excitations of two particles from different valley, is positive semi-definite.
Thus there is no inter-valley pairing superconductivity of the PSDH $H_I$ at the integer fillings $\nu$ of the ground states in \cref{p5:eq:U(4)-GSMT,p5:eq:U(4)U(4)-GSMT}. We expect this property to hold slightly away from integer fillings. Since TBG shows superconductivity at $\nu=2$ or slightly away from integer fillings, our results show that either kinetic energy or phonons are responsible for pairing. 

Here we briefly sketch the proof.
We consider the expectation value of $T^{(\eta,-\eta)\prpr}(\kk+\qq,\kk;\pp)$ on an arbitrary complex function $\phi_{m_2,m_1}(\kk)$: 
\begin{align}
\inn{T''}^\eta_\phi(\pp) =& \sum_{\kk_1,\kk_2} \sum_{mm'm_2m_2}
     T^{(\eta,-\eta)\prpr}(\kk_2,\kk_1;\pp) \nono\\
     &\times \phi^*_{mm'}(\kk_2) \phi_{m_2,m_1}(\kk_1).
\end{align}
As detailed in \cref{p5:Cooperpairappendix}, substituting the definition of the $M$ matrix (\cref{p5:eq:M-defMT}) into \cref{p5:neweq:Tpp}, we can rewrite the expectation value as 
\begin{align}
\inn{T''}^\eta_\phi(\pp) =& \frac{2}{N_\GG} \sum_{\kk_1\kk_2\GG_1\GG_2} \Tr[ W^\dagger(\kk_2+\GG_2) W(\kk_1+\GG_1) ] \nono\\
& \times V(\kk_2+\GG_2-\kk_1-\GG_1),
\end{align}
where
{\small
\begin{equation}
W(\kk+\GG) = \sum_{m_2,m_1,\GG} u_{m_2,\eta}(\kk+\pp+\GG) \phi_{m_2,m_1}(\kk) u_{m_1,\eta}^\dagger(\kk+\GG).
\end{equation}}%
Here $u_{m_2,\eta}(\kk+\GG)$ is the $2 N_\QQ\times 1$ vector $u_{\QQ-\GG,\alpha;m_2,\eta}(\kk)$ and $W(\kk+\GG)$ is a $2N_\QQ\times 2N_\QQ$ matrix, with $N_\QQ$ being the $\QQ$ lattice size (see App. \ref{p5:singpartRevappendix} for definition of the $\QQ$ lattice).
For simplicity, we use $a$ and $b$ to represent the composite indices $(\QQ,\alpha)$.
Then $\inn{T''}^\eta_\phi(\pp)$ can be written as $\sum_{ab} W_{ab}^\dagger V W_{ab}$, where now $W_{ab}$ is viewed as an $N_M\times1$ vector and $V$ an $N_M\times N_M$ matrix.
Since $V$ is positive semi-definite, for each pair of $a,b$, the summation over $\kk_1,\kk_2,\GG_1,\GG_2$ is non-negative. 
Thus $T''$ is positive semi-definite since $\inn{T''}^\eta_\phi(\pp)\ge 0$ for arbitrary $\phi$.

In \cref{p5:Cooperpairappendix} we also proved that, for $\eta_2=\eta_1=\eta$, $T^{(\eta,\eta)\prpr}(\kk+\qq,\kk;\pp)$ is also positive semi-definite due to the symmetry $PC_{2z}T$, with $P$ being the unitary single-body PH symmetry of TBG \cite{song_all_2019,ourpaper2}.
Therefore, neither the inter-valley pairing nor the intra-valley Cooper pair has binding energy in the projected Coulomb Hamiltonian for any integer fillings $\nu$ in the chiral-flat limit, and for any even fillings $\nu=0,\pm2,\pm4$ in the nonchiral-flat limit.

\section{Conclusions}

In this paper, we have calculated the excitation spectra of a series positive semi-definite Hamiltonians (PSDHs) initially introduced by Kang and Vafek \cite{kang_strong_2019} that generically appear \cite{ourpaper3} in projected Coulomb Hamiltonians to bands with nonzero Berry phases and which exhibit ferromagnetic states as ground states, under weak assumptions \cite{ourpaper4, ourpaper3} . These assumptions were also used by Kang and Vafek \cite{kang_strong_2019} to find the $\nu=2$ ground states in TBG. In this paper, we show that not only the ground states, but a large number of low-energy excited states can be obtained in PSDHs. We obtain the general theory for the charge $\pm 1, \pm 2$ and neutral excitations energies and eigenstates and particularize it to the case of TBG insulating states. We find that charge $+1$ excitations are gapped, with the smallest gap at the $\Gamma$ point. 
In both the (first) chiral-flat limit and the nonchiral-flat limit, we find the Goldstone stiffness of the ferromagnetic state, as well as the Cooper pairing binding at integer fillings. In particular, we proved by the Richardson criterion \cite{richardson,richardson1964,richardson_numerical_1966,richardson1977} that Cooper pairing is not favored at integer fillings (even fillings when nonchiral) in the flat band limit. 
Since superconductivity has been observed in experiments with screened Coulomb potentials \cite{stepanov_interplay_2020,saito_independent_2020,liu2020tuning} (such as at $\nu=2$), we conjecture the origin of superconductivity in TBG is not Coulomb, but is contributed by other mechanisms, \eg the electron-phonon interaction \cite{Lian2019TBG,Wu2018TBG-BCS,lewandowski2020pairing}, or due to kinetic terms. In particular, our theorem shows that the Luttinger-Kohn mechanism of creating attractive interactions out of repulsive Coulomb forces is ineffective for flat bands. A similar statement can be made for the super-exchange interaction. A finite kinetic energy is hence required for these mechanisms.

In future work, the charge excitation energies of these Hamiltonians will be obtained in perturbation theory with the kinetic terms. 
A further question, of whether there are other further eigenstates of the PSDHs, remains unsolved. 

B.A.B thanks Oskar Vafek for fruitful discussions, and for sharing their similar results on this problem before publication \cite{vafek2020hidden}, where they also compute the Goldstone and charge 1 excitation spectrum, which agrees with ours.  B.A.B also thanks Pablo Jarillo-Herrero for discussions and for pointing out Ref. \cite{park2020flavour}. This work was supported by the DOE Grant No. DE-SC0016239, the Schmidt Fund for Innovative Research, Simons Investigator Grant No. 404513, the Packard Foundation, the Gordon and Betty Moore Foundation through Grant No. GBMF8685 towards the Princeton theory program, and a Guggenheim Fellowship from the John Simon Guggenheim Memorial Foundation. Further support was provided by the NSF-EAGER No. DMR 1643312, NSF-MRSEC No. DMR-1420541 and DMR-2011750, ONR No. N00014-20-1-2303, Gordon and Betty Moore Foundation through Grant GBMF8685 towards the Princeton theory program, BSF Israel US foundation No. 2018226, and the Princeton Global Network Funds. B.L. acknowledge the support of Princeton Center for Theoretical Science at Princeton University during the early stage of this work.

\bibliography{ref, HexalogyInternalRefs}

\onecolumngrid
\tableofcontents
\appendix
\section{Review of notation: single particle and interacting Hamiltonians} \label{p5:newapp:Ham}

For completeness, we here briefly review the notations used for the single-particle and interacting Hamiltonians. Their properties and (explicit and hidden) symmetries of both the single particle and the interacting problems are detailed at length in our recent papers \cite{ourpaper1,ourpaper2,ourpaper3,ourpaper4}.

\subsection{Single particle Hamiltonian: short review of notation}\label{p5:singpartRevappendix}

The single-particle Hamiltonian, symmetries, and properties of the wavefunctions have been discussed at length in Ref. \cite{ourpaper1,ourpaper2,song_all_2019}. 
For completeness of notation, we give its expression here, for completeness, but we skip all details. The total single particle Hamiltonian is  
\begin{equation}
\hat{H}_{0}=\sum_{\mathbf{k}\in\mathrm{MBZ}}\sum_{\eta\alpha\beta s}\sum_{\mathbf{Q}\mathbf{Q}^{\prime}}\Big[h_{\mathbf{Q}\mathbf{Q}^{\prime}}^{\left(\eta\right)}\left(\mathbf{k}\right)\Big]_{\alpha\beta}c_{\mathbf{k},\mathbf{Q},\eta,\alpha s}^{\dagger}c_{\mathbf{k},\mathbf{Q}^{\prime},\eta,\beta s}.\label{p5:eq:H0}
\end{equation} 
where $c_{\mathbf{k},\mathbf{Q},\eta,\alpha s}^{\dagger}$ is the creation operator at momentum $\kk$ (in  the moir\'e BZ - MBZ) in valley $\eta$ ($\pm$), sublattice $\alpha$ (1,2), spin $s$ ($\up\down$), and moir\'e momentum lattice $\QQ$. 
The Hamiltonians in the two valleys are 
\begin{align}
&h_{\mathbf{Q}\mathbf{Q}^{\prime}}^{\left(+\right)}\left(\mathbf{k}\right)=\delta_{\mathbf{Q},\mathbf{Q}^{\prime}}v_{F}\left(\mathbf{k}-\mathbf{Q}\right)\cdot\boldsymbol{\sigma}+\sum_{j=1}^3\left(\delta_{\mathbf{Q}-\mathbf{Q}^{\prime},\mathbf{q}_{j}}+\delta_{\mathbf{Q}^{\prime}-\mathbf{Q},\mathbf{q}_{j}}\right)T_{j}  \nonumber \\ &
h_{\mathbf{Q}\mathbf{Q}^{\prime}}^{\left(-\right)}\left(\mathbf{k}\right)  =\delta_{\mathbf{Q},\mathbf{Q}^{\prime}}v_{F}\left(\mathbf{k}-\mathbf{Q}\right)\cdot\boldsymbol{\sigma}^*+\sum_{j=1}^3\left(\delta_{\mathbf{Q}-\mathbf{Q}^{\prime},\mathbf{q}_{j}}+\delta_{\mathbf{Q}^{\prime}-\mathbf{Q},\mathbf{q}_{j}}\right)\sigma_{x}T_j\sigma_{x},
\end{align}
where $\boldsymbol{\sigma}=(\sigma_{x},\sigma_{y})$, $\boldsymbol{\sigma}^*=(-\sigma_{x},\sigma_{y})$ are Pauli matrices, $\small T_j=w_{0}\sigma_{0}+w_{1}(\cos\frac{2\pi}{3}(j-1) \sigma_{x} + \sin\frac{2\pi}{3}(j-1)  \sigma_y)$, and $\mathbf{q}_j=k_\theta C_{3z}^{j-1}(0,1)^T$ ($j=1,2,3$) with $k_\theta=2|K|\sin\frac{\theta}2$ being the distance of the Graphene $K$ momenta from the top layer and bottom layer, $\theta$ the twist angle, $w_{0}$ the interlayer AA-hopping, and $w_{1}$ the interlayer AB-hopping. $\mathbf{Q}$ belongs to a hexagonal momentum space lattice, $\QQ\in \mathcal{Q}_\pm$, where $\mathcal{Q}_\pm=\mathcal{Q}_0\pm\mathbf{q}_1$. 
The eigenstates of  \cref{p5:eq:H0} take the form
\begin{equation}
c_{\mathbf{k}n\eta s}^{\dagger}=\sum_{\mathbf{Q}\alpha}u_{\mathbf{Q}\alpha;n\eta}\left(\mathbf{k}\right)c_{\mathbf{k},\mathbf{Q},\eta,\alpha s}^{\dagger}. \label{p5:eq:solution}
\end{equation} where $ c_{\mathbf{k}+\GG,\mathbf{Q},\eta\alpha s}^{\dagger}=c_{\mathbf{k},\mathbf{Q}-\GG,\eta\alpha s}^{\dagger}$, for any moir\'e reciprocal wavevector $\GG$, and hence
\begin{equation}
u_{\mathbf{Q}\alpha; n\eta}\left(\mathbf{k}+\GG\right)=u_{\mathbf{Q}-\GG,\alpha;n\eta}\left(\mathbf{k}\right) \label{p5:neweq:u-period}
\end{equation}
such that $c_{\mathbf{k}+\GG,n\eta s}^{\dagger}=c_{\mathbf{k}n\eta s}^{\dagger}$. This is the MBZ periodic gauge. 

In the numerical calculations, we take the parameters $\theta=1.05^\circ$, $|K|=1.703\mathrm{\mathring{A}}^{-1}$, $v_F=5.944{\rm eV \cdot \mathring{ A} }$, $w_1=110$meV. 

The projected kinetic Hamiltonian in the flat bands will be denoted by $H_0$ (without hat), which is given in Eq. (\ref{p5:eq:HHMT}).

\subsection{Interaction Hamiltonian: short review of notation}\label{p5:interactionRevappendix}

The many-body Hamiltonian, symmetries, and properties of the wavefunctions, as well as the derivations, have been discussed at length in Refs.~\cite{ourpaper3, ourpaper4}. For completeness of notation, we give its expression here, for completeness, but we skip all details. The Hamiltonian before projection was derived to be (denoted by a hat)  \cite{ourpaper3, ourpaper4}: 

\beq
\hat{H}_{I}=\frac{1}{2\Omega_{\mathrm{tot}}}\sum_{\mathbf{G}}\sum_{\mathbf{q}\in\mathrm{MBZ}}V\left(\mathbf{G}+\mathbf{q}\right) 
\delta \rho_{-\mathbf{G}-\mathbf{q}}
\delta \rho_{\mathbf{G}+\mathbf{q}};\;\;\;\;  V\left(\mathbf{r}\right)=\frac{1}{\Omega_{\mathrm{tot}}}\sum_{\mathbf{G}}\sum_{\mathbf{q}\in\mathrm{MBZ}}e^{-i\left(\mathbf{q}+\mathbf{G}\right)\cdot\mathbf{r}}V\left(\mathbf{q}+\mathbf{G}\right)  \label{p5:eq:HI-0}
\eeq
where 
\beq
\delta\rho_{\mathbf{q+G}}=\sum_{\eta,\alpha,s,\mathbf{k},\mathbf{Q}} (c_{\mathbf{k+q},\mathbf{Q-G},\eta,\alpha, s}^\dag c_{\mathbf{k},\mathbf{Q},\eta,\alpha, s}-\frac{1}{2}\delta_{\mathbf{q,0}}\delta_{\mathbf{G,0}}) 
\eeq
is the total electron density at momentum $\mathbf{q+G}$ relative to the charge neutral point. 
$\Omega_{\mathrm{tot}}$ is the total area of the moir\'e lattice, $\mathbf{G}$ sums over moir\'e reciprocal lattice, and $\mathbf{q}$
sums over momenta in MBZ zone.

For the analytic derivations in the current paper, we keep $V(\rr)$ generic. For the numerical plots of the energy dispersion and other properties, we use twisted bilayer graphene Coulomb interactions screened by
the electrons in the two planar conducting gates \cite{throckmorton_fermions_2012,kang_strong_2019}:  $V\left(\mathbf{r}\right)=U_{\xi}\sum_{n=-\infty}^{\infty}\left(-1\right)^{n}/{\sqrt{\left(\mathbf{r}/\xi\right)^{2}+n^{2}}}$ with $\xi=10\mathrm{nm}$ being the distance between the two gates,
$U_{\xi}=e^{2}/\left(\epsilon\xi\right)=26\mathrm{meV}$ (in Gauss units), $\epsilon\approx6$
the dielectric constant of boron nitride. The derivation of this interaction was explained at length in Ref. \cite{ourpaper3,ourpaper4}. It was also showed that the interaction has non-vanishing Fourier component only for intra-valley scattering to give
\begin{equation}\label{szdeq:Vq}
V\left(\mathbf{q}\right)=\left(\pi\xi^{2}U_{\xi}\right)\frac{\tanh\left(\xi q/2\right)}{\xi q/2}.
\end{equation} For the given values of the parameters, $V\left(\mathbf{q}\right)$ was plotted in Ref. \cite{ourpaper2} and is a slowly decreasing function of $|\mathbf{q}|$ in the BZ, reaching (around the magic angle)  about half of its maximal  value as  $|\mathbf{q}|$ spans the whole MBZ around.

\subsubsection{Gauge fixing and the projected interaction}\label{p5:gaugefixingRevappendix}

We define for our many-body Hamiltonian the form-factors, also called the overlap matrix of a set of bands $m,n$ as
\begin{equation}
M_{m,n}^{\left(\eta\right)}\left(\mathbf{k},\mathbf{q}+\mathbf{G}\right)=\sum_{\alpha}\sum_{\QQ}
u_{\mathbf{Q}-\mathbf{G},\alpha;m\eta}^{*}\left(\mathbf{k}+\mathbf{q}\right)u_{\mathbf{Q},\alpha;n\eta}\left(\mathbf{k}\right) , \label{p5:eq:M-def}
\end{equation}
In terms of which the projected density operator interaction and Hamiltonian to a set of  bands denoted by $m, n$ can be written as
\beq
\begin{split}
&H_{I}=\frac{1}{2\Omega_{\mathrm{tot}}}\sum_{\mathbf{G}}\sum_{\mathbf{q}\in\mathrm{MBZ}}V\left(\mathbf{G}+\mathbf{q}\right) 
 \overline{\delta\rho}_{-\mathbf{G}-\mathbf{q}}
 \overline{\delta\rho}_{\mathbf{G}+\mathbf{q}} \nonumber \\ & \overline{\delta\rho}_{\GG+\qq} = \sum_{\eta s} \sum_{mn\in\mrm{proj}} \sum_{\kk} M_{m,n}^\eta (\kk,\qq+\GG) 
	\pare{ c^\dagger_{\kk+\qq,m,\eta,s} c_{\kk,n,\eta,s} 
	- \frac12 \delta_{\qq,0} \delta_{mn}}.
\end{split}
\eeq
For our working convenience, we then define the operator 
\begin{equation}
O_{\qq,\GG}=\sqrt{V(\qq+\GG)}\overline{\delta\rho}_{\mathbf{G}+\mathbf{q}}. 
\end{equation}
This allows us to rewrite the projected interaction Hamiltonian $H_I$ into the form of Eqs. (\ref{p5:eq:nonnegative-HintMT}) and (\ref{p5:Oqdef1MT}). 
While most of the projected Hamiltonian properties are valid for any number of projected bands that respect the symmetries of the system (including PH), in TBG at the first magic angle we usually are interested in the projection of  the Hamiltonian onto \emph{the lowest two flat bands} per spin per valley of TBG (8 bands in total). An important step in any calculations - especially numerical - is the gauge-fixing procedure. Different gauges for the wavefunctions, that make different symmetries of the form factors \cref{p5:eq:M-def} more explicit, can be chosen. 
This is explained at length in our manuscript Ref. \cite{ourpaper3}, but for completeness we briefly mention them here. 
We consider only $2$ active bands, the general gauge-fixing mechanism for projection in more than $2$ bands is found in Ref. \cite{ourpaper3}. 
To fix the gauge of the Bloch wavefunctions in Eq. (\ref{p5:eq:solution}), $\ket{\psi_{\kk,n,\eta,s}} = \sum_{\QQ,\alpha} u_{\QQ,\alpha; m\eta}(\kk) c_{\mathbf{k},\mathbf{Q},\eta,\alpha,s}^{\dagger} \ket{0}$, where $u_{\QQ,\alpha;m\eta}(\kk)$ is the solution of the single-particle Hamiltonian $h_{\QQ\QQ'}^{(\eta)}(\kk)$, in each $\eta=\pm$, $s=\up\down$ sector, we label the higher energy band by $m=+$ and the lower band by $m=-$. 
Due to the spin-SU(2) symmetry, we set the real space wavefunctions for $s=\up\down$ to be identical and omit the index $s$ for the single-particle states. For a symmetry operation $g$, the sewing matrices $B^g_{n^\pr\eta^\pr, n\eta}(\kk) = \bra{\psi_{g\kk,n^\pr,\eta^\pr}} g \ket{\psi_{\kk,n,\eta}}$, which relate states at momentum $\kk$ with states at the transformed momentum $g \kk$ can be consistently chosen to be
\beq
B^{C_{2z}T}(\kk) = \zeta^0 \tau^0,\qquad
B^{C_{2z}}(\kk) = \zeta^0 \tau^x,\qquad
B^{P}(\kk) = i\zeta^y \tau^z, \qquad
B^{C_{2z}P}(\kk) = \zeta^y \tau^y
\label{p5:eq:gauge-0}
\eeq
for the  $C_{2z}T$, $C_{2z}$, $P$ symmetries of TBG where $\zeta^a,\tau^a$ ($a=0,x,y,z$) are Pauli-matrices acting on the band and valley indices, respectively \cite{ourpaper3}. 
Here $P$ is a unitary single-body PH symmetry that transforms $\kk$ to $-\kk$ \cite{song_all_2019,ourpaper2}.
We leave the other sewing matrices - for $C_{3z}$, $C_{2x}$ - unfixed. With these sewing matrices, once we obtain, by diagonalizing the single particle Hamiltonian, the  wavefunctions in the valley $\eta=+$ for band $m=+$, then we first fix the $C_{2z}T$  (in TBG, this is done at the detriment of the wavefunction being continuous), then we use of the PH to generate the $m=-$ band, while finally, using $C_{2z}$ symmetry to generate the wavefunctions in the valley $\eta=-$. 
In Ref.~\cite{ourpaper3}, we use the above gauge to determine the most generic form of the $M$ matrix coefficient (\cref{p5:eq:M-def}). 
We found that - away from the point $w_1=0$ in parameter space which we call ``second chiral limit" -  we can decompose the $M$ matrix  into four terms: \cite{ourpaper2,ourpaper3}
\beq 
M(\kk,\qq+\GG) = \zeta^0\tau^0 \alpha_0(\kk,\qq+\GG) + \zeta^x\tau^z \alpha_1(\kk,\qq+\GG) + i\zeta^y\tau^0 \alpha_2(\kk,\qq+\GG) + \zeta^z\tau^z \alpha_3 (\kk,\qq+\GG).  \label{p5:eq:M-para}
\eeq where $\alpha_{0,1,2,3}$ are real functions which satisfy the following symmetry conditions.
\beq
\quad 
\alpha_a(\kk,\qq+\GG) = \alpha_a(\kk+\qq,-\qq-\GG)\quad \text{for }a=0,1,3,\qquad
\alpha_2(\kk,\qq+\GG) =-\alpha_2(\kk+\qq,-\qq-\GG), \label{p5:eq:alpha-cond1}
\eeq
\beq
\quad
\alpha_a(\kk,\qq+\GG) = \alpha_a(-\kk,-\qq-\GG)\quad \text{for }a=0,2,\qquad
\alpha_a(\kk,\qq+\GG) =-\alpha_a(-\kk,-\qq-\GG)\quad \text{for }a=1,3.  \label{p5:eq:alpha-cond2}
\eeq
In particular, the combination of Eqs. (\ref{p5:eq:alpha-cond1}) and (\ref{p5:eq:alpha-cond2}) implies that at $\qq=\mathbf{0}$, we have
\begin{equation} \label{szdeq:alpha-cond3}
\alpha_0(\kk,\GG)=\alpha_0(-\kk,\GG)\ ,\qquad \alpha_j(\kk,\GG)=-\alpha_j(-\kk,\GG),\quad (j=1,2,3).
\end{equation}
Note that the same gauge fixing can be found with projection in a larger number of bands \cite{ourpaper3}.

A further simplification occurs in the first chiral limit $w_0=0$ of the single-particle Hamiltonian due to the presence of an extra symmetry \cite{bultinck_ground_2020,ourpaper2,ourpaper3}. 
(A similar simplification takes place in the second chiral limit $w_1=0$, found in Refs.~\cite{ourpaper2,ourpaper3}). 
In this limit there is another (chiral) symmetry $C$  of the one-body first-quantized Hamiltonian $h_{QQ'}(k)$, which in band space is defined by its sewing matrix:
\beq
B^{C}_{m\eta^\pr, n\eta} (\kk) = \bra{\psi_{\kk,m,\eta^\pr}} C \ket{\psi_{\kk,n,\eta}} 
\propto \delta_{m,-n} \delta_{\eta^\pr,\eta}.
\eeq
A gauge choice in which $B^{C}_{m\eta^\pr, n\eta} (\kk)$ is $\kk$ independent is possible
\begin{equation}
 B^C(\kk) = \zeta^y\tau^z. 
\end{equation} This allows us to find the wavefunction of the $-$ band at $\kk$ from the $+$ band at $\kk$, in the same valley. In Ref.~\cite{ourpaper3} we prove that the $M$ matrix in the interaction satisfies the chiral symmetry,  $B^{C\dg} M(\kk,\qq+\GG) B^C = M(\kk,\qq+\GG)$.
Thus with the chiral symmetry, $M$ takes the form
\begin{equation} 
M(\kk,\qq+\GG) = \zeta^0\tau^0 \alpha_0(\kk,\qq+\GG) + i\zeta^y\tau^0 \alpha_2(\kk,\qq+\GG).  \label{p5:eq:M-para-chiral}
\end{equation}

\subsubsection{Chern band basis} \label{p5:chernbasisRevappendix}
In Ref. \cite{ourpaper2} we have shown that the two flat bands can be recombined as two Chern bands:
\beq 
d^\dagger_{\kk,e_Y,\eta,s} = \frac{1}{\sqrt2} ( c^\dg_{\kk,+,\eta,s} + i e_Y c^\dg_{\kk,-,\eta,s}).
 \label{p5:eq:Chern-band}
\eeq
Their corresponding Berry curvatures are continuous in the MBZ and yield Chern numbers $e_Y=\pm1$, respectively. 
We fixed the ambiguities by requiring that the Berry's curvature of each Chern band basis $d^\dg_{\kk,e_Y,\eta,s}$ is continuous, or, equivalently,
\begin{equation}
\lim_{\qq\to 0} |\inn{u_{\kk+\qq,e_Y,\eta,s}^\pr | u_{\kk,e_Y',\eta,s}^\pr }| = \delta_{e_Y,e_Y'}, \label{p5:eq:Chern-band-gauge}
\end{equation}
where $\ket{u^\pr_{\kk,e_Y,\eta,s}}$ is the periodic part of the Bloch wavefunction for the operator $d^\dg_{\kk,e_Y,\eta,s}$. In this gauge the band $d^\dg_{\kk,e_Y,\eta,s}$ has nonzero Chern number $C_{e_Y,\eta,s} =e_Y$ \cite{ourpaper2,hejazi2020hybrid}. The Chern numbers of $d^\dg_{\kk,e_Y,-,s}$ equals to the Chern numbers of $d^\dg_{\kk,e_Y,+,s}$, because they are related by $C_{2z}$ rotation.

The $M$ matrix in the Chern band basis becomes
\begin{equation} \label{p5:eq:chiral-MqG1}
M_{e_Y,e_Y}^{(\eta)}(\kk,\qq+\GG) =M_{e_Y}(\kk,\qq+\GG) = \alpha_0(\mathbf{k,q+G})+ie_Y\alpha_2(\mathbf{k,q+G}),
\end{equation}
\begin{equation} \label{p5:eq:chiral-MqG2}
M_{-e_Y,e_Y}^{(\eta)}(\kk,\qq+\GG) = \eta F_{e_Y} (\kk,\qq+\GG),\quad
F_{e_Y} (\kk,\qq+\GG)=\alpha_1(\mathbf{k,q+G})+ie_Y\alpha_3(\mathbf{k,q+G}).
\end{equation}
For later convenience, we have introduced the factors $M_{e_Y}(\kk,\qq+\GG)$ and $F_{e_Y}(\kk,\qq+\GG)$ to represent the diagonal element and off-diagonal element in the Chern band basis, respectively.
In the first chiral limit, where $\alpha_1=\alpha_3=0$, we have $F_{e_Y}=0$.

\subsubsection{Many-body charge-conjugation symmetry of the projected interaction and kinetic Hamiltonian} \label{p5:projectedinteractionRevappendix}

In Ref. \cite{ourpaper3}, we showed that the full projected Hamiltonian $H_0+H_I$ has a many-body charge-conjugation symmetry, $\CC_c$ defined  as the single-particle transformation $C_{2z}TP$ followed by an interchange between electron annihilation operators $c$ and creation operators $c^\dag$: $\CC_c c_{\kk,n,\eta,s}^\dg \CC_c^{-1}= c_{-\kk,m,\eta^\pr,s} B^{C_{2z}TP}_{m\eta^\pr,n\eta}(\kk)$:

\beq
\CC_c H_0 \CC_c^{-1} =  H_0 + const. \;\;\;\; \CC_c \delta\rho_{\GG+\qq} \CC_c^{-1} =  - \delta \rho_{\GG+\qq}.
\eeq
The interaction has the charge-conjugation symmetry, and the many-body physics is PH symmetric in this limit.

\subsubsection{The U(4) symmetry of the projected Hamiltonian in the flat-band limit}

Using the unitary PH symmetry $P$ introduced in Ref. \cite{song_all_2019}, we demonstrated in Ref. \cite{ourpaper3} that the projected TBG Hamiltonian has a $U(4)$ symmetry if the kinetic energy is set to zero (flat band limit), for \emph{any} number of projected bands. 
This generalizes the $U(4)$ symmetry introduced in Ref. \cite{bultinck_ground_2020} for the two active bands. 
Using the continuous symmetry operator notation
\beq
[e^{i \gamma_{ab} S^{ab}},H_I] =0,\;\; \forall \gamma_{ab}\in\mathbb{R};\;\;\;\; S^{ab}= \sum_{\kk, m, m', \eta, \eta', s, s'}  
c_{\kk, m, \eta, s}^\dagger  s^{ab}_{m, \eta, s; m', \eta', s'} c_{\kk, m^\pr, \eta^\pr, s^\pr},
\eeq 
we can generate a full set of $U(4)$ generators given by \cite{ourpaper3}
\begin{equation}\label{p5:eq:U4-generator}
\boxed{s^{ab}=\{\zeta^0\tau^0 s^b,\ \zeta^y\tau^x s^b,\ \zeta^y\tau^y s^b,\ \zeta^0\tau^z s^b\}, \qquad (a, b=0,x,y,z)\ .}
\end{equation}
where  the Pauli matrices $\zeta^a, \tau^a, s^a$ with $a=0,x,y,z$ are identity and $x,y,z$ Pauli matrices in band, valley and spin-space respectively. 
For $2N_1,\; N_1\in \mbb{Z}$ projected bands, the generators would be identical, with the $\zeta$ representing the $+ = \{1,\ldots, N_1\}$ and $- = \{N_1+1, \ldots, 2N_1\}$ bands.

The kinetic plus the projected interaction term exhibit the Cartan symmetry $U(2)\times U(2)$ subgroup of the $U(4)$ symmetry group of the projected interaction, which can be most naturally chosen as the valley spin and charge: $ \text{Cartan}$: $\zeta^0 \tau^0 s^0, \quad \zeta^0 \tau^0 s^z,\quad \zeta^0 \tau^z s^0, \quad \zeta^0 \tau^z s^z$.

\subsubsection{Enhanced \texorpdfstring{U(4)$\times$U(4)}{U(4)xU(4)} symmetries in the first chiral limit \texorpdfstring{$w_0=0$}{w0=0}}\label{p5:enhancedU4U4symmetryRevappendix}

In Ref. \cite{ourpaper3}  we demonstrated in detail the presence of two enhanced unitary U(4)$\times$U(4) symmetry in two limits of the single-particle parameter space the first and second chiral limits $w_0=0<w_1$ and $w_1=0<w_0$. For the first chiral limit $w_0=0$, this symmetry was presented in  Ref. \cite{bultinck_ground_2020} for the case of two projected bands, but we find that it is maintained, in both chiral limits, for a projection in any number of bands. 
For the matrix elements in \cref{p5:eq:M-para-chiral}, the interaction commutes with the following matrices, which form the U(4)$\times$U(4) generators  \cite{ourpaper3}:
\begin{equation} \label{p5:eq:U4xU4-generator1}
\zeta^0 \tau^a s^b,\qquad \zeta^y \tau^a s^b, \qquad (a,b=0,x,y,z). 
\end{equation}
The Cartan subalgebra of the chiral U(4)$\times$U(4) is the Cartan subalgebra of the  $U(2)_{non-chiral}\times U(2)_{non chiral}\times U(2)_{chiral}\times U(2)_{chiral}$: $\zeta^0 \tau^0s^0, \quad  \zeta^0 \tau^0 s^z, \quad \zeta^0 \tau^z s^0,\quad  \zeta^0 \tau^z s^z,\quad \zeta^y \tau^0 s^0,\quad  \zeta^y \tau^0 s^z,\quad  \zeta^y \tau^z s^0,\quad \zeta^y \tau^z s^z$. The U(4) implied by $C_{2z}P$ (\cref{p5:eq:U4-generator}) is a subgroup of this U(4)$\times$U(4), but not one of the U(4) factors \cite{ourpaper3}.

In Ref. \cite{ourpaper2} we found that there exists a further more convenient gauge choice for the wavefunctions in the chiral limit $w_0=0$, called the Chern basis, (an extension to many bands of the Chern basis in Ref.~\cite{bultinck_ground_2020} to many-bands), in which we choose the single-particle representations of U(4)$\times$U(4) generators as
\begin{equation}
s^{ab}_+=\frac{1}{2}(\zeta^0+\zeta^y) \tau^a s^b,\qquad
s^{ab}_-=\frac{1}{2}(\zeta^0-\zeta^y) \tau^a s^b,\qquad
(a,b=0,x,y,z), \label{p5:eq:U4xU4-generator}
\end{equation}
which correspond to the first U(4) and second U(4), respectively. 

Adding the kinetic term in the chiral limit breaks the U(4)$\times$U(4) symmetry of the projected interaction,  to a U(4) subset $\zeta^0 \tau^a s^b$, $(a,b=0,x,y,z)$. 

We note that the nonchiral-flat U(4) symmetry and the first chiral-flat U(4)$\times$U(4) symmetry are first identified by \cite{bultinck_ground_2020}. A similar U(4) symmetry is proposed in \cite{kang_strong_2019}, the difference and similarity between which and the symmetries reviewed here is studied in \cite{ourpaper3}.

\subsubsection{U(4) irrep of electrons in the nonchiral-flat case and the Chern  basis in the \texorpdfstring{U(4)$\times$U(4)}{U(4)xU(4)}  chiral limit}\label{p5:sec:U(4)irrep}

In Ref. \cite{ourpaper3}, we showed that the $8$ single-particle basis of the nonchiral-flat U(4) symmetry generators given in Eq. (\ref{p5:eq:U4-generator}) can be decomposed into two $4$-dimensional fundamental irreps of the U(4) group, which have $\zeta_y$ eigenvalues $e_Y=\pm1$, respectively, for each momentum $\mathbf{k}$ (\cref{p5:eq:Chern-band}).
The U(4) generators in the Chern band basis are $s^{ab}(e_Y)=e_Y\tau^{x}s^a,\  e_Y\tau^{y}s^a,\ \tau^{0}s^a,\ \tau^{z}s^a$,
respectively.
We also showed \cite{ourpaper3} that the $e_Y=+1$ irrep and the $e_Y=-1$ irrep  are the same - and not conjugate - irrep: the $4$-dimensional fundamental U(4) irrep represented by a one-box Young tableau labeled by $[1]_4$. 
We presented a detailed review of the U(4) representations related to TBG in Ref. \cite{ourpaper3}, but for the purpose of the current paper, the notation adopted for irreps is the standard  Young tableau, conveniently denoted by $[\lambda_1,\lambda_2,\cdots]_N$, where $\lambda_i$ is the number of boxes in row $i$ ($\lambda_i\ge\lambda_{i+1}$).  
The number of boxes in the $i$-th row is no smaller than that in the $(i+1)$-th row.  The \emph{Hook rule} then provides the dimensions of each of these irreps. In particular, $[1^N]_N$ is an SU($N$) singlet state.

In the first chiral limit $w_0=0$,  $d^\dagger_{\kk,e_Y,\eta,s}$ defined in Eq. (\ref{p5:eq:Chern-band}) gives the single-particle basis irrep U(4)$\times$U(4) of Eq. (\ref{p5:eq:U4xU4-generator}). 
We proved in Ref. \cite{ourpaper3} that $d^\dagger_{\kk,+1,\eta,s}$ generates the $([1]_4,[0]_4)$ irrep of U(4)$\times$U(4), while $d^\dagger_{\kk,-1,\eta,s}$ generates the $([0]_4,[1]_4)$ irrep of U(4)$\times$U(4) with generators  

$s_\pm^{ab}=\frac{1}{2}(1\pm e_Y) \tau^a s^b$, respectively.  A similar discussion is provided for the second chiral limit $w_1=0$ in Ref. \cite{ourpaper3}. 

In the chiral-nonflat case, the generators U(4) symmetry is given by generators $\zeta^0\tau^a s^b$ and either the original band basis $c^\dagger_{\kk,m,\eta,s}$ for a fixed band index $m (=\pm)$ or the Chern basis $d^\dagger_{\kk,e_Y,\eta,s}$ with a fixed $e_Y(=\pm1)$ form a fundamental of U(4).

\subsection{Exact ground states in different limits: review of notation}\label{p5:sec:exact-GS}

In Ref. \cite{ourpaper4} we  have examined in detail the exact ground states in the nonchiral-flat U(4) symmetric limit and in the chiral-flat U(4)$\times$U(4) symmetric limit. For completeness we briefly review the results. 
In Ref. \cite{kang_strong_2019} Kang and Vafek first introduced a type of Hamiltonians which have, under some conditions, exact ground states. 
We have found \cite{ourpaper4} that \emph{any} translationally invariant interaction Hamiltonian projected to some active bands can be written in a form of Ref. \cite{kang_strong_2019}, and that, under some conditions, exact eigenstates and ground states can be found. 
The key idea of Kang and Vafek for obtaining exact ground states is to rewrite the interacting Hamiltonian into a non-negative form. 
A state with eigenvalue zero is then ensured to be the ground state.

In Ref. \cite{ourpaper3}, we proved that the projected Coulomb Hamiltonian can be written as 
\begin{equation}
H_{I}=\frac{1}{2\Omega_{\text{tot}}}\sum_{\mathbf{G}}\sum_{\mathbf{q}} O_{\mathbf{q,G}} O_{\mathbf{-q,-G}}=\frac{1}{2}\int_{\mathbf{r}\in \Omega_{\text{tot}}} d^2\mathbf{r} O(\mathbf{r})^2\ ,\;\;\; O(\mathbf{r})=\frac{1}{\Omega_{\text{tot}}}\sum_{\mathbf{G}}\sum_{\mathbf{q}} O_{\mathbf{q,G}} e^{i(\mathbf{q}+\mathbf{G})\cdot\mathbf{r}},
\label{p5:eq:nonnegative-Hint}
\end{equation}
which is non-negative, with
\begin{equation}
O_{\mathbf{q,G}}=\sum_{\mathbf{k},m,n,\eta,s} \sqrt{V(\mathbf{G}+\mathbf{q})} M_{m,n}^{\left(\eta\right)}\left(\mathbf{k},\mathbf{q}+\mathbf{G}\right) \left(\rho_{\mathbf{k,q},m,n,s}^\eta-\frac{1}{2}\delta_{\mathbf{q,0}}\delta_{m,n}\right), \;\; \; O_{\mathbf{q,G}}^\dag=O_{\mathbf{-q,-G}}\label{p5:Oqdef1}
\end{equation}
where $\rho_{\mathbf{k,q},m,n,s}^\eta=c^\dag_{\mathbf{k+q},m,\eta,s}c_{\mathbf{k},n,\eta,s}$ is the density operator in band basis. Note that $O(\mathbf{r})$ and $O(\mathbf{r}')$ generically do not commute and hence the Hamiltonian is not solvable. The interaction can in general be rewritten as
\begin{equation}
H_{I}=\frac{1}{2\Omega_{\text{tot}}}\sum_{\mathbf{G}}\left[\sum_{\mathbf{q}} (O_{\mathbf{q,G}}-A_{\GG} N_M\delta_{\mathbf{q},0}) (O_{\mathbf{-q,-G}}-A_{-\GG} N_M\delta_{-\mathbf{q},0}) + 2A_{-\GG} N_M O_{\mathbf{0,G}} -A_{-\GG}A_\GG N_M^2 \right]\ ,
\label{p5:eq:shifted-HI}
\end{equation}
where $N_M$ is the number of moir\'e unit cells, and $A_{\GG}$ is some arbitrarily chosen $\GG$ dependent coefficient satisfying $A_\GG=A_{-\GG}^*$. Note that the first term in Eq. (\ref{p5:eq:shifted-HI}) is nonnegative. 

In Ref. \cite{ourpaper4} we found an important condition which can show that eigenstates of $H_I$ are in fact ground states of $H_I$. 
If the $\qq=0$ component of the matrix element in \cref{p5:eq:M-def} $M_{m,n}^{\left(\eta\right)}\left(\mathbf{k},\mathbf{G}\right)$ is not dependent on $k$ for all $\mathbf{G}$'s, \ie
\begin{equation}\label{p5:eqn-condition-at-nu}
\text{Flat Metric Condition:}\;\; M_{m,n}^{\left(\eta\right)}\left(\mathbf{k},\mathbf{G}\right)=\xi(\mathbf{G})\delta_{m,n}
\end{equation} 
then much more information about \cref{p5:eq:nonnegative-Hint} can be obtained. 
This condition is always true for $\mathbf{G}=0$, for which
$M_{m,n}^{\left(\eta\right)}\left(\mathbf{k},0\right) = \delta_{mn}$ from wavefunction normalization. 
In Ref. \cite{ourpaper1} we have showed that, around the first magic angle,  $M_{m,n}^{\left(\eta\right)}\left(\mathbf{k},\mathbf{G}\right) \approx 0$ for, $|\mathbf{G}|>\sqrt{3}k_\theta$ for $i=1,2$. 
Hence, the condition \cref{p5:eqn-condition-at-nu} is valid for all $\mathbf{G}$ with the exception of $\mathbf{G}$ for which $|\mathbf{G}|= \sqrt{3}k_\theta$. 
Hence, the condition is largely valid, and the numerical analysis \cite{ourpaper6} confirms its validity for a large part of the MBZ. 
In the below, we will always specify when the condition  \cref{p5:eqn-condition-at-nu} is used. 
If the \cref{p5:eqn-condition-at-nu} is satisfied, one has $O_{\mathbf{0,G}}$ proportional to the total electron number $N$ and the second term in  Eq. (\ref{p5:eq:shifted-HI}) is simply a chemical potential term
\begin{equation}\label{p5:eq:chemical-potential-shift}
\mu=\frac{1}{N_M\Omega_M}\sum_{\GG}A_{-\GG} \sqrt{V(\mathbf{G})}\sum_{\kk}M_{+1,+1}^{\left(\eta\right)}\left(\mathbf{k},\mathbf{G}\right)=\sum_{\GG}A_{-\GG} \sqrt{V(\mathbf{G})}\xi(\GG)/\Omega_M\ ,
\end{equation}
where $\Omega_M=\Omega_{\text{tot}}/N_M$ is the area of moir\'e unit cell. For a fixed total number of electrons, $N=\sum_{\kk,m,\eta,s} c^\dag_{\mathbf{k},m,\eta,s}c_{\mathbf{k},m,\eta,s} =(\nu+4) N_M$ is a constant, where $\nu$ is the filling fraction (number of doped electrons per moir\'e unit cell) relative to the charge neutrality point, thus the ground state at finite filling is solely determined by the first term which is non-negative.

\subsubsection{Exact ground states in the first chiral-flat \texorpdfstring{U(4)$\times$U(4)}{U(4)xU(4)} limit}\label{p5:exactgroundstatechirallimitRevappendix}

To build the excitations around a ground state, we review the ground states found in Ref. \cite{ourpaper4} of the projected Hamiltonian \cref{p5:eq:nonnegative-Hint}.  
We proved that in the Chern basis of \cref{p5:eq:chiral-MqG1} and \cref{p5:eq:Chern-band}, diagonal in the valley index $\eta$, spin index $s$ and Chern band index $e_Y$, the projected Hamiltonian \cref{p5:eq:nonnegative-Hint} has as eigenstates at integer filling $\nu$ the \emph{filled band} wavefunctions (without assuming condition \cref{p5:eqn-condition-at-nu}):
\begin{equation}\label{p5:eq:U(4)U(4)-GS}
|\Psi_{\nu}^{\nu_+,\nu_-}\rangle =\prod_{\mathbf{k}} \left(\prod_{j_1=1}^{\nu_+}d^\dag_{\mathbf{k},+1,\eta_{j_1},s_{j_1}} \prod_{j_2=1}^{\nu_-}d^\dag_{\mathbf{k},-1,\eta_{j_2},s_{j_2}}\right)|0\rangle,
\end{equation}
\begin{equation}\label{p5:eq:U(4)U(4)-GSenergy}
H_I|\Psi_{\nu}^{\nu_+,\nu_-}\rangle=\frac{1}{2\Omega_{\text{tot}}}\sum_{\qq,\GG}O_{\mathbf{-q,-G}}O_{\mathbf{q,G}} |\Psi_{\nu}^{\nu_+,\nu_-}\rangle = \frac{\nu^2}{2\Omega_{\text{tot}}}\sum_{\GG} V(\mathbf{G}) \Big(\sum_\kk\alpha_0(\mathbf{k,G})\Big)^2 |\Psi_{\nu}^{\nu_+,\nu_-}\rangle,
\end{equation}
where $\nu_+-\nu_-=\nu_C$ is the total Chern number of the state, and $\nu_++\nu_-=\nu+4$ is the total number of electrons per moir\'e unit cell in the projected bands, with $0\le \nu_\pm\le 4$, $\kk$ running over the entire MBZ. 
The occupied spin/valley indices $\{\eta_{j_1},s_{j_1}\}$ and $\{\eta_{j_2},s_{j_2}\}$ can be arbitrarily chosen. 
These eigenstates of \cref{p5:eq:nonnegative-Hint} are moreover eigenstates of the $O_{\mathbf{q,G}}$ operator in  \cref{p5:Oqdef1} \cite{ourpaper4}:

\begin{equation}\label{p5:eq:U(4)U(4)-OqG-action}
\begin{split}
&O_{\mathbf{q,G}}|\Psi_{\nu}^{\nu_+,\nu_-}\rangle =\delta_{\mathbf{q,0}} A_\mathbf{G}N_M |\Psi_{\nu}^{\nu_+,\nu_-}\rangle; \;\; \;A_\mathbf{G}=\frac{\sqrt{V(\mathbf{G})}}{N_M}\sum_\kk \nu\alpha_0(\mathbf{k,G});\;\; 
\end{split}
\end{equation}
In Ref. \cite{ourpaper4}, we found that the U(4)$\times$U(4) irrep of this multiplet is labeled by $\left([N_M^{\nu_+}]_4,[N_M^{\nu_-}]_4\right)$. For a fixed filling factor $\nu$, from Eq. (\ref{p5:eq:U(4)U(4)-GSenergy}) we found \cite{ourpaper4} that the states with different Chern number $\nu_C$ are all degenerate.

At charge neutrality $\nu=0$, the U(4)$\times$U(4) multiplet of eigenstate state $|\Psi_0^{\nu_+,\nu_-}\rangle$ with Chern number $\nu_C=\nu_+-\nu_-=0,\pm 2,\pm 4$ has exactly zero energy and hence are exact degenerate ground states.  At nonzero fillings $\nu$, we cannot guarantee that the $\nu\ne 0$ eigenstates are ground states (without condition \cref{p5:eqn-condition-at-nu}).

Assuming the flat metric condition Eq. (\ref{p5:eqn-condition-at-nu}), we showed \cite{ourpaper4} that we can rewrite the interaction into the form of Eq. (\ref{p5:eq:shifted-HI}), with the coefficient $A_\GG =\nu\sqrt{V(\mathbf{G})}\xi(\mathbf{G})$ in Eq. (\ref{p5:eq:U(4)U(4)-OqG-action}). By Eq. (\ref{p5:eq:U(4)U(4)-OqG-action}), we showed that $(O_{\mathbf{q,G}}-A_{\GG} N_M\delta_{\mathbf{q},0})$ annihilates $|\Psi_{\nu}^{\nu_+,\nu_-}\rangle$ 
for any $\nu_C=\nu_+-\nu_-$ and thus all the eigenstates $|\Psi_{\nu}^{\nu_+,\nu_-}\rangle$ with any Chern number $\nu_C=\nu_+-\nu_-$ are degenerate ground states at filling $\nu$ \cite{ourpaper4}.

\subsubsection{Exact ground states in the nonchiral-flat U(4) limit}\label{p5:exactgroundstatenonchirallimitRevappendix}

Without chiral symmetry, with U(4) symmetry Eq. (\ref{p5:eq:U4-generator}), $O_{\mathbf{q,G}}$ is no longer diagonal in any band basis (such as the Chern basis). Nevertheless, $O_{\mathbf{q,G}}$ is still diagonal in $\eta$ and $s$ and hence filling both $m=\pm$ bands, of any valley/spin is still an exact, Chern number $0$ eigenstates \cite{ourpaper4}: 
\begin{equation}\label{p5:eq:U(4)-GS}
|\Psi_\nu\rangle=\prod_{\mathbf{k}} \left(\prod_{j=1}^{(\nu+4)/2}c^\dag_{\mathbf{k},+,\eta_{j},s_{j}} c^\dag_{\mathbf{k},-,\eta_{j},s_{j}}\right)|0\rangle =\prod_{\mathbf{k}} \left(\prod_{j=1}^{(\nu+4)/2}d^\dag_{\mathbf{k},+1,\eta_{j},s_{j}} d^\dag_{\mathbf{k},-1,\eta_{j},s_{j}}\right)|0\rangle\ ,
\end{equation} for even fillings $\nu=0,\pm2,\pm4$, where $\{\eta_j,s_j\}$ are distinct valley-spin flavors which are fully occupied. 
With $M_{m,n}^{\left(\eta\right)}\left(\mathbf{k},\mathbf{q}+\mathbf{G}\right)$ in Eq. (\ref{p5:eq:M-para}), we have the same eigenvalue expression as in \cref{p5:eq:U(4)U(4)-OqG-action}, $ O_{\mathbf{q,G}}|\Psi_\nu\rangle=\nu \sqrt{V(\mathbf{G})}\delta_{\mathbf{q,0}} \sum_{\kk,m,\eta,s} \alpha_0\left(\mathbf{k},\mathbf{G}\right) |\Psi_\nu\rangle$. 
Along with any U(4) rotation, it is an eigenstate of $H_I$, without using condition Eq. (\ref{p5:eqn-condition-at-nu}). 
Moreover, for  $\nu=0$, the state (\ref{p5:eq:U(4)-GS}) is always a ground state with or without condition Eq. (\ref{p5:eqn-condition-at-nu}) \cite{ourpaper4}. Furthermore if the condition Eq. (\ref{p5:eqn-condition-at-nu}) is satisfied, by choosing $A_\GG =\nu\sqrt{V(\mathbf{G})}\xi(\mathbf{G})$ we have showed in Ref.~\cite{ourpaper4} that the states in Eq. (\ref{p5:eq:U(4)-GS}) are Chern number zero exact ground states. 
The multiplet of states forms a U(4) irrep  $[(2N_M)^{(\nu+4)/2}]_4$ \cite{ourpaper4}.

\section{Charge commutation relations} \label{p5:ChargeCommutationRelationsAppendix}
In order to compute the charge 0, $\pm1$, $\pm2$ excitations, a series of commutators are needed. We provide their expressions here. 

\subsection{The non-chiral case} \label{p5:ChargeCommutationRelationsNonChiralAppendix}

In the non-chiral case of \cref{p5:eq:M-para} we have:
\begin{equation}
\begin{split}
&[O_{\mathbf{q,G}}, c_{\mathbf{k},n,\eta,s}^\dag]= \sum_{\mathbf{k}',m,n',\eta',s'} \sqrt{V(\mathbf{G}+\mathbf{q})} M_{m,n'}^{\left(\eta\right)}\left(\mathbf{k}',\mathbf{q}+\mathbf{G}\right)[\rho_{\mathbf{k',q},m,n',s'}^{\eta'}, c_{\mathbf{k},n,\eta,s}^\dag]\\
=&\sum_{\mathbf{k}',m,n',\eta',s'} \sqrt{V(\mathbf{G}+\mathbf{q})} M_{m,n'}^{\left(\eta\right)}\left(\mathbf{k}',\mathbf{q}+\mathbf{G}\right) c_{\mathbf{k'+q},m,\eta',s'}^\dag \{c_{\mathbf{k'},n',\eta',s'}, c_{\mathbf{k},n,\eta,s}^\dag\}\\
=&\sum_{m}\sqrt{V(\mathbf{G}+\mathbf{q})} M_{m,n}^{\left(\eta\right)}\left(\mathbf{k},\mathbf{q}+\mathbf{G}\right) c_{\mathbf{k+q},m,\eta,s}^\dag \ .
\end{split}
\end{equation}
and 
\begin{equation}
\begin{split}
&[O_{\mathbf{q,G}}, c_{\mathbf{k},n,\eta,s}]= \sum_{\mathbf{k}',m,n',\eta',s'} \sqrt{V(\mathbf{G}+\mathbf{q})} M_{m,n'}^{\left(\eta\right)}\left(\mathbf{k}',\mathbf{q}+\mathbf{G}\right)[\rho_{\mathbf{k',q},m,n',s'}^{\eta'}, c_{\mathbf{k},n,\eta,s}]\\
=&- \sum_{\mathbf{k}',m,n',\eta',s'} \sqrt{V(\mathbf{G}+\mathbf{q})} M_{m,n'}^{\left(\eta\right)}\left(\mathbf{k}',\mathbf{q}+\mathbf{G}\right) \{c_{\mathbf{k'+q},m,\eta',s'}^\dag , c_{\mathbf{k},n,\eta,s}\} c_{\mathbf{k'},n',\eta',s'}\\
=&-\sum_{m}\sqrt{V(\mathbf{G}+\mathbf{q})} M_{n,m}^{\left(\eta\right)}\left(\mathbf{k}-\mathbf{q},\mathbf{q}+\mathbf{G}\right) c_{\mathbf{k-q},m,\eta,s} \\
=&-\sum_{m}\sqrt{V(\mathbf{G}+\mathbf{q})} M_{m,n}^{\left(\eta \right)* }\left(\mathbf{k},-\mathbf{q}-\mathbf{G}\right) c_{\mathbf{k-q},m,\eta,s} \ .
\end{split}
\end{equation} 
where we have used the property \cite{ourpaper3} $M_{m,n}^{\left(\eta \right)* }\left(\mathbf{k},-\mathbf{q}-\mathbf{G}\right) = M_{n,m}^{\left(\eta\right)}\left(\mathbf{k}-\mathbf{q},\mathbf{q}+\mathbf{G}\right) $. 
From these basic equations, we further find
\begin{equation}\label{p5:eqncommOOd}
\begin{split}
&\ [O_{\mathbf{-q,-G}}O_{\mathbf{q,G}}, c_{\mathbf{k},n,\eta,s}^\dag]=O_{-\mathbf{q,-G}}[O_{\mathbf{q,G}}, c_{\mathbf{k},n,\eta,s}^\dag]+ [O_{-\mathbf{q,-G}}, c_{\mathbf{k},n,\eta,s}^\dag] O_{\mathbf{q,G}} \\
=&O_{-\mathbf{q,-G}}\sum_{m}\sqrt{V(\mathbf{G}+\mathbf{q})} M_{m,n}^{\left(\eta\right)}\left(\mathbf{k},\mathbf{q}+\mathbf{G}\right) c_{\mathbf{k+q},m,\eta,s}^\dag + \sum_{m}\sqrt{V(\mathbf{G}+\mathbf{q})} M_{m,n}^{\left(\eta\right)}\left(\mathbf{k},-\mathbf{q}-\mathbf{G}\right) c_{\mathbf{k-q},m,\eta,s}^\dag O_{\mathbf{q,G}} \\
=&\sum_{m',m}V(\mathbf{G}+\mathbf{q}) M_{m',m}^{\left(\eta\right)}\left(\mathbf{k+q},-\mathbf{q}-\mathbf{G}\right) M_{m,n}^{\left(\eta\right)}\left(\mathbf{k},\mathbf{q}+\mathbf{G}\right) c_{\mathbf{k},m',\eta,s}^\dag \\
&+\sum_{m}\sqrt{V(\mathbf{G}+\mathbf{q})} M_{m,n}^{\left(\eta\right)}\left(\mathbf{k},\mathbf{q}+\mathbf{G}\right) c_{\mathbf{k+q},m,\eta,s}^\dag O_{-\mathbf{q,-G}} + \sum_{m}\sqrt{V(\mathbf{G}+\mathbf{q})} M_{m,n}^{\left(\eta\right)}\left(\mathbf{k},-\mathbf{q}-\mathbf{G}\right) c_{\mathbf{k-q},m,\eta,s}^\dag O_{\mathbf{q,G}}\ .
\end{split}
\end{equation}
and
\begin{equation}\label{p5:eqncommOOdminus}
\begin{split}
&\ [O_{\mathbf{-q,-G}}O_{\mathbf{q,G}}, c_{\mathbf{k},n,\eta,s}]=O_{-\mathbf{q,-G}}[O_{\mathbf{q,G}}, c_{\mathbf{k},n,\eta,s}]+ [O_{-\mathbf{q,-G}}, c_{\mathbf{k},n,\eta,s}] O_{\mathbf{q,G}} \\
=&-O_{-\mathbf{q,-G}}\sum_{m}\sqrt{V(\mathbf{G}+\mathbf{q})} M_{m,n}^{\left(\eta\right)*}\left(\mathbf{k},-\mathbf{q}+- \mathbf{G}\right) c_{\mathbf{k+q},m,\eta,s} - \sum_{m}\sqrt{V(\mathbf{G}+\mathbf{q})} M_{m,n}^{\left(\eta\right)}\left(\mathbf{k},\mathbf{q}+\mathbf{G}\right) c_{\mathbf{k+q},m,\eta,s} O_{\mathbf{q,G}} \\
=&\sum_{m',m}V(\mathbf{G}+\mathbf{q}) M_{m',m}^{\left(\eta\right)*}\left(\mathbf{k-q},\mathbf{q}+\mathbf{G}\right) M_{m,n}^{\left(\eta\right)*}\left(\mathbf{k},-\mathbf{q}-\mathbf{G}\right) c_{\mathbf{k},m',\eta,s}\\
&-\sum_{m}\sqrt{V(\mathbf{G}+\mathbf{q})} M_{m,n}^{\left(\eta\right)*}\left(\mathbf{k},-\mathbf{q}-\mathbf{G}\right) c_{\mathbf{k-q},m,\eta,s} O_{-\mathbf{q,-G}} - \sum_{m}\sqrt{V(\mathbf{G}+\mathbf{q})} M_{m,n}^{\left(\eta\right)*}\left(\mathbf{k},\mathbf{q}+\mathbf{G}\right) c_{\mathbf{k+q},m,\eta,s} O_{\mathbf{q,G}}\ .
\end{split}
\end{equation} Using  $M_{m',m}^{\left(\eta\right)}\left(\mathbf{k+q},-\mathbf{q}-\mathbf{G}\right)= M_{m, m'}^{\left(\eta\right) *}\left(\mathbf{k},\mathbf{q}+\mathbf{G}\right) $ and $M_{m',m}^{\left(\eta\right)*}\left(\mathbf{k-q},+\mathbf{q}+\mathbf{G}\right)=  M_{m,m'}^{\left(\eta\right)}\left(\mathbf{k},-\mathbf{q}-\mathbf{G}\right) $, we have

\begin{equation}
\begin{split}
\ [O_{\mathbf{-q,-G}}O_{\mathbf{q,G}}, c_{\mathbf{k},n,\eta,s}^\dag] =& \sum_{m}P_{mn}^{\left(\eta\right)}\left(\mathbf{k},\mathbf{q}+\mathbf{G}\right)c_{\mathbf{k},m,\eta,s}^\dag +\sum_{m}\sqrt{V(\mathbf{G}+\mathbf{q})} M_{m,n}^{\left(\eta\right)}\left(\mathbf{k},\mathbf{q}+\mathbf{G}\right) c_{\mathbf{k+q},m,\eta,s}^\dag O_{-\mathbf{q,-G}} \\
& + \sum_{m}\sqrt{V(\mathbf{G}+\mathbf{q})} M_{m,n}^{\left(\eta\right)}\left(\mathbf{k},-\mathbf{q}-\mathbf{G}\right) c_{\mathbf{k-q},m,\eta,s}^\dag O_{\mathbf{q,G}}\\
[O_{\mathbf{-q,-G}}O_{\mathbf{q,G}}, c_{\mathbf{k},n,\eta,s}] =& \sum_{m}  P_{mn}^{\left(\eta\right)*}\left(\mathbf{k},\mathbf{q}+\mathbf{G}\right) c_{\mathbf{k},m,\eta,s}-\sum_{m}\sqrt{V(\mathbf{G}+\mathbf{q})} M_{m,n}^{\left(\eta\right)*}\left(\mathbf{k},-\mathbf{q}-\mathbf{G}\right) c_{\mathbf{k-q},m,\eta,s} O_{-\mathbf{q,-G}} \\
&- \sum_{m}\sqrt{V(\mathbf{G}+\mathbf{q})} M_{m,n}^{\left(\eta\right)*}\left(\mathbf{k},+\mathbf{q}+\mathbf{G}\right) c_{\mathbf{k+q},m,\eta,s} O_{\mathbf{q,G}}\ .
\end{split}
\end{equation}
where we define the new matrix element $P=V M^\dagger M$, the convolution of the Coulomb potential and the form factor matrices
\begin{equation}
 \boxed{ P_{mn}^{\left(\eta\right)}\left(\mathbf{k},\mathbf{q}+\mathbf{G}\right)  =\sum_{m'} V(\mathbf{G}+\mathbf{q}) M_{m',m}^{\left(\eta\right)*}\left(\mathbf{k},\mathbf{q}+\mathbf{G}\right) M_{m',n}^{\left(\eta\right)}\left(\mathbf{k},\mathbf{q}+\mathbf{G}\right)  =  V(\mathbf{G}+\mathbf{q}) (M^{\left(\eta\right)\dagger }M^{\left(\eta\right)})_{mn} \left(\mathbf{k},\mathbf{q}+\mathbf{G}\right) }
\end{equation}
These are the commutators needed to obtain the wavefunctions and energies of the excitations in the non-chiral limit.

\subsection{The first chiral limit} \label{p5:ChargeCommutationRelationsChiralAppendix}

In the first chiral limit, we can use the Chern band basis \cref{p5:eq:chiral-MqG1}, where the $O_{\mathbf{q,G}}$ is diagonal in the Chern basis.
Its form factors do not depend on the valley $\eta$, and spin $s$:
\begin{align} \label{p5:OqGOperatorInChiralLimit}
&O_{\mathbf{q,G}}=\sum_{k}  \sum_{e_Y = \pm} \sqrt{V(\mathbf{G}+\mathbf{q})} M_{e_Y}(\kk,\qq+\GG) \sum_{\eta, s} (d^\dagger_{\kk + \qq,e_Y,\eta,s} d_{\kk,e_Y,\eta,s}-\frac{1}{2}\delta_{\mathbf{q,0}})
\end{align}
In this limit, the commutators between  $O_{\mathbf{q,G}}$ and the Chern number $e_Y=\pm 1$ band creation operators   become simpler

\begin{equation}
\begin{split}
&[O_{\mathbf{q,G}}, d_{\mathbf{k},e_Y,\eta,s}^\dag]= \sqrt{V(\mathbf{G}+\mathbf{q})} M_{e_Y}(\kk,\qq+\GG) d^\dagger_{\kk + \qq,e_Y,\eta,s} \ .
\end{split}
\end{equation}
and 
\begin{equation}
\begin{split}
&[O_{\mathbf{q,G}}, d_{\mathbf{k},e_Y,\eta,s}]=-\sqrt{V(\mathbf{G}+\mathbf{q})}M^*_{e_Y}(\kk,-\qq-\GG)d_{\kk - \qq,e_Y,\eta,s}    .
\end{split}
\end{equation} leading to the commutators

\begin{equation}\label{p5:chirallimitOqGcommutators}
\begin{split}
& [O_{\mathbf{-q,-G}}O_{\mathbf{q,G}}, d_{\mathbf{k},e_Y,\eta,s}^\dag]\\
=& P_{e_Y}\left(\mathbf{k},\mathbf{q}+\mathbf{G}\right)d_{\mathbf{k},m,\eta,s}^\dag +\sqrt{V(\mathbf{G}+\mathbf{q})}  (M_{e_Y}\left(\mathbf{k},\mathbf{q}+\mathbf{G}\right) d_{\mathbf{k+q},e_Y,\eta,s}^\dag O_{-\mathbf{q,-G}} + M_{e_Y}\left(\mathbf{k},-\mathbf{q}-\mathbf{G}\right) d_{\mathbf{k-q},e_Y,\eta,s}^\dag O_{\mathbf{q,G}})\\
& [O_{\mathbf{-q,-G}}O_{\mathbf{q,G}}, d_{\mathbf{k},e_Y,\eta,s}] \\
=& P_{e_Y}^* \left(\mathbf{k},\mathbf{q}+\mathbf{G}\right) d_{\mathbf{k},e_Y,\eta,s}-\sqrt{V(\mathbf{G}+\mathbf{q})}( M_{e_Y}^*\left(\mathbf{k},-\mathbf{q}-\mathbf{G}\right) d_{\mathbf{k-q},e_Y,\eta,s} O_{-\mathbf{q,-G}} +M_{e_Y}^{*}\left(\mathbf{k},\mathbf{q}+\mathbf{G}\right) d_{\mathbf{k+q},e_Y,\eta,s} O_{\mathbf{q,G}}) \ .
\end{split}
\end{equation}
where $P=V M^\dagger M$, the convolution of the Coulomb potential and the form factor matrices, takes the chiral limit form \begin{equation} \label{p5:Pinchirallimit}
 \boxed{ P_{e_Y}\left(\mathbf{k},\mathbf{q}+\mathbf{G}\right)  = V(\mathbf{G}+\mathbf{q})|M_{e_Y}\left(\mathbf{k},\mathbf{q}+\mathbf{G}\right)|^2  =   V(\mathbf{G}+\mathbf{q})(\alpha^2_0(\mathbf{k,q+G})+\alpha^2_2(\mathbf{k,q+G}) = P\left(\mathbf{k},\mathbf{q}+\mathbf{G}\right) ), }
\end{equation} 
where $\alpha_0(\mathbf{k,q+G}),\alpha_2(\mathbf{k,q+G})$ are the decomposition of the form factors in \cref{p5:eq:chiral-MqG1}. 
Notice in the Chern basis, $P_{e_Y}\left(\mathbf{k},\mathbf{q}+\mathbf{G}\right) $ does not depend on $e_Y$, so we just denote it as $P\left(\mathbf{k},\mathbf{q}+\mathbf{G}\right) $.
These are the commutators needed to obtain the wavefunctions and energies of the excitations in the first chiral limit. 

\section{Charge \texorpdfstring{$\pm1$}{+-1} excitations of the exact ground states}\label{p5:charge1excitationappendix}

Remarkably, the existence of exact ground states and/or eigenstates \cref{p5:eq:U(4)U(4)-GS,p5:eq:U(4)-GS} allows for the presence of more eigenstates. 
In fact, a part of the low energy spectrum can be computed with polynomial efficiency. 
In this appendix, we give the exact charge $\pm1$ excitations on top of the exact (ground) states given in Ref. \cite{ourpaper4} and reviewed in \cref{p5:sec:exact-GS}.

\subsection{Exact charge +1 excitations in the nonchiral-flat U(4) limit} \label{p5:charge1excitationBandappendix}
 
To look for the charge one excitations (adding an electron into the system), we sum the commutators in \cref{p5:eqncommOOd,p5:chirallimitOqGcommutators} over $\qq, \GG$ and use the shifted Hamiltonian in \cref{p5:eq:shifted-HI}.  
For a generic exact eigenstate $|\Psi\rangle$ at chemical potential $\mu$ satisfying $(O_{\mathbf{q,G}}-A_{\GG} N_M\delta_{\mathbf{q},0})|\Psi \rangle =0$ for some coefficient $A_\GG$, we find
\begin{equation}\label{p5:eq:excitation-HIv3}
\left[H_I-\mu N, c_{\mathbf{k},n,\eta,s}^\dag \right] |\Psi\rangle =\frac{1}{2\Omega_{\text{tot}}} \sum_{m} R_{mn}^\eta(\mathbf{k}) c_{\mathbf{k},m,\eta,s}^\dag|\Psi\rangle\ ,
\end{equation}
where $N$ is the electron number operator, and the matrix
\begin{align}\label{p5:eq:excitation-Rmn}
R_{mn}^\eta(\mathbf{k}) =& \sum_{\GG}\left[\left(\sum_{\qq, m'}V(\mathbf{G}+\mathbf{q}) M_{m',m}^{\left(\eta\right)*}\left(\mathbf{k},\mathbf{q}+\mathbf{G}\right) M_{m',n}^{\left(\eta\right)}\left(\mathbf{k},\mathbf{q}+\mathbf{G}\right)\right) +2N_M A_{-\GG} \sqrt{V(\mathbf{G})} M_{m,n}^{\left(\eta\right)}\left(\mathbf{k},\mathbf{G}\right)\right]-\mu\delta_{mn} \nonumber \\
=& \frac{1}{2\Omega_{\text{tot}}}\sum_{\GG}\left[\left(\sum_{\qq}     P_{mn}^{\left(\eta\right)}\left(\mathbf{k},\mathbf{q}+\mathbf{G}\right) \right)  +2N_M A_{-\GG} \sqrt{V(\mathbf{G})} M_{m,n}^{\left(\eta\right)}\left(\mathbf{k},\mathbf{G}\right)\right]-\mu\delta_{mn}\ .
\end{align} 
We hence see that , if $|\Psi\rangle$  is  one of the $|\Psi_{\nu}^{\nu_+,\nu_-}\rangle$  (\cref{p5:eq:U(4)U(4)-GS}) or $|\Psi_\nu\rangle$ (\cref{p5:eq:U(4)-GS}) eigenstates of $H_I$, then $c_{\mathbf{k},m,\eta,s}^\dag|\Psi\rangle\ $ is also an eigenstate of $H_I$ with eigenvalues obtained by diagonalizing the $2 \times 2$ matrix $R_{mn}^\eta(\mathbf{k})$. 
In the case of TBG, this is a $2\times 2$ matrix, hence the diagonalization can be done by hand, providing a band of excitations. We note that, the expression $c_{\mathbf{k},n,\eta,s}^\dag|\Psi\rangle=0$ may vanish and give no charge excitation, for instance, if valley $\eta$ and spin $s$ is fully occupied. We now delve more into the energies and eigenstates $c_{\mathbf{k},m,\eta,s}^\dag|\Psi\rangle$. 

Due to the symmetry $C_{2z}P$ (\cref{p5:eq:gauge-0}), the $M$ matrix (\cref{p5:eq:M-para}) satisfies $M^{(\eta)}_{m,n}(\kk,\qq+\GG)=mnM^{(-\eta)}_{-m,-n}(\kk,\qq+\GG)$.
Correspondingly, the $R$ matrix satisfies
\begin{equation} \label{p5eq:Reta-eta}
R^{\eta}_{m,n}(\kk) = mn R^{-\eta}_{-m,-n} (\kk).
\end{equation}
Since $R^{+}(\kk)$ and $R^{-}(\kk)$ are related by a unitary transformation, they must have the same spectrum.

\subsubsection{Band of charge 1 excitation in the nonchiral-flat U(4) limit}\label{p5:charge1excitationBandNonChiralappendix}

In the nonchiral limit, the eigenstates $|\Psi_\nu\rangle$ we found in Ref. \cite{ourpaper4} (and re-written in Eq. (\ref{p5:eq:U(4)-GS}))  have only fully occupied or fully empty valley $\eta$ and spin $s$ flavors. 
For TBG, this means that both active bands $m=\pm$ are either full or empty for each valley $\eta$ and spin $s$.  
In this case we can only obtain exact charge +1 excitation at even fillings, \ie $\nu=0,\pm2$.
We can consider two charge $+1$ states $c_{\mathbf{k},n,\eta,s}^\dag|\Psi\rangle$ ($n=\pm$) at a fixed $\kk$ in a fully empty valley $\eta$ and spin $s$. These two states then form a closed subspace with a $2\times2$ subspace Hamiltonian $R^\eta(\mathbf{k})$ defined by Eq. (\ref{p5:eq:excitation-Rmn}). 
Diagonalizing the matrix $R^\eta(\mathbf{k})$ then gives the excitation eigenstates and excitation energies. 
It is worth noting that, due to \cref{p5eq:Reta-eta}, the spectrum of $R^\eta(\kk)$ does {\it not} depend on $\eta$. 
Since the U(4) irrep of the ground state is $|\Psi_\nu\rangle$ is $[(2N_M)^{(\nu+4)/2}]_4$, the U(4) irrep of the charge 1 excited state is given by $[(2N_M)^{(\nu+4)/2},1]_4$. 
Furthermore, at $\nu=0$, the state $|\Psi_{\nu=0}\rangle$ in Eq. (\ref{p5:eq:U(4)-GS}) is the \emph{ground state} of the interaction Hamiltonian $H_I$ and hence $c_{\mathbf{k},n,\eta,s}^\dag|\Psi_{\nu=0} \rangle$ is the charge excitation above the ground state.  
Note that this does not assume the ``flat metric condition" (\ref{p5:eqn-condition-at-nu}) and is hence fully generic.

If we further assume the flat band condition Eq.~(\ref{p5:eqn-condition-at-nu}), the eigenstates $|\Psi_\nu\rangle$ become exact ground states, and the chemical potential is given by Eq. (\ref{p5:eq:chemical-potential-shift}). In this case, the $2\times2$ excitation sub-Hamiltonian $R^\eta(\mathbf{k})$ takes a simpler form:
\begin{equation}\label{p5:eq:excitation-Rmn-GS}
R_{mn}^\eta(\mathbf{k})=   \sum_{q, G}    P_{mn}^{\left(\eta\right)}\left(\mathbf{k},\mathbf{q}+\mathbf{G}\right)   =   \sum_{\GG,\qq, m'}V(\mathbf{G}+\mathbf{q}) M_{m',m}^{\left(\eta\right)*}\left(\mathbf{k},\mathbf{q}+\mathbf{G}\right) M_{m',n}^{\left(\eta\right)}\left(\mathbf{k},\mathbf{q}+\mathbf{G}\right)\ ,
\end{equation}
which can be diagonalized to give the band excitation eigenstates and energies  above the ground state at each momentum $\kk$.

\subsubsection{Spectrum properties of a generic charge 1 excitation in the nonchiral-flat U(4) limit} \label{p5:charge1excitationSpectrumPropertiesappendix}

The spectrum at every $\kk$ is obtained from diagonalizing the matrix $R^\eta(\mathbf{k}) = \sum_{\GG,\qq}V(\mathbf{G}+\mathbf{q}) M^{\left(\eta\right)\dagger}\left(\mathbf{k},\mathbf{q}+\mathbf{G}\right) M^{\left(\eta\right)}\left(\mathbf{k},\mathbf{q}+\mathbf{G}\right)$, which depends (up to a convolution with the Coulomb potential), only on the projected band wavefunctions. 
This is clearly a sum (over $\qq, \GG$) of positive semidefinite matrices (remember that $ V(\mathbf{G}+\mathbf{q})>0$). 
Hence $R^\eta(\mathbf{k})$ is a positive semidefinite matrix, whose eigenvalues are non-negative (expected, since we proved that these are excitations \emph{above} the ground state). 
We now find conditions that these excitations are gapped, \ie that the matrix $R^\eta(\mathbf{k}) $ is positive \emph{definite} at each $\kk$. 
We now show this by re-writing the $R^\eta(\mathbf{k}) $ as
\begin{equation}
R_{mn}^\eta(\mathbf{k}) =     (M^{\left(\eta\right)\dagger}(\kk) V  M^{\left(\eta\right)}(\kk))_{mn}, 
\end{equation}
where now $ M^{\left(\eta\right)}(k)$ is a matrix of $(2N_M \cdot N_{\GG}) \times 2$ matrix (with $2$ because we are projecting into the two active TBG bands), where $N_M$ is number of moir\'e unit cells, $N_\GG$ is the number of plane waves (MBZs) taken into consideration. 
In Ref. \cite{ourpaper1} we have showed that the number of plane waves needed is very small: the matrix elements fall off exponentially with $|\GG|$ and any contribution above $|\GG|=\sqrt{3}k_\theta$ is negligible. 
The matrix elements read $  M^{\left(\eta\right)}_{\{m\qq\GG\},n} (\kk)= M_{m,n}^{\left(\eta\right)}\left(\mathbf{k},\mathbf{q}+\mathbf{G}\right)$. 
$V$ is a  $2N_M\cdot N_{\GG} \times 2N_M \cdot N_{\GG}  $ diagonal matrix with elements $V_{\{m\qq\GG\},\{m'\qq'\GG'\}  } = \delta_{m,m'}\delta_{\qq,\qq'}  \delta_{\GG,\GG'} V(\qq+\GG)$. 
Since $V(\qq+\GG)\ge0$ and diagonal, we can re-write $R^\eta(\kk)= (\sqrt{V} M^\eta(\kk))^\dagger \sqrt{V} M^\eta(\kk)$, and its rank is equal to $\mrm{Rank}(\sqrt{V} M^\eta(\kk))\le2$.
We show the rank has to be $2$ (or in general the number of occupied bands), by the simple argument
\begin{align}
R_{mn}^\eta(\mathbf{k}) &=  \sum_{m} V(0) M_{m',m}^{\left(\eta\right)}\left(\mathbf{k}, 0\right) M_{m,n}^{\left(\eta\right)}\left(\mathbf{k}, 0\right) +   \sum_{\{\qq,\GG\} \ne \{0,0\}} \sum_{p,p'=1,2}  M^{\left(\eta\right)\dagger}_{\{p\qq\GG\},m} (\kk) V_{\{p \qq\GG\},\{p'\qq'\GG'\}  } M^{\left(\eta\right)}_{\{p'\qq'\GG'\},n} (\kk) \nonumber \\ 
&=      V(0) \delta_{mn} +    \sum_{\{\qq,\GG\} \ne \{0,0\}} \sum_{p,p'=1,2}  M^{\left(\eta\right)\dagger}_{\{p\qq\GG\},m} (\kk) V_{\{p \qq\GG\},\{p'\qq'\GG'\}  }  M^{\left(\eta\right)}_{\{p'\qq'\GG'\},n} (\kk). \end{align} 
The second term is still a positive semidefinite matrix, while the first term is diagonal and has eigenvalues  $V(0)/2\Omega_{\text{tot}}$. 
Hence by Weyl's theorem, the energies of the excited states are  $\ge V(0)/2\Omega_{\text{tot}}$. 
In general, our discussion  shows that the states $c_{\mathbf{k},n,\eta,s}^\dag|\Psi\rangle$ are not degenerate to the ground state $|\Psi\rangle$ (note that we did not prove these are the unique ground states). 
However, we cannot exclude a gapless excitation.

\subsubsection{ Spectrum relation to the quantum distance and  generic argument for the existence of a charge gap}\label{p5:charge1excitationGapappendix}

In general, however, it seems that this method gives rise to finite gap charge 1 excitations.  
The largest gap happens in the atomic limit or a material, where $\langle{ u_{\kk+\qq}^m} | u_\kk^n\rangle = \delta_{mn} $, for which $R_{mn} =\delta_{mn} \sum_{\qq,\GG} V(\qq + \GG) = \delta_{mn} \Omega_{\rm tot} V(\bf{r}=0)$. 
Since we know that TBG is far away from an atomic limit - the bands being topological, we expect a reduction in this gap. However, we argue  that this type of charge excitation is always gapped. We perform a different decomposition of the matrix $R_{mn}^\eta$:
\begin{align}
R_{mn}^\eta(\mathbf{k})  &=  \sum_{\qq} \sum_{p,p'=1,2}  
    M^{\left(\eta\right)\dagger}_{\{p\qq0 \},m} (\kk) V_{\{p \qq 0 \},\{p'\qq' 0 \}  }  M^{\left(\eta\right)}_{\{p'\qq'0\},n} (\kk)  \nonumber \\ 
& +  \sum_{\{\qq,\GG\} \ne \{0,0\}} \sum_{p,p'=1,2}  M^{\left(\eta\right)\dagger}_{\{p\qq\GG\},m} (\kk) V_{\{p \qq\GG\},\{p'\qq'\GG'\}  }  M^{\left(\eta\right)}_{\{p'\qq'\GG'\},n} (\kk) \ . 
\end{align} 
Since the second term is still a semi positive definite matrix by construction, the eigenvalues of $R^\eta(\kk)$ will be greater or larger than the eigenvalues of the first term. 
In fact, due to \cite{ourpaper1}, we know that the eigenvalues of the second term are negligible for $|\GG| \ge 2|\bb_{M1}|$. 
Hence, using the math notation of  $A \ge B$ for $A-B$ positive semidefinite, we have
\begin{align}
R_{mn}^\eta(\mathbf{k}) \ge &    \sum_{\qq} \sum_{p,p'=1,2}  M^{\left(\eta\right)\dagger}_{\{p\qq0 \},m} (\kk) V_{\{p \qq 0 \},\{p'\qq' 0 \}  }  M^{\left(\eta\right)}_{\{p'\qq'0\},n} (\kk)    \nonumber \\ 
&=  \sum_\qq V(\qq) M_{m',m}^{\left(\eta\right)*}\left(\mathbf{k},\mathbf{q}\right) M_{m',n}^{\left(\eta\right)}\left(\mathbf{k},\mathbf{q}\right)=    \sum_\qq V(\qq) (\delta_{mn} - 	\mathfrak{G}_\eta^{mn}(\kk,\qq)) \ . 
\end{align}
We call  $\mathfrak{G}^{mn}(\kk,\qq)$  the  generalized ``quantum geometric tensor'', whose trace is the generalized Fubini-Study Metric.  
The property of the generalized quantum geometric tensor/Fubini Study metric is that they become the conventional quantum geometric tensor/Fubini Study metric for small transfer momentum $\bf{q}$. 
The tensor quantifies the distance between two eigenstates in momentum space. The conventional quantum geometric tensor is defined as: \begin{equation}
	\mathfrak{G}^{mn}_{ij}(\mathbf{k}) = \sum_{a,b=1}^N \partial_{k_i} u^*_{a,m}(\mathbf{k})\left(\delta_{a,b} - \sum_l^{n_\text{occupied}} u_{a,l}(\mathbf{k})u^*_{b,l}(\mathbf{k}) \right)\partial_{k_j}u_{b,n}(\mathbf{k})\,,
\end{equation}
in which $m,n$ are energy band indices and $i,j$ are spatial direction indices of $n_{occupied}$ orthonormal vectors $u_m(\mathbf{k})$ in a $N$ dimensional Hilbert space, where $\mathbf{k}$ is some parameter. We can show that:
\begin{equation}
M_{m',m}^{\left(\eta\right)*}\left(\mathbf{k},\mathbf{q}\right) 
=\delta_{mn}- \mathfrak{G}_\eta^{mn}(\kk,\qq)
= M_{m',n}^{\left(\eta\right)}\left(\mathbf{k},\mathbf{q}\right) = \delta_{mn}- q_{i} q_{j} \mathfrak{G}^{mn}_{ij}(\mathbf{k}) + \mcl{O}(q^2)\ .  
\end{equation} 
Generically, we expect \cite{xie_superfluid_2020} that the overlap between two functions at $\kk$ and $\kk +\bf{q}$ to fall off as $\bf{q}$ increases, leaving a finite term in $R_{mn}^\eta(\mathbf{k})$, the electron gap, at every $\bf{k}$.

\subsection{Exact charge +1 excitations in the (first) chiral-flat  \texorpdfstring{U(4)$\times$U(4)}{U(4)xU(4)} limit}\label{p5:charge1excitationChiralLimitappendix}

For an eigenstate $|\Psi_{\nu}^{\nu_+,\nu_-}\rangle$ defined by Eq. (\ref{p5:eq:U(4)U(4)-GS}) in the (first) chiral-flat U(4)$\times$U(4) limit, one has the coefficients $M(\kk,\qq+\GG) = \zeta^0\tau^0 \alpha_0(\kk,\qq+\GG) + i\zeta^y\tau^0 \alpha_2(\kk,\qq+\GG)$. Without condition Eq.~(\ref{p5:eqn-condition-at-nu}), the eigenstate $|\Psi_{\nu}^{\nu_+,\nu_-}\rangle$ (which is not necessarily the ground state) satisfies Eq. (\ref{p5:eq:U(4)U(4)-OqG-action}), which is equivalent to choosing $N_MA_\GG=\nu\sqrt{V(\mathbf{G})} \sum_\kk \alpha_0(\mathbf{k,G})$ for Eq. (\ref{p5:eq:excitation-Rmn}). 
Using the relation (\ref{p5:eq:alpha-cond1}), we can simplify the matrix $R^\eta_{mn}(\kk)$ defined in Eq. (\ref{p5:eq:excitation-Rmn}) as
\begin{equation}
R^\eta(\kk)= R_0(\kk)\zeta^0\ ,
\end{equation}
where
\begin{equation} \label{p5:R0InTheChiralLimit}
R_0(\kk)= 
\sum_{\GG} \left[\left(\sum_\qq V(\GG+\qq) [\alpha_0 (\kk,\qq+\GG)^2 +\alpha_2 (\kk,\qq+\GG)^2] \right)+2N_M A_{-\GG}\sqrt{V(\mathbf{G})} \alpha_0 (\kk,\GG) \right]-\mu\ .
\end{equation}
Therefore, $R^\eta(\kk)$ is proportional to the identity matrix. Since the state $|\Psi_{\nu}^{\nu_+,\nu_-}\rangle$ is written in the Chern band basis defined in Eq. (\ref{p5:eq:Chern-band}), it is more convenient to work in the Chern band basis. We then find the charge excitation eigenstates with the corresponding excitation energy $R_0^\eta(\kk)$ given by
\begin{equation}\label{p5:eq:excitation-HIv2}
d^\dag_{\kk,e_Y,\eta,s}|\Psi_{\nu}^{\nu_+,\nu_-}\rangle\ , \;\;\; 
\left[H_I-\mu N, d_{\mathbf{k},e_Y,\eta,s}^\dag \right] |\Psi\rangle = \frac{1}{2\Omega_{\text{tot}}}R_0(\kk)  d_{\mathbf{k},e_Y,\eta,s}^\dag|\Psi\rangle\ ,
\end{equation}
provided the Chern band $e_Y (=\pm1)$ at valley $\eta$ and spin $s$ is fully empty. 
With condition Eq.~(\ref{p5:eqn-condition-at-nu}) assumed, the states $|\Psi_{\nu}^{\nu_+,\nu_-}\rangle$ become ground states. At the same time, with the chemical potential is given by Eq. (\ref{p5:eq:chemical-potential-shift}), we can simplify the excitation energy $R_0(\kk)$ into
\begin{equation} \label{p5:R0InTheChiralLimitWithFlatCondition}
R_0(\kk)=  
\sum_{\GG,\qq} V(\GG+\qq) [\alpha_0 (\kk,\qq+\GG)^2 +\alpha_2 (\kk,\qq+\GG)^2] =   \sum_{\GG,\qq} P(\kk, \qq+\GG)\ .
\end{equation}
independent on $\eta$, and with $P(\kk, \qq+\GG)$ defined in \cref{p5:Pinchirallimit}. 
Since the U(4)$\times$ U(4)  irrep of the ground state is $\left([N_M^{\nu_+}]_4,[N_M^{\nu_-}]_4\right)$, the U(4)$\times$U(4) irrep of the charge 1 excited states with $e_Y=1$ and $e_Y=-1$ are given by $\left([N_M^{\nu_+},1]_4,[N_M^{\nu_-}]_4\right)$ and $\left([N_M^{\nu_+}]_4,[N_M^{\nu_-},1]_4\right)$, respectively.

\subsection{Charge -1 excitations} \label{p5:charge-1excitationappendix}

The charge $-1$ excitations are obtained in a similar manner as the charge $+1$ excitations. Charge -1 excitations can be obtained by considering states $c_{\mathbf{k},n,\eta,s}|\Psi\rangle$ ($n=\pm$) at a fixed $\kk$ in a fully filled valley $\eta$ and spin $s$. By Hermitian conjugate of Eq. (\ref{p5:eqncommOOd}), we find similar to the charge 1 excitation,
\begin{equation}\label{p5:eq:excitation-HI-hole}
\left[H_I-\mu N, c_{\mathbf{k},n,\eta,s} \right] |\Psi\rangle =\frac{1}{2\Omega_{\text{tot}}} \sum_{m} \widetilde{R}_{mn}^\eta(\mathbf{k}) c_{\mathbf{k},m,\eta,s}|\Psi\rangle\ ,
\end{equation}
where $N$ is the electron number operator, and the matrix
\begin{align}\label{p5:eq:excitation-Rmn-hole}
& \widetilde{R}_{mn}^\eta(\mathbf{k})= \sum_{\GG}\left[\left(\sum_{\qq, m'}V(\mathbf{G}+\mathbf{q}) M_{m',m}^{\left(\eta\right)*}\left(\mathbf{k},\mathbf{q}+\mathbf{G}\right) M_{m',n}^{\left(\eta\right)}\left(\mathbf{k},\mathbf{q}+\mathbf{G}\right)\right)^* -2N_M A_{-\GG} \sqrt{V(\mathbf{G})} M_{m,n}^{\left(\eta\right)*}\left(\mathbf{k},\mathbf{G}\right)\right]+\mu\delta_{mn} \nonumber\\ & =\sum_{\GG}\left[ 
\left(\sum_{\qq}     P_{nm}^{\left(\eta\right)}\left(\mathbf{k},\mathbf{q}+\mathbf{G}\right) \right) -2N_M A_{-\GG} \sqrt{V(\mathbf{G})} M_{m,n}^{\left(\eta\right)*}\left(\mathbf{k},\mathbf{G}\right)\right]+\mu\delta_{mn}\ .
\end{align}
Note that $\widetilde{R}_{mn}^\eta(\mathbf{k})$ differs from Eq. (\ref{p5:eq:excitation-Rmn}) by a sign in the last two terms as well as by the complex conjugation of the first term. Diagonalizing $\widetilde{R}_{mn}^\eta(\mathbf{k})$ gives the charge $-1$ (hole) excitations. The chemical potential is the one  given in Eq. (\ref{p5:eq:chemical-potential-shift}).
In the (first) chiral limit, $\td{R}_mn^\eta$ becomes a two-by-two identity at each $\kk$ and independent with $\eta$, \ie $\td{R}^\eta_{mn}=\td{R}_0(\kk) \delta_{mn}$. 
The $R_0(\kk)$ function is given by
\begin{equation}
\td{R}_0(\kk)= 
\sum_{\GG} \left[\left(\sum_\qq V(\GG+\qq) [\alpha_0 (\kk,\qq+\GG)^2 +\alpha_2 (\kk,\qq+\GG)^2] \right)-2N_M A_{-\GG}\sqrt{V(\mathbf{G})} \alpha_0 (\kk,\GG) \right]+\mu\ .
\end{equation}

 If condition Eq.~(\ref{p5:eqn-condition-at-nu}) is satisfied, and $\mu$ is given in Eq. (\ref{p5:eq:chemical-potential-shift}), the charge $-1$ excitations have identical dispersion as that of the charge +1 excitations (Eq. (\ref{p5:eq:excitation-Rmn})). 
 If condition Eq.~(\ref{p5:eqn-condition-at-nu}) is not satisfied and $\nu\neq0$, the charge $+1$ and $-1$ excitations will have different dispersions. 
 The dispersions will also depend on filling $\nu$, since $A_\mathbf{G}$ defined in \cref{p5:eq:U(4)U(4)-OqG-action} depends on $\nu$.

Since U(4) irrep of the ground state $|\Psi_\nu\rangle$ is $[(2N_M)^{(\nu+4)/2}]_4$, in the non-chiral limit, the U(4) irrep of the charge $-1$ excited state is given by $[(2N_M)^{(\nu+2)/2},2N_M-1]_4$. Since the U(4)$\times$ U(4)  irrep of the ground state is $\left([N_M^{\nu_+}]_4,[N_M^{\nu_-}]_4\right)$, the U(4)$\times$U(4) irrep of the charge -1 excited states with $e_Y=1$ and $e_Y=-1$ are given by $\left([N_M^{\nu_+-1},N_M-1]_4,[N_M^{\nu_-}]_4\right)$ and $\left([N_M^{\nu_+}]_4,[N_M^{\nu_--1},N_M-1]_4\right)$, respectively. 

The computed charge gaps for the TBG Hamiltonian are of order $10$meV.

\begin{figure}[t]
\centering
\includegraphics[width=0.8\linewidth]{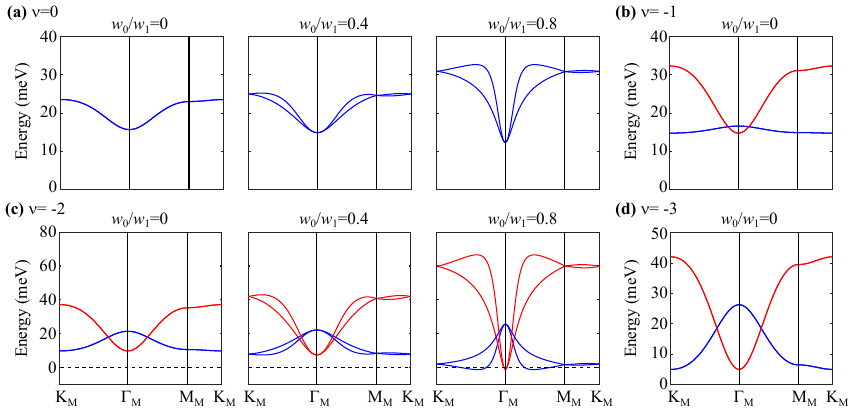}
\caption{Exact charge $+1$ (blue) and $-1$ (red) excitations at $\theta=1.05^\circ$. The flat metric condition is {\it not} imposed. 
In this plot we have used the parameters defined in \cref{p5:newapp:Ham}: $v_F=5.944{\rm eV \cdot \mathring{ A} }$, $|K|=1.703\mathring{\rm A}^{-1}$, $w_1=110{\rm meV}$,  $U_\xi=26$meV, $\xi=10$nm. Note that the excitation gap is largely reduced from the flat-condition limit.
}
\label{p5:fig:E1appendix}
\end{figure}

\begin{figure}
\centering
\includegraphics[width=0.8\linewidth]{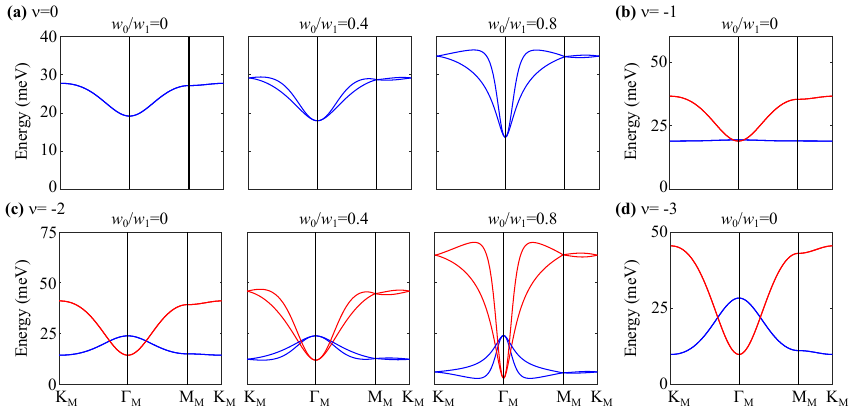}
\caption{Exact charge $+1$ (blue) and $-1$ (red) excitations at $\theta=1.05^\circ$. The flat metric condition is {\it not} imposed. 
In this plot we set the screening length as $\xi=20$nm and accordingly the interaction strength as $U_\xi=13$meV.
The other parameters are same as in \cref{p5:newapp:Ham}: $v_F=5.944{\rm eV \cdot \mathring{ A} }$, $|K|=1.703\mathring{\rm A}^{-1}$, $w_1=110{\rm meV}$.
}
\label{p5:fig:E1appendix20}
\end{figure}

\subsection{Charge \texorpdfstring{$\pm1$}{+-1} excitation spectra for different parameters}\label{newapp:plot1}

Before we present the numerical results, let us first explain how we choose the chemical potential in the cases without assuming flat-metric-condition. 
First, the sum of the lowest charge +1 ($\Delta_+$) and charge -1 ($\Delta_-$) gaps do not depend on the chemical potential since upon adding a $-\delta \mu N$ term the two gaps change by $-\delta \mu$ and $\delta \mu$, respectively. Then we choose the chemical potential such that $\Delta_+=\Delta_-$. 
In a band structure picture, $\Delta_+$ is the lowest conduction band energy and $-\Delta_-$ is the highest valence band energy. The condition $\Delta_+=\Delta_-$ simply means that the chemical potential locates at the middle of conduction and valence bands. 

In \cref{p5:fig:E1appendix,p5:fig:E1appendix20} the charge $\pm1$ excitations are plotted at different fillings and $w_0/w_1$'s for two different screening lengths of the Coulomb interaction (\cref{szdeq:Vq}), \ie $\xi$=10nm, 20nm, respectively.
The corresponding interaction strengths are $U_{\rm 10nm}=$26meV, $U_{\rm 20nm}$=13meV.
We have used $w_1$=110meV in all the calculations and $w_0/w_1=$0, 0.4, 0.8 for $\nu=0,-2$ and $w_0/w_1=0$ for $\nu=-1,-3$.

For $\nu=0$ and $\xi=$10nm, 20nm, the charge $\pm1$ gaps are at the $\Gamma_M$ momentum and are always larger than 10meV for different $w_0/w_1$'s. 
For $\nu=-2$ and $\xi=$20nm, the charge $\pm1$ gaps are always larger than 5meV for different $w_0/w_1$'s. 
For $\nu=-2$ and $\xi=$10nm, the charge $\pm1$ gaps are finite for $w_0/w_1=0,0.4$ but become negative at $w_0/w_1=0.8$ (\cref{p5:fig:E1appendix}c).
We find that the gaps close around $w_0/w_1\approx0.75$, implying that, with $\xi$=10nm and the other parameters we have used, the ground states in \cref{p5:eq:U(4)-GSMT} become unstable for $w_0/w_1\approx0.75$. 

The instability shown in \cref{p5:fig:E1appendix}c will lead to a metallic phase at $\nu=-2$ in the nonchiral-flat limit with strong chiral symmetry breaking ($w_0/w_1=0.8$).
In the band structure picture, we can understand the charge +1 excitation as the conduction band and the charge -1 excitation as the reverted valence band. The negativities of both imply that the energy of conduction band overlaps with the energy of valence band. Since we have chosen to fully occupy the valence bands, the ground state energy is not minimized in this case: One can move one particle from the top of valence band to bottom of conduction band to lower the energy. The ground state energy will be minimized by redistributing electrons to occupy the bands below the chemical potential. Due to the overlap between conduction and valence bands, the redistributed band structure will have electron and hole pockets and hence is a metallic state. 
In the resulted metallic phase at $\nu=-2$, there are two fully empty valley-spin sectors and two partially filled valley-spin sectors.
The partially filled sectors contribute to the electron and hole pockets.  
This state is still invariant under a U(2) subgroup within the fully empty valley-spin sectors. 
Since U(4) and U(2) have 16 and 4 parameters, there will be $\frac{16-4}{2}=6$ Goldstone modes.

\section{Charge neutral excitations and the Goldstone stiffness}\label{p5:NeutralExcitationappendix}

While the charge $\pm 1$ excitations can be obtained by diagonalizing a $2\times 2$ matrix, we can obtain the charge neutral, 2-body excitations above the ground state. 
These can be obtained by diagonalizing a $2N_M \times 2N_M$ matrix, or a \emph{one-body} problem, despite the state having a thermodynamic number of particles. Due to the fact that we know the exact eigenstates  (or ground states) of the system, building excitations of the Hamiltonian on top of these eigenstates (or ground states) becomes a problem of diagonalizing a basis formed only from the excitations. 
We now obtain the charge neutral excitations, show that  they exhibit Goldstone modes with quadratic dispersion - as required by U(4) (or U(4) $\times$ U(4)) ferromagnetism, and obtain the stiffness of the Goldstone dispersion in the first chiral limit. 

\subsection{Exact charge neutral excitations in the nonchiral-flat U(4)  limit}\label{p5:NeutralExcitationNonChiralappendix}

We choose a basis for the neutral excitations 
\begin{eqnarray}
c_{\kk_2, m_2, \eta_2, s_2}^\dagger c_{\kk_1, m_1, \eta_1, s_1}|\Psi \rangle
\end{eqnarray} where $|\Psi \rangle$ is any of  the exact ground states and/or eigenstates in \cref{p5:eq:U(4)U(4)-GS,p5:eq:U(4)-GS}. 
The scattering matrix of this basis can be solved as easily as a one-body problem, despite the fact that \cref{p5:eq:U(4)U(4)-GS,p5:eq:U(4)-GS} hold a thermodynamic number of particles.  
We first have to compute the commutators:
\begin{align}
& [O_{\mathbf{-q,-G}}O_{\mathbf{q,G}},c_{\kk_2, m_2, \eta_2, s_2}^\dagger c_{\kk_1, m_1, \eta_1, s_1}] \nono\\
=& [O_{\mathbf{-q,-G}}O_{\mathbf{q,G}}, c_{\kk_2, m_2, \eta_2, s_2}^\dagger]  c_{\kk_1, m_1, \eta_1, s_1}+ c_{\kk_2, m_2, \eta_2, s_2}^\dagger [O_{\mathbf{-q,-G}}O_{\mathbf{q,G}},  c_{\kk_1, m_1, \eta_1, s_1}],
\end{align}
which, in detail reads:
\begin{align}
& [O_{\mathbf{-q,-G}}O_{\mathbf{q,G}}, c_{\kk_2, m_2, \eta_2, s_2}^\dagger c_{\kk_1, m_1, \eta_1, s_1}]  \nonumber \\ &= \sum_{m}P_{mm_2}^{\left(\eta_2\right)}\left(\mathbf{k}_2,\mathbf{q}+\mathbf{G}\right)c_{\mathbf{k}_2,m,\eta_2,s_2}^\dag c_{\kk_1, m_1, \eta_1, s_1}  + \sum_{m}  P_{m_1 m}^{\left(\eta_1\right)}\left(\mathbf{k}_1,-\mathbf{q}-\mathbf{G}\right)  c_{\kk_2, m_2, \eta_2, s_2}^\dagger c_{\kk_1, m, \eta_1, s_1}\nonumber \\ & + \sqrt{V(\mathbf{G}+\mathbf{q})} \sum_{m} \left( M_{m,m_2}^{\left(\eta_2\right)}\left(\mathbf{k}_2,\mathbf{q}+\mathbf{G}\right) c_{\kk_2+ \qq, m, \eta_2, s_2}^\dagger c_{\kk_1, m_1, \eta_1, s_1} O_{-\mathbf{q,-G}} + (\qq,\GG \leftrightarrow -\qq, -\GG) \right) \nonumber \\ &-\sqrt{V(\mathbf{G}+\mathbf{q})}\sum_{m} \left(  M_{m,m_1}^{\left(\eta_1\right)*}\left(\mathbf{k}_1,-\mathbf{q}-\mathbf{G}\right) c_{\kk_2, m_2, \eta_2, s_2}^\dagger c_{\kk_1-\qq, m, \eta_1, s_1} O_{-\mathbf{q,-G}}   + (\qq,\GG \leftrightarrow -\qq, -\GG)  \right)\nonumber \\ &-
V(\mathbf{G}+\mathbf{q}) \sum_{m,m'}  \left(M_{m,m_2}^{\left(\eta_2\right)}\left(\mathbf{k}_2,\mathbf{q}+\mathbf{G}\right)  M_{m',m_1}^{\left(\eta_1\right)*}\left(\mathbf{k}_1,\mathbf{q}+ \mathbf{G}\right)  c_{\kk_2+\qq, m, \eta_2, s_2}^\dagger c_{\kk_1+\qq, m', \eta_1, s_1}+ (\qq,\GG \leftrightarrow -\qq, -\GG)  \right)
\end{align}
By rewriting $\kk_2=\kk+\pp$ and $\kk_1=\kk$, we can write the scattering equation as 
\begin{equation}\label{p5:eq:neutralexcitation-HI}
\left[H_I-\mu N,c_{\kk+\pp, m_2, \eta_2, s_2}^\dagger c_{\kk, m_1, \eta_1, s_1} \right] |\Psi\rangle 
= \frac{1}{2\Omega_{\text{tot}}}\sum_{m,m'}\sum_\qq S^{(\eta_2,\eta_1)} _{m m';m_2 m_1}(\kk+\qq,\kk;\pp) c_{\kk+\pp+\qq, m, \eta_2, s_2}^\dagger c_{\kk+\qq, m', \eta_1, s_1}|\Psi\rangle\ .
\end{equation} 
The $ |\Psi\rangle$ are the states  $|\Psi_\nu\rangle$ in \cref{p5:eq:U(4)-GS}, and hence $\eta_1, s_1$ belong to the valley-spin flavor/s which are fully occupied, while  $\eta_2, s_2$ belong to the valley/spin flavor which are not  occupied.

For a generic exact eigenstate $|\Psi\rangle$ at chemical potential $\mu$ satisfying $(O_{\mathbf{q,G}}-A_{\GG} N_M\delta_{\mathbf{q},0})|\Psi \rangle =0$ for some coefficient $A_\GG$, we find that the scattering matrix reads
\begin{equation}\label{p5:neutralexcitations1}
\begin{split}
S^{(\eta_2,\eta_1)}_{m,m';m_2,m_1}(\mathbf{k}+\qq,\kk;\pp) 
=& \delta_{\qq,\mathbf{0}}(\delta_{m,m_2} \widetilde{R}_{m'm_1}^{\eta_1}(\mathbf{k}) +\delta_{m',m_1} R_{mm_2}^{\eta_2}(\mathbf{k}+\pp) ) \nono\\
& - 2\sum_\GG V(\mathbf{G}+\mathbf{q})   M_{m,m_2}^{\left(\eta_2\right)}\left(\mathbf{k}+\pp,\mathbf{q}+\mathbf{G}\right)  M_{m',m_1}^{\left(\eta_1\right)*}\left(\mathbf{k},\mathbf{q}+ \mathbf{G}\right),
\end{split}
\end{equation}
where $R_{mn}^\eta(\mathbf{k}), \widetilde{R}_{mn}^\eta(\mathbf{k})$ are the $\pm 1$-excitation matrices in \cref{p5:eq:excitation-Rmn,p5:eq:excitation-Rmn-hole}. We see that the neutral energy is a sum of the two single-particle energies (first row of \cref{p5:neutralexcitations1}) plus an interaction energy (second row of \cref{p5:neutralexcitations1}).

The exact expression of the PH excitation spectrum allows for the determination of the Goldstone stiffness. 
The Goldstone of the U(4) and U(4)$\times$U(4) ferromagnetic ground states is part of the spectrum of the neutral excitation \cref{p5:eq:neutralexcitation-HI}, and is the state at small momentum $\pp=\kk_1-\kk_2$. 
We will solve this in the simpler, chiral limit, but it can be obtained in the general, non-chiral limit \cref{p5:eq:neutralexcitation-HI}.

\subsection{Exact charge neutral excitations in the (first) chiral-flat \texorpdfstring{U(4)$\times$U(4)}{U(4)xU(4)} limit}\label{p5:NeutralExcitationChiralappendix}

We now consider the charge neutral excited states reachable by creating one electron-hole pair with total momentum $\mathbf{p}$ on the chiral-flat limit eigenstate $|\Psi_{\nu}^{\nu_+,\nu_-}\rangle$. Assume the valley-spin flavor $\{\eta_1,s_1\}$ has Chern band basis $e_{Y1}$ fully occupied and the valley-spin flavor  $\{\eta_2,s_2\}$ Chern band basis $e_{Y2}$ fully empty. We consider the Hilbert space of the following sets of states of momentum quantum number  $\mathbf{p}$
\begin{equation}
|\mathbf{k+p,k},e_{Y1}, e_{Y2},\eta_2, \eta_1 ,s_2,s_1,\Psi_{\nu}^{\nu_+,\nu_-}\rangle=d^\dag_{\mathbf{k+p},e_{Y2},\eta_2,s_2} d_{\mathbf{k},e_{Y1},\eta_1,s_1}|\Psi_{\nu}^{\nu_+,\nu_-}\rangle\ ,
\end{equation}

The $O_{\qq,\GG}$ operators in the chiral limit have the simple, diagonal expression of \cref{p5:OqGOperatorInChiralLimit}, which leads to the scattering equation. 

\begin{equation}\label{p5:eq:neutralexcitationflatband-HI}
\left[H_I-\mu N,d^\dag_{\mathbf{k+p},e_{Y2},\eta_2,s_2} d_{\mathbf{k},e_{Y1},\eta_1,s_1}\right] |\Psi\rangle =\frac{1}{2\Omega_{\text{tot}}}\sum_\qq S _{e_{Y2};e_{Y1}}(\mathbf{k+q},\mathbf{k};\pp)d^\dag_{\mathbf{k+p+q},e_{Y2},\eta_2,s_2} d_{\mathbf{k+q},e_{Y1},\eta_1,s_1}|\Psi\rangle\ .
\end{equation} 
The $ |\Psi\rangle$ are the states $|\Psi_{\nu}^{\nu_+,\nu_-}\rangle$  in \cref{p5:eq:U(4)U(4)-GS}, and hence $e_{Y1}, \eta_1, s_1$ belong to the valley-spin flavor/s which are fully occupied, while  $e_{Y2},\eta_2, s_2$ belong to the valley/spin flavor which are not  occupied. The scattering matrix in the chiral limit does not depend on $\eta_1, \eta_2$:
\begin{eqnarray}
& S _{e_{Y2};e_{Y1}}(\mathbf{k+q},\mathbf{k};\pp) =  \delta_{\qq,\mathbf{0}}(  R_0(\mathbf{k+p})+ \widetilde{R}_{0}(\mathbf{k}) ) -2\sum_\GG V(\mathbf{G}+\mathbf{q}) M_{e_{Y2}}(\kk+\mathbf{p},\qq+\GG)  M^*_{e_{Y1}}(\kk,\qq+\GG)\ ,
\end{eqnarray}
where $M_{e_Y}(\kk,\qq+\GG)$ is given in \cref{p5:eq:chiral-MqG1} and $R_0^\eta(\kk)$ is given in \cref{p5:R0InTheChiralLimit}. 

If condition \cref{p5:eqn-condition-at-nu} is satisfied, the eigenstates $|\Psi_{\nu}^{\nu_+,\nu_-}\rangle$ are the ground states, and hence the states \cref{p5:eq:neutralexcitationflatband-HI} are the neutral excitations on top of the ground states. 
Without the condition Eq.~(\ref{p5:eqn-condition-at-nu}), only $\nu=0$ states are guaranteed to be the ground states, although the others are still eigenstates. With condition Eq.~(\ref{p5:eqn-condition-at-nu}), we have
\begin{align}\label{p5:scatteringmatrixneutralchiral}
& S_{e_{Y2};e_{Y1}}(\mathbf{k+q},\mathbf{k};\pp) =  \delta_{\qq,\mathbf{0}}\sum_{\GG,\qq'} V(\GG+\qq') [\alpha_0 (\kk,\qq'+\GG)^2 +\alpha_2 (\kk,\qq'+\GG)^2+\alpha_0 (\kk+\mathbf{p},\qq'+\GG)^2 +\alpha_2 (\kk+\mathbf{p},\qq'+\GG)^2] \nonumber \\ &-2\sum_\GG V(\mathbf{G}+\mathbf{q}) (\alpha_0(\mathbf{k+p,q+G})+ie_{Y2}\alpha_2(\mathbf{k+p,q+G})   )  (\alpha_0(\mathbf{k,q+G})- ie_{Y1}\alpha_2(\mathbf{k,q+G}))
\end{align}
Solving  \cref{p5:eq:neutralexcitationflatband-HI} provides us with the expression for the neutral excitations at momentum $\mathbf{p}$ on top of the TBG ground states.

\subsection{Goldstone mode in the first chiral limit and the Goldstone stiffness}\label{p5:Goldstoneappendix}

We show that the Goldstone mode of the ferromagnetic ground states are included in the neutral excitations of \cref{p5:eq:neutralexcitationflatband-HI} and we obtain their dispersion relation, in terms of the quantum geometry factors of the TBG. 
We are able to analytically obtain the Goldstone mode if the condition \cref{p5:eqn-condition-at-nu} holds.
We first show the presence of an exact zero eigenstate of \cref{p5:eq:neutralexcitationflatband-HI}.

\subsubsection{Exact zero energy neutral mode Eigenstate}\label{p5:zeroenergyneutralappendix}

We now show that \cref{p5:eq:neutralexcitationflatband-HI} has an exact zero energy eigenstate.  
In order to see this, we remark that the $\mbf{p}=0, e_{Y1}= e_{Y2}$ state  \cref{p5:scatteringmatrixneutralchiral} has a scattering matrix 
\begin{align} \label{p5:neutralmodescatteringforp=0}
&S_{e_{Y};e_{Y}}(\mathbf{k+q},\kk;\mathbf{0}) =  2 \delta_{\qq,\mathbf{0}}\sum_{\GG,\qq'} V(\GG+\qq') [\alpha_0 (\kk,\qq'+\GG)^2 +\alpha_2 (\kk,\qq'+\GG)^2]     \nonumber \\ &-2\sum_\GG V(\mathbf{G}+\mathbf{q})   (\alpha_0(\mathbf{k,q+G})^2+ \alpha_2(\mathbf{k,q+G})^2) \,
\end{align}
whose elements in every row sums to zero (irrespective of $\eta_{1,2},s_{1,2}$):
\begin{equation}
\sum_{\qq} S _{e_{Y};e_{Y}}(\mathbf{k+q},\kk;\mathbf{0}) =0\ .
\end{equation} 
This guarantees that the rank of the scattering matrix is not maximal, and that there is at least one zero energy eigenstate, with  equal  amplitude on every $|\mathbf{k,k},e_{Y}, e_{Y},\eta_2, \eta_1 ,s_2,s_1,\Psi_{\nu}^{\nu_+,\nu_-}\rangle$

\begin{equation} \label{p5:zeromomentumgoldstone}
 |e_{Y}, e_{Y},\eta_2, \eta_1 ,s_2,s_1\rangle =\sum_{\qq} d^\dag_{\mathbf{k+q},e_{Y},\eta_2,s_2} d_{\mathbf{k+q },e_{Y},\eta_1,s_1}|\Psi_{\nu}^{\nu_+,\nu_-}\rangle\ , \;\; 
 (H_I-\mu N) |e_{Y}, e_{Y},\eta_2, \eta_1 ,s_2,s_1\rangle= 0
\end{equation} A U(4)$\times$U(4) multiplet of this state is also at zero energy. 
Moreover, the scattering matrix $S_{e_{Y};e_{Y}}(\mathbf{k+q},\kk;0)$ is positive semidefinite (as it should, since these eigenvalues are energies of excitations on top of the ground states). 
For the matrix $S_{e_{Y};e_{Y}}(\mathbf{k+q},\kk;0)$, we prove that its negative, $- S _{e_{Y};e_{Y}}(\mathbf{k+q},\kk;0)$, has only non-positive eigenvalues, and hence  $S_{e_{Y};e_{Y}}(\mathbf{k+q},\kk;0)$ has only non-negative eigenvalues. 
For $-S_{e_{Y};e_{Y}}(\mathbf{k+q},\kk;0)$, the diagonal elements $ -S _{e_{Y};e_{Y}}(\mathbf{k},\kk;0)$ are non-positive, while the off-diagonal elements $- S_{e_{Y};e_{Y}}(\mathbf{k+q},\kk;0)$ ($\qq\neq0$) are non-negative. Hence by the Gershgorin circle theorem, all eigenvalues lie in at least one of the Gershgorin disks (which, due to the fact that the matrix is Hermitian, are intervals) centered at $-S_{e_{Y};e_{Y}}(\mathbf{k},\kk; 0)$ and with radius  
$- \sum_{\qq\neq0} S_{e_{Y};e_{Y}}(\mathbf{k+q},\kk;0)$. 
These intervals are:
\begin{align}
& [-  S _{e_{Y};e_{Y}}(\mathbf{k},\kk;0)  +\sum_{\qq\neq0} S _{e_{Y};e_{Y}}(\kk+\qq,\kk;0), -  S_{e_{Y};e_{Y}}(\mathbf{k},\kk;0)  -\sum_{\qq\neq0} S _{e_{Y};e_{Y}}(\mathbf{k+q},\kk; 0)]= \nonumber \\ & =  [2 \sum_{\qq\neq0} S _{e_{Y};e_{Y}}(\mathbf{k+q},\kk; 0), 0].
\end{align} 
Since $ \sum_{\qq\neq0} S_{e_{Y};e_{Y}}(\mathbf{k+q},\kk; 0) \le 0, \;\; \forall \kk$, all eigenvalues of  $- S _{e_{Y};e_{Y}}(\mathbf{k+q},\kk;0)$ are non-positive, and hence all eigenvalues of $ S _{e_{Y};e_{Y}}(\mathbf{k+q},\kk;0)$ are non-negative.

\subsubsection{Goldstone stiffness }\label{p5:GoldstoneStiffnessappendix}
Since the $\pp=0$ state has zero energy, for small $\pp$, there will be low-energy states in the neutral continuum. 
By using the $\pp=0$ states in \cref{p5:zeromomentumgoldstone}, one can compute their dispersion. 
First, we write the Hamiltonian matrix elements acting on the two particle states
\begin{align}
&\langle \mathbf{k'+p,k'},e_{Y1}', e_{Y2}',\eta_2', \eta_1' ,s_2',s_1',\Psi_{\nu}^{\nu_+,\nu_-}|H_I |\mathbf{k+p,k},e_{Y1}, e_{Y2},\eta_2, \eta_1 ,s_2,s_1,\Psi_{\nu}^{\nu_+,\nu_-}\rangle \nonumber \\
&=\delta_{ e_{Y1}', e_{Y2}',\eta_2', \eta_1' ,s_2',s_1; e_{Y1}, e_{Y2},\eta_2, \eta_1 ,s_2,s_1} \frac{1}{2\Omega_{\text{tot}}} (\delta_{\kk',\kk} S _{e_{Y2};e_{Y1}}(\mathbf{k},\mathbf{k};\pp)  ) - S _{e_{Y2};e_{Y1}}(\mathbf{k}',\mathbf{k}; \pp)  )\ . 
\end{align} 
Hence, for small $\pp$, the energy of the Goldstone mode is given by the expectation value
\begin{align}
E_{\text{Goldstone}} (\mathbf{p}) =
& \sum_{\kk,\kk'} \langle \mathbf{k'+p,k'},e_{Y}, e_{Y},\eta_2, \eta_1 ,s_2,s_1,\Psi_{\nu}^{\nu_+,\nu_-}|H_I |\mathbf{k+p,k},e_{Y}, e_{Y},\eta_1, \eta_1 ,s_2,s_1,\Psi_{\nu}^{\nu_+,\nu_-}\rangle \nonumber \\ 
 =&  \frac{1}{2\Omega_{\text{tot}}} \sum_k \left(S_{e_{Y};e_{Y}}(\mathbf{k},\mathbf{k};\pp) - \sum_\qq S _{e_{Y};e_{Y}}(\mathbf{k+q},\mathbf{k}; \pp)\right) \ .
\end{align} 
As expected for the Goldstone of a ferromagnet, the linear term in $\pp$ vanishes; by using $\alpha_a(\kk,\qq+\GG) = \alpha_a(-\kk,-\qq-\GG)\quad \text{for }a=0,2$ of \cref{p5:eq:alpha-cond2}, we find can prove that the linear terms vanish exactly. 
To second order in $\pp$, we find the Goldtone stiffness
\begin{equation}
E_{\text{Goldstone}} (\mathbf{p}) = \frac{1}{2} m_{ij} p_i p_j,
\end{equation}
\begin{align}
m_{ij} =& \frac{1}{2\Omega_{\text{tot}}}\sum_{\kk, \qq,\GG} V(\GG+\qq) [\alpha_0 (\kk,\qq+\GG) \partial_{k_i}\partial_{k_j} \alpha_0 (\kk,\qq+\GG)  +\alpha_2 (\kk,\qq+\GG) \partial_{k_i}\partial_{k_j} \alpha_2 (\kk,\qq+\GG) \nonumber \\ &+ 2\partial_{k_i} \alpha_0 (\kk,\qq+\GG) \partial_{k_j} \alpha_0 (\kk,\qq+\GG) +2\partial_{k_i} \alpha_2 (\kk,\qq+\GG) \partial_{k_j} \alpha_2 (\kk,\qq+\GG)] \ .
\end{align}

\begin{figure}[t]
\centering
\includegraphics[width=0.8\linewidth]{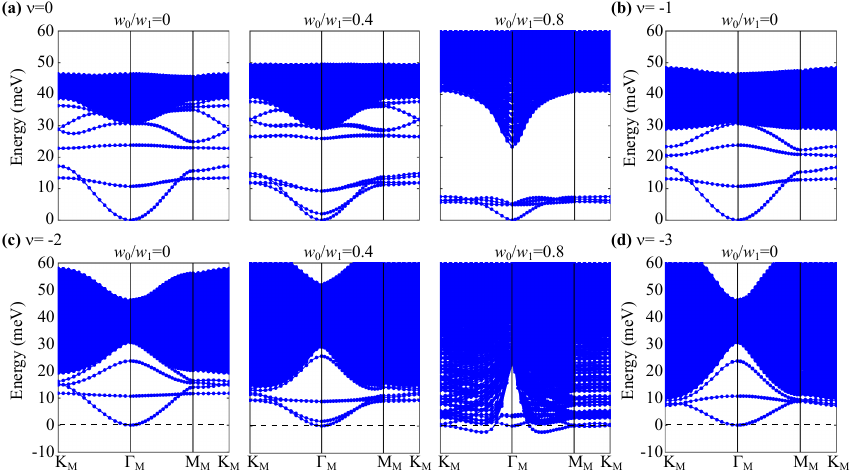}
\caption{Exact charge neutral excitations with $\theta=1.05^\circ$. The flat metric condition is {\it not} imposed. 
In this plot we have used the parameters defined in \cref{p5:newapp:Ham}: $v_F=5.944{\rm eV \cdot \mathring{ A} }$, $|K|=1.703\mathring{\rm A}^{-1}$, $w_1=110{\rm meV}$,  $U_\xi=26$meV, $\xi=10$nm.
}
\label{p5:fig:E0appendix}
\end{figure}

\begin{figure}[t]
\centering
\includegraphics[width=0.8\linewidth]{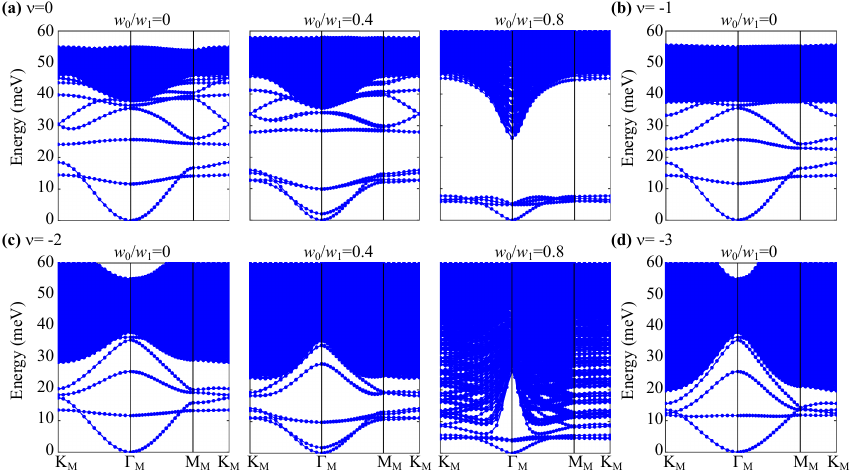}
\caption{Exact charge neutral excitations with $\theta=1.05^\circ$. The flat metric condition is {\it not} imposed. 
In this plot we use the screening length $\xi$=20nm and $U_\xi=13$meV accordingly. The other parameters are same as in \cref{p5:newapp:Ham}, \ie $v_F=5.944{\rm eV \cdot \mathring{ A} }$, $|K|=1.703\mathring{\rm A}^{-1}$, $w_1=110{\rm meV}$.
}
\label{p5:fig:E0appendix20}
\end{figure}

\subsection{Charge neutral excitation spectra for different parameters}\label{newapp:plot0}

In \cref{p5:fig:E0appendix,p5:fig:E0appendix20} the charge neutral excitations are plotted at different fillings and $w_0/w_1$'s for two different screening lengths of the Coulomb interaction (\cref{szdeq:Vq}), \ie $\xi$=10nm, 20nm, respectively.
The corresponding interaction strengths are $U_{\rm 10nm}=$26meV, $U_{\rm 20nm}$=13meV.
We have used $w_1$=110meV in all the calculations and $w_0/w_1=$0, 0.4, 0.8 for $\nu=0,-2$ and $w_0/w_1=0$ for $\nu=-1,-3$.

For the parameters we used, the $\nu=0$ states with $w_0/w_1=0,0.4,0.8$ and $\nu=-1,-3$ states with $w_0=0$ have non-negative excitations and hence are stable. 
The $\nu=-2$ states with $w_0/w_1=0,0.4,0.8$ are also stable for $\xi=$20nm.
However, the $\nu=-2$ states for $\xi=$10nm become unstable at $w_0/w_1=0.8$ (\cref{p5:fig:E0appendix}c).
The instability can be understood from the instability of the charge $\pm1$ excitations shown in \cref{p5:fig:E1appendix}c. 
From \cref{p5:fig:E1appendix}c we can see that the charge +1 excitation is negative at some momenta between $\Gamma_M$ and $K_M$ and the charge -1 excitation is negative at $\Gamma_M$. Combining a pair of these negative particle and hole one obtains a negative charge neutral excitation. 
As discussed in the end of \cref{newapp:plot1}, this instability will lead to a metallic phase.

For the stable ground states, where the spectrum is non-negative, the charge neutral spectrum consists of a particle-hole continuum (the blue area in \cref{p5:fig:E0appendix,p5:fig:E0appendix20}) and a set of gapped collective modes. 
In \cref{p5:fig:E0appendix,p5:fig:E0appendix20} we only plot the eigenvalues of the scattering matrix. 
In practice, the existence and degeneracy of an excitation mode also depend on the occupied U(4) flavors (and U(4)$\times$U(4) flavors in the first chiral limit) of the ground state, as discussed in \cref{p5:sec:method-charge1}. 
In particular, the number of Goldstone modes for different ground states are given in \cref{p5:tab:Goldstone-U4,p5:tab:Goldstone-U4xU4}. 
In general, the states $\ket{\Psi_{\nu}^{\nu_+,\nu_-}}$ (\cref{p5:eq:U(4)U(4)-GSMT}) in the (first) chiral-flat U(4)$\times$U(4) limit with $\nu=0,\pm2$ and $\nu_+=\nu_-=\frac{\nu+4}{2}$ (such that it has vanishing Chern number) has more Goldstone modes than the state $\ket{\Psi_\nu}$ (\cref{p5:eq:U(4)-GSMT}) with the same filling $\nu$.
From $w_0/w_1=0.4$ to $w_0/w_1=0$ for the states at $\nu=0,-2$ we can clearly see in \cref{p5:fig:E0appendix,p5:fig:E0appendix20} that a collective mode is soften and become Goldstone mode, consistent with the theoretical analysis.

\section{Charge \texorpdfstring{$\pm 2$}{+-2} excitations and conditions on Cooper pair instability}\label{p5:charge2appendix}

The charge $\pm 1$ excitations can be obtained by diagonalizing a $2\times 2$ matrix; the charge neutral above the ground state can be obtained by diagonalizing a $2N_M \times 2N_M$ matrix, or a \emph{one-body} problem, despite the state having a thermodynamic number of particles, due to the fact that we know the exact eigenstates  (or ground states) of the system. We now show that the charge $+2$ excitations can also be obtained by diagonalizing a $2N_M\times 2N_M$ matrix. 
The conditions for which Cooper pairing occurs are also obtained. 

\subsection{Charge \texorpdfstring{$+2$}{+2} excitations in the nonchiral-flat U(4)  limit}\label{p5:charge2NonChiralappendix}

We choose a basis for the neutral excitations 
\begin{eqnarray}
c_{\kk_2, m_2, \eta_2, s_2}^\dagger c^\dagger_{\kk_1, m_1, \eta_1, s_1}|\Psi \rangle
\end{eqnarray} 
where $|\psi \rangle$ is any of  the exact ground states/eigenstates \cref{p5:eq:U(4)U(4)-GS,p5:eq:U(4)-GS}. 
The scattering matrix of these basis can be solved as easily as a one-body problem. 
We first have to compute the commutators:
\begin{align}
& [O_{\mathbf{-q,-G}}O_{\mathbf{q,G}},c_{\kk_2, m_2, \eta_2, s_2}^\dagger c^\dagger_{\kk_1, m_1, \eta_1, s_1}] \nono\\
=& [O_{\mathbf{-q,-G}}O_{\mathbf{q,G}}, c_{\kk_2, m_2, \eta_2, s_2}^\dagger]  c^\dagger_{\kk_1, m_1, \eta_1, s_1}+ c_{\kk_2, m_2, \eta_2, s_2}^\dagger [O_{\mathbf{-q,-G}}O_{\mathbf{q,G}},  c^\dagger_{\kk_1, m_1, \eta_1, s_1}] \ ,
\end{align}
which, in detail reads:
\begin{align}
& [O_{\mathbf{-q,-G}}O_{\mathbf{q,G}}, c_{\kk_2, m_2, \eta_2, s_2}^\dagger c^\dagger_{\kk_1, m_1, \eta_1, s_1}]  \nonumber \\ 
&= \sum_{m}P_{mm_2}^{\left(\eta_2\right)}\left(\mathbf{k}_2,\mathbf{q}+\mathbf{G}\right)c_{\mathbf{k}_2,m,\eta_2,s_2}^\dag c^\dagger_{\kk_1, m_1, \eta_1, s_1}  + \sum_{m}  P_{m m_1 }^{\left(\eta_1\right)}\left(\mathbf{k}_1,\mathbf{q}+\mathbf{G}\right)  c_{\kk_2, m_2, \eta_2, s_2}^\dagger c^\dagger_{\kk_1, m, \eta_1, s_1} \nonumber \\ 
& + \sqrt{V(\mathbf{G}+\mathbf{q})} \sum_{m} \left( M_{m,m_2}^{\left(\eta_2\right)}\left(\mathbf{k}_2,\mathbf{q}+\mathbf{G}\right) c_{\kk_2+ \qq, m, \eta_2, s_2}^\dagger c^\dagger_{\kk_1, m_1, \eta_1, s_1} O_{-\mathbf{q,-G}} + (\qq,\GG \leftrightarrow -\qq, -\GG) \right) \nonumber \\
&+\sqrt{V(\mathbf{G}+\mathbf{q})}\sum_{m} \left(  M_{m,m_1}^{\left(\eta_1\right)}\left(\mathbf{k}_1,\mathbf{q}+\mathbf{G}\right) c_{\kk_2, m_2, \eta_2, s_2}^\dagger c^\dagger_{\kk_1+\qq, m, \eta_1, s_1} O_{-\mathbf{q,-G}}   + (\qq,\GG \leftrightarrow -\qq, -\GG)  \right)\nonumber \\ 
&+ V(\mathbf{G}+\mathbf{q}) \sum_{m,m'}  \left(M_{m,m_2}^{\left(\eta_2\right)}\left(\mathbf{k}_2,\mathbf{q}+\mathbf{G}\right)  M_{m',m_1}^{\left(\eta_1\right)}\left(\mathbf{k}_1,-\mathbf{q}- \mathbf{G}\right)  c_{\kk_2+\qq, m, \eta_2, s_2}^\dagger c^\dagger_{\kk_1-\qq, m', \eta_1, s_1}+ (\qq,\GG \leftrightarrow -\qq, -\GG)  \right)
\end{align}
By rewriting $\kk_2=\kk+\pp$ and $\kk_1=-\kk$, we can write the scattering equation as 
\begin{equation}\label{p5:eq:charge2excitation-HI}
\left[H_I-\mu N,c_{\kk+\pp, m_2, \eta_2, s_2}^\dagger c^\dagger_{-\kk, m_1, \eta_1, s_1} \right] |\Psi\rangle =\frac{1}{2\Omega_{\text{tot}}}\sum_{m,m'}\sum_\qq T^{(\eta_2,\eta_1)} _{m m';m_2 m_1}(\kk+\qq,\kk;\pp) c_{\kk+\pp+\qq, m, \eta_2, s_2}^\dagger c^\dagger_{-\kk-\qq, m', \eta_1, s_1}|\Psi\rangle\ ,
\end{equation} 
The $ |\Psi\rangle$ are the states  $|\Psi_\nu\rangle$ in \cref{p5:eq:U(4)-GS}, and hence $\eta_1, s_1,\eta_2, s_2$ belong to the valley/spin flavor which are not  occupied. 
For a generic exact eigenstate $|\Psi\rangle$ at chemical potential $\mu$ satisfying $(O_{\mathbf{q,G}}-A_{\GG} N_M\delta_{\mathbf{q},0})|\Psi \rangle =0$ for some coefficient $A_\GG$, we find that the $T^{(\eta_2,\eta_1)} _{m_2,m;m_1m'}(\mathbf{k}_1,\kk_2;\qq)$ matrix reads
\begin{align}\label{p5:charge2excitations1}
T^{(\eta_2,\eta_1)} _{m m'; m_2 m_1}(\kk+\qq,\kk;\pp) =&  \delta_{\qq,\mathbf{0}}(\delta_{m,m_2} R_{m'm_1}^{\eta_1}(-\mathbf{k}) +\delta_{m',m_1} R_{mm_2}^{\eta_2}(\mathbf{k}+\pp) ) +\nono \\
&+2\sum_\GG V(\mathbf{G}+\mathbf{q})    M_{m,m_2}^{\left(\eta_2\right)}\left(\mathbf{k}+\pp,\mathbf{q}+\mathbf{G}\right)  M_{m',m_1}^{\left(\eta_1\right)}\left(-\mathbf{k},-\mathbf{q}- \mathbf{G}\right) \ ,
\end{align}
where $R_{mn}^\eta(\mathbf{k}), R_{mn}^\eta(\mathbf{k})$ are the $+1$ excitation matrices in \cref{p5:eq:excitation-Rmn,p5:eq:excitation-Rmn-hole}. 
We see that the charge $+2$ energy is a sum of the two single-particle energies (first row of \cref{p5:neutralexcitations1}) plus an interaction energy (second row of \cref{p5:neutralexcitations1}). 

The exact expression of the charge $+2$ excitation spectrum allows for the determination of the Cooper pair binding energy (if any). 

\subsection{Charge \texorpdfstring{$+2$}{+-2} excitations in the (first) chiral-flat \texorpdfstring{U(4)$\times$U(4)}{U(4)xU(4)} limit} \label{p5:charge2Chiralappendix}

We now consider the charge $+2$ excited states reachable by creating two electron pair with total momentum $\mathbf{p}$ on the chiral-flat limit eigenstate $|\Psi_{\nu}^{\nu_+,\nu_-}\rangle$. Assume the valley-spin flavor $\{\eta_{1,2},s_{1,2}\}$ has Chern band basis $e_{Y1}, e_{Y2}$ fully empty. We consider the Hilbert space of the following sets of states of momentum quantum number  $\mathbf{p}$ ($\kk_2=-\kk_1+\mathbf{p},\; \kk_1= \kk$):

\begin{equation} \label{p5:charge2excitation1}
|\mathbf{k+p,-k},e_{Y1}, e_{Y2},\eta_2, \eta_1 ,s_2,s_1,\Psi_{\nu}^{\nu_+,\nu_-}\rangle=d^\dag_{\mathbf{k+p},e_{Y2},\eta_2,s_2} d^\dag_{-\mathbf{k},e_{Y1},\eta_1,s_1}|\Psi_{\nu}^{\nu_+,\nu_-}\rangle\ ,
\end{equation}
The $O_{\qq,\GG}$ operators in the chiral limit have the simple, diagonal expression of \cref{p5:OqGOperatorInChiralLimit}, which leads to the scattering equation. 

\begin{equation}\label{p5:eq:charge2excitationflatband-HI}
\left[H_I-\mu N,d^\dag_{\mathbf{k+p},e_{Y2},\eta_2,s_2} d^\dag_{-\mathbf{k},e_{Y1},\eta_1,s_1}\right] |\Psi\rangle =\frac{1}{2\Omega_{\text{tot}}}\sum_\qq T _{e_{Y2};e_{Y1}}(\mathbf{k+q},\mathbf{k};\pp)d^\dag_{\mathbf{k+p+q},e_{Y2},\eta_2,s_2} d^\dag_{\mathbf{-k-q},e_{Y1},\eta_1,s_1}|\Psi\rangle\ .
\end{equation} 
The $ |\Psi\rangle$ are the states  $|\Psi_{\nu}^{\nu_+,\nu_-}\rangle$  in \cref{p5:eq:U(4)U(4)-GS}, and hence $e_{Y1}, \eta_1, s_1$, $e_{Y2},\eta_2, s_2$ belong to the valley/spin flavor which are not  occupied. 
The scattering matrix in the first chiral limit does not depend on $\eta_1, \eta_2$
\begin{align}
&T_{e_{Y2};e_{Y1}}(\mathbf{k+q},\mathbf{k};\pp) =  \delta_{\qq,0}(  R_0(\mathbf{k+p})+ R_{0}(-\mathbf{k}) ) +2\sum_\GG V(\mathbf{G}+\mathbf{q}) M_{e_{Y2}}(\kk+\mathbf{p},\qq+\GG)  M_{e_{Y1}}(-\kk,-\qq-\GG)  \ ,
\end{align}
where $M_{e_Y}(\kk,\qq+\GG)$ is given in \cref{p5:eq:chiral-MqG1} and $R_0^\eta(\kk)$ is given in \cref{p5:R0InTheChiralLimit}. 

If condition \cref{p5:eqn-condition-at-nu} is satisfied, the eigenstates $|\Psi_{\nu}^{\nu_+,\nu_-}\rangle$ are the ground states, and hence the states \cref{p5:eq:charge2excitationflatband-HI} are the neutral excitations on top of the ground states. 
Without \cref{p5:eqn-condition-at-nu}, only $\nu=0$ states are guaranteed to be the ground states, although the others are still eigenstates. 
With condition \cref{p5:eqn-condition-at-nu}, we have
\begin{align}\label{p5:scatteringmatrixcharge2chiral}
&T_{e_{Y2};e_{Y1}}(\mathbf{k+q},\mathbf{k};\pp) =  \delta_{\qq,0}\sum_{\GG,\qq} V(\GG+\qq) [\alpha_0 (\kk,\qq+\GG)^2 +\alpha_2 (\kk,\qq+\GG)^2+\alpha_0 (\kk+\mathbf{p},\qq+\GG)^2 +\alpha_2 (\kk+\mathbf{p},\qq+\GG)^2]  \nonumber \\ &+2\sum_\GG V(\mathbf{G}+\mathbf{q}) (\alpha_0(\mathbf{k+p,q+G})+ie_{Y2}\alpha_2(\mathbf{k+p,q+G})   )  (\alpha_0(\mathbf{k,q+G})+ ie_{Y1}\alpha_2(\mathbf{k,q+G})) \ ,
\end{align} 
where we have used the $\alpha_{0,2}(\kk , \qq+ \GG) = \alpha_{0,2}(-\kk , -\qq- \GG)$.
Solving \cref{p5:eq:charge2excitationflatband-HI} provides us with the expression for the charge $+2$ excitations at momentum $\mathbf{p}$ on top of the TBG ground states. 

\subsection{Absence of Cooper pairing in the projected Coulomb Hamiltonian}\label{p5:Cooperpairappendix}

In \cref{p5:newsec:noCooper}, we have derived the sufficient conditions for the existence (\cref{p5:neweq:Cooper}) and absence (\cref{p5:neweq:noCooper}) of Cooper pairing binding energies.
Now we prove that, in the projected Coulomb Hamiltonian with time-reversal symmetry $T$ and the combined symmetry $PC_{2z}T$, where $P$ is the unitary PH symmetry \cite{song_all_2019,ourpaper2},  $T^{(\eta_2,\eta_1)\prpr}_{m,m';m_2,m_1}(\kk+\qq,\kk;\pp)$ is guaranteed to be positive semi-definite. 
Thus the condition Eq.~(\ref{p5:neweq:noCooper}) for the absence of Cooper pairing binding energy is always satisfied.
We write $T^{(\eta_2,\eta_1)\prpr}_{m,m';m_2,m_1}(\kk+\qq,\kk;\pp)$, which is defined as the third term in \cref{p5:charge2excitations1}, as
\begin{equation}
 T^{(\eta_2,\eta_1)\prpr}_{m,m';m_2,m_1}(\kk+\qq,\kk;\pp) 
= 2\sum_\GG V(\mathbf{G}+\mathbf{q}) M_{m m_2}^{\left(\eta_2\right)}\left(\kk+\pp,\mathbf{q}+\mathbf{G}\right)  M_{m' m_1}^{\left(\eta_1\right)}\left(-\kk,-\mathbf{q}-\mathbf{G}\right).
\end{equation}

We first consider the case $\eta_2=-\eta_1=\eta$. 
Due to the time-reversal symmetry $T \ket{u_{n,\eta}(-\kk)} = \ket{u_{n,-\eta}(\kk)}$ in the gauge \cref{p5:eq:gauge-0} \cite{ourpaper3}, where $\ket{u_{n,\eta}(\kk)}$ is the $2N_\QQ\times 1$ vector $u_{\QQ,\alpha;n\eta}(\kk)$ in \cref{p5:eq:solution}, and the definition of the $M$ matrix (\cref{p5:eq:M-def}), we have
\begin{align}
& M_{mn}^{(\eta)}(\kk,\qq+\GG) = \inn{u_{m,\eta}(\kk+\qq+\GG)| u_{n,\eta}(\kk)} =
 \inn{T u_{m,-\eta}(-\kk-\qq-\GG)| T u_{n,-\eta}(-\kk)} \nono\\
=& \inn{u_{n,-\eta}(-\kk) |u_{m,-\eta}(-\kk-\qq-\GG)} = \inn{u_{m,-\eta}(-\kk-\qq-\GG)|u_{n,-\eta}(-\kk)}^* = M_{mn}^{(-\eta)}(-\kk,-\qq-\GG).
\end{align}
Thus we can rewrite $T^{(\eta,-\eta)\prpr}_{m,m';m_2,m_1}(\kk+\qq,\kk;\pp)$ as
\begin{equation}
T^{(\eta,-\eta)\prpr}_{m,m';m_2,m_1}(\kk+\qq,\kk;\pp) = 2\sum_\GG V(\mathbf{G}+\mathbf{q}) 
    M_{m m_2}^{\left(\eta\right)}\left(\kk+\pp,\mathbf{q}+\mathbf{G}\right)  
    M_{m' m_1}^{\left(\eta\right)*}\left(\kk,\mathbf{q}+\mathbf{G}\right).
\end{equation}
We consider the expectation value of $T^{(\eta,-\eta)\prpr}(\kk+\qq,\kk;\pp)$ on a complex function $\phi_{m_2,m_1}(\kk)$: 
\begin{align} \label{p5:neweq:T-expectation}
& \inn{T''}^\eta_\phi(\pp) = \sum_{\kk_1,\kk_2} \sum_{mm'm_2m_2}
     \phi^*_{mm'}(\kk_2) T^{(\eta,-\eta)\prpr}(\kk_2,\kk_1;\pp) \phi_{m_2,m_1}(\kk_1) \nono\\
=& 2\sum_{\kk_1,\kk_2,\GG} V(\kk_2+\mathbf{G}-\kk_1) 
    \phi^*_{mm'}(\kk_2) \inn{u_{m\eta}(\kk_2+\pp+\GG)|u_{m_2\eta}(\kk_1+\pp)} \phi_{m_2,m_1}(\kk_1)
    \inn{u_{m_1}(\kk_1) |u_{m'\eta}(\kk_2+\GG)}.
\end{align}
Using \cref{p5:neweq:u-period}, we have $\inn{u_{m_1}(\kk) |u_{m'\eta}(\kk+\qq+\GG)} = \inn{u_{m_1}(\kk+\GG') |u_{m'\eta}(\kk+\qq+\GG+\GG')}$ and hence 
\begin{align}
\inn{T''}^\eta_\phi(\pp) =& \frac{2}{N_\GG} \sum_{\kk_1,\kk_2,\GG_1,\GG_2}
    \sum_{mm' m_2m_1} V(\kk_2+\GG_2-\kk_1-\GG_1 )  \nono\\
&\times    \phi^*_{mm'}(\kk_2)
    \inn{u_{m\eta}(\kk_2+\pp+\GG_2)|u_{m_2\eta}(\kk_1+\pp+\GG_1)} \phi_{m_2,m_1}(\kk_1)
    \inn{u_{m_1}(\kk_1+\GG_1) |u_{m'\eta}(\kk_2+\GG_2)}.
\end{align}
We then define the matrix 
\begin{equation}
W(\kk_1+\GG_1) = \sum_{m_2 m_1} \ket{u_{m_2\eta}(\kk_1+\pp+\GG_1)} \phi_{m_2,m_1}(\kk_1)
    \bra{u_{m_1}(\kk_1+\GG_1)} 
\end{equation}
such that $\inn{T''}^\eta_\phi(\pp)$ can be written as
\begin{align}
\inn{T''}^\eta_\phi(\pp) =& \frac{2}{N_\GG} \sum_{\kk_1,\kk_2,\GG_1,\GG_2} \Tr[ W^\dagger(\kk_2+\GG_2) W(\kk_1+\GG_1) ] V(\kk_2+\GG_2-\kk_1-\GG_1 ) \nono\\
=& \frac{2}{N_\GG} \sum_{ab}\sum_{\kk_1,\kk_2,\GG_1,\GG_2} W_{ab}^*(\kk_2+\GG_2) W_{ab}(\kk_1+\GG_1) V(\kk_2+\GG_2-\kk_1-\GG_1), \label{p5:neweq:T-positive-TRS}
\end{align}
where $a,b$ are the indices of the matrix $W(\kk+\GG)$.
For each term with given $a,b$ in \cref{p5:neweq:T-positive-TRS}, we can view the summation over $\kk_1,\kk_2,\GG_1,\GG_2$ as $W_{ab}^\dagger V W_{ab}$, where now $W_{ab}(\kk+\GG)$ is viewed as a vector with the index $\kk+\GG$. 
Since $V(\kk_2+\GG_2-\kk_1-\GG_1)$ is a positive semi-definite matrix, $W_{ab}^\dagger V W_{ab}$ must be non-negative, and hence $\inn{T''}^\eta_\phi(\pp)\ge 0$ for arbitrary $\phi$. 
Therefore $T^{(\eta,-\eta)\prpr}_{m,m';m_2,m_1}(\kk+\qq,\kk;\pp)$ is positive semi-definite at every $\pp$. 

Then we prove that $T^{(\eta_2,\eta_1)\prpr}_{m,m';m_2,m_1}(\kk+\qq,\kk;\pp)$ with $\eta_2=\eta_1=\eta$ is also positive semi-definite. 
Due to the symmetry $PC_{2z}T \ket{u_{n,\eta}(\kk)} = n\ket{u_{-n,\eta}(-\kk)}$ in the gauge \cref{p5:eq:gauge-0} \cite{ourpaper3} and the definition of the $M$ matrix (\cref{p5:eq:M-def}), we have
\begin{align}
& M_{mn}^{(\eta)}(\kk,\qq+\GG) = \inn{u_{m,\eta}(\kk+\qq+\GG)| u_{n,\eta}(\kk)} =
 nm\inn{PC_{2z}T u_{-m,\eta}(-\kk-\qq-\GG)| PC_{2z}T u_{-n,\eta}(-\kk)} \nono\\
=& nm\inn{u_{-n,\eta}(-\kk) |u_{-m,\eta}(-\kk-\qq-\GG)} =  nm M_{-m,-n}^{(\eta)*}(-\kk,-\qq-\GG).
\end{align}
Thus we can rewrite $T^{(\eta,\eta)\prpr}_{m,m';m_2,m_1}(\kk+\qq,\kk;\pp)$ as
\begin{equation}
T^{(\eta,\eta)\prpr}_{m,m';m_2,m_1}(\kk+\qq,\kk;\pp) = 2\sum_\GG V(\mathbf{G}+\mathbf{q}) 
    M_{m m_2}^{\left(\eta\right)}\left(\kk+\pp,\mathbf{q}+\mathbf{G}\right)  
    m' m_1 M_{-m',-m_1}^{\left(\eta\right)*}\left(\kk,\mathbf{q}+\mathbf{G}\right).
\end{equation}
Repeating the calculations starting from \cref{p5:neweq:T-expectation}, one can show that $T^{(\eta,\eta)\prpr}_{m,m';m_2,m_1}(\kk+\qq,\kk;\pp)$ must be positive semi-definite. 
The only difference with the above proof is that the definition of the $W$ matrix becomes 
\begin{equation}
W(\kk_1+\GG_1) = \sum_{m_2 m_1} \ket{u_{m_2\eta}(\kk_1+\pp+\GG_1)} \phi_{m_2,m_1}(\kk_1)
    \bra{u_{-m_1}(\kk_1+\GG_1)} m_1.
\end{equation}

\begin{figure}[t]
\centering
\includegraphics[width=0.8\linewidth]{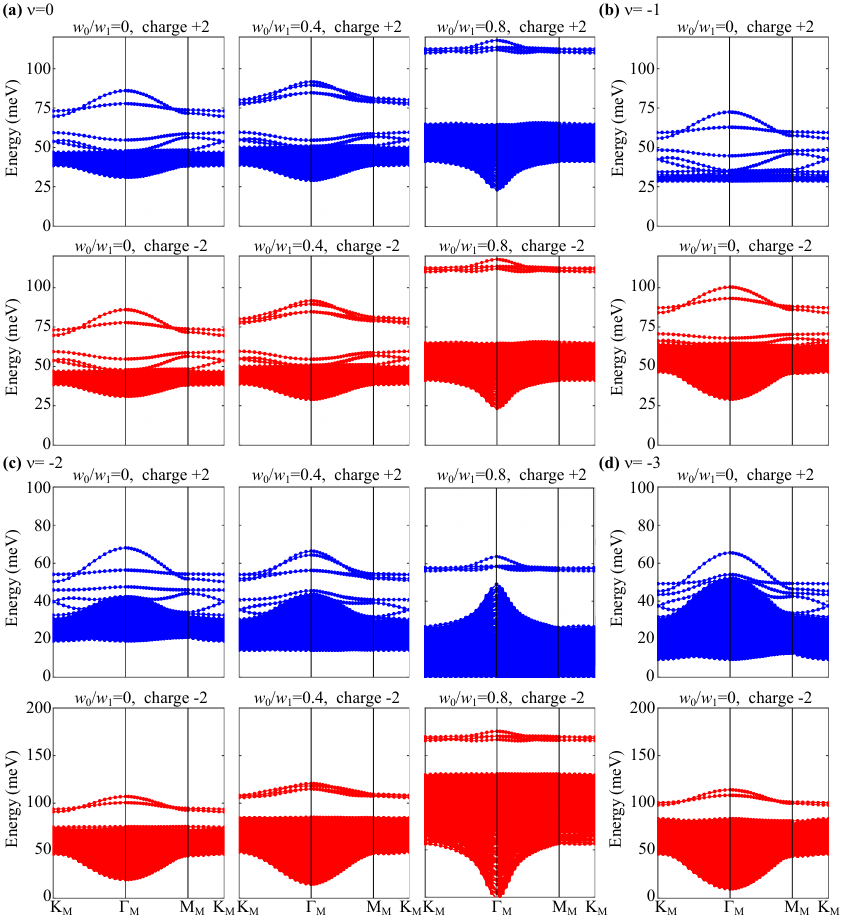}
\caption{Exact charge $+2$ (blue) and $-2$ (red) excitations at $\theta=1.05^\circ$. The flat metric condition is {\it not} imposed. 
In this plot we have used the parameters defined in \cref{p5:newapp:Ham}: $v_F=5.944{\rm eV \cdot \mathring{ A} }$, $|K|=1.703\mathring{\rm A}^{-1}$, $w_1=110{\rm meV}$,  $U_\xi=26$meV, $\xi=10$nm.
}
\label{p5:fig:E2appendix}
\end{figure}

\begin{figure}[t]
\centering
\includegraphics[width=0.8\linewidth]{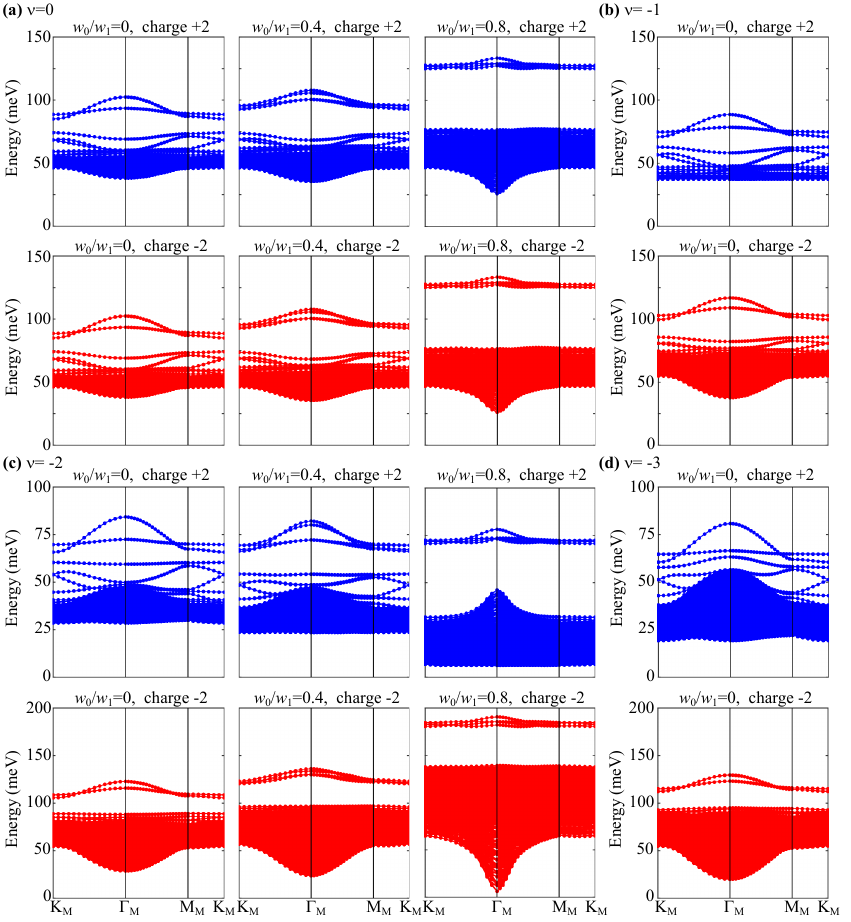}
\caption{Exact charge $+2$ (blue) and $-2$ (red) excitations at $\theta=1.05^\circ$. The flat metric condition is {\it not} imposed. 
In this plot we use the screening length $\xi=$20nm and the interaction strength $U_\xi$=13meV accordingly.
The other parameters are same as in \cref{p5:newapp:Ham}: $v_F=5.944{\rm eV \cdot \mathring{ A} }$, $|K|=1.703\mathring{\rm A}^{-1}$, $w_1=110{\rm meV}$.
}
\label{p5:fig:E2appendix20}
\end{figure}

\subsection{Charge \texorpdfstring{$-2$}{-2} excitations}\label{p5:charge-2excitations} 

Based on the above, the charge $-2$ excitations are trivial to obtain.  We do not give the details, but just the expression for the scattering elements
\begin{equation}\label{p5:eq:charge-2excitation-HI}
\left[H_I-\mu N,c_{\kk+\pp, m_2, \eta_2, s_2} c_{-\kk, m_1, \eta_1, s_1} \right] |\Psi\rangle =\frac{1}{2\Omega_{\text{tot}}}\sum_{m,m'}\sum_\qq \widetilde{T}^{(\eta_2,\eta_1)} _{m m';m_2 m_1}(\mathbf{k+q},\kk;\pp) c_{\kk+\pp+\qq, m, \eta_2, s_2} c_{-\kk-\qq, m', \eta_1, s_1}|\Psi\rangle\ ,
\end{equation} 
The $ |\Psi\rangle$ are the states  $|\Psi_\nu\rangle$ in \cref{p5:eq:U(4)-GS}, and hence $\eta_1, s_1,\eta_2, s_2$ belong to the valley/spin flavor which are not  occupied. 
For a generic exact eigenstate $|\Psi\rangle$ at chemical potential $\mu$ satisfying $(O_{\mathbf{q,G}}-A_{\GG} N_M\delta_{\mathbf{q},0})|\Psi \rangle =0$ for some coefficient $A_\GG$, we find that the $\widetilde{T}^{(\eta_2,\eta_1)} _{m_2,m;m_1m'}(\mathbf{k}_1,\kk_2;\qq)$ matrix reads
\begin{align}\label{p5:charge-2excitations1}
\widetilde{T}^{(\eta_2,\eta_1)} _{m m';m_2 m_1}(\mathbf{k+q},\kk;\pp) = & \delta_{\qq,0}(\delta_{m,m_2} \widetilde{R}_{m'm_1}^{\eta_1}(-\mathbf{k}) +\delta_{m',m_1} \widetilde{R}_{mm_2}^{\eta_2}(\mathbf{k}+\pp) ) \nono\\ &+2\sum_\GG V(\mathbf{G}+\mathbf{q})   M_{m,m_2}^{\left(\eta_2\right)*}\left(\mathbf{k}+\pp,\mathbf{q}+\mathbf{G}\right)  M_{m',m_1}^{\left(\eta_1\right)*}\left(-\mathbf{k},-\mathbf{q}- \mathbf{G}\right)  \ ,
\end{align}
where $\widetilde{R}_{mn}^\eta(\mathbf{k})$ are the $-1$ excitation matrices in \cref{p5:eq:excitation-Rmn-hole}. 
We see that the charge $-2$ energy is a sum of the two single-particle energies (first row of \cref{p5:charge-2excitations1}) plus an interaction energy (second row of \cref{p5:charge-2excitations1}). 
In particular, for the chiral limit, the scattering matrix elements are identical to those of charge $+2$, \ie \cref{p5:scatteringmatrixcharge2chiral}.

\subsection{Charge \texorpdfstring{$\pm2$}{+-2} excitation spectra for different parameters}\label{newapp:plot2}

In \cref{p5:fig:E2appendix,p5:fig:E2appendix20} the charge $\pm2$ excitations are plotted at different fillings and $w_0/w_1$'s for two different screening lengths of the Coulomb interaction (\cref{szdeq:Vq}), \ie $\xi$=10nm, 20nm, respectively.
The corresponding interaction strengths are $U_{\rm 10nm}=$26meV, $U_{\rm 20nm}$=13meV.
We have used $w_1$=110meV in all the calculations and $w_0/w_1=$0, 0.4, 0.8 for $\nu=0,-2$ and $w_0/w_1=0$ for $\nu=-1,-3$.

The charge $+2$ ($-2$) spectrum consists of a two-particle (two-hole) continuum (the blue (red) area in \cref{p5:fig:E2appendix,p5:fig:E2appendix20}) and a set of gapped charge $+2$ ($-2$) collective modes. 
The energies in the two-particle (hole) continuum are just sums of two charge $+1$ ($-1$) excitation energies. 
Note that all the charge  $+2$ ($-2$) collective modes appear above the two-particle (two-hole) continuum, implying the absence of Cooper pairing binding energy, as proved in \cref{p5:newsec:noCooper}. 

\clearpage

\section{Approximate charge \texorpdfstring{$\pm1$}{+-1} and neutral exitations at odd fillings in the nonchiral-flat U(4) limit}

\subsection{Approximate charge +1 excitations at odd fillings in the nonchiral-flat U(4) limit}\label{p5app:charge1-approx}

In the nonchiral-flat limit the states $\ket{\Psi_{\nu}^{\nu_+,\nu_-}}$ at odd fillings $\nu=\pm1,\pm3$ are no longer exact eigenstates of the Hamiltonian.
However, in Ref. \cite{ourpaper4} we have shown that the states $\ket{\Psi_{-3}^{1,0}}$ (or its U(4) rotations) and $\ket{\Psi_{-1}^{2,1}}$ (or its U(4) rotations) are the lowest states at fillings $\nu=-3,-1$ to first order perturbation of the (first) chiral symmetry breaking. 
In this subsection, we derive the approximate charge 1 excitations above these perturbative lowest states. 

Since these states have fully occupied Chern bands, it is convenient to work in the Chern band basis. 
We write the $O_{\qq,\GG}$ operator as a sum of a (first) chiral preserving term 
\begin{align}
O_{\qq,\GG}^0 =&
\sum_{\kk,e_Y,\eta,s} \sqrt{V(\qq+\GG)} M_{e_Y,e_Y}^{(\eta)}(\kk,\qq+\GG) (d_{\kk+\qq,e_Y,\eta,s}^\dagger d_{\kk,e_Y,\eta,s} - \frac12 \delta_{\qq,0}) \nono\\
=&
\sum_{\kk,e_Y,\eta,s} \sqrt{V(\qq+\GG)} M_{e_Y}(\kk,\qq+\GG) (d_{\kk+\qq,e_Y,\eta,s}^\dagger d_{\kk,e_Y,\eta,s} - \frac12 \delta_{\qq,0}),
\end{align}
and a chiral breaking term 
\begin{align}
O_{\qq,\GG}^1 =&  
\sum_{\kk,e_Y,\eta,s} \sqrt{V(\qq+\GG)} M_{-e_Y,e_Y}^{(\eta)}(\kk,\qq+\GG) d_{\kk+\qq,-e_Y,\eta,s}^\dagger d_{\kk,e_Y,\eta,s}, \nono\\
=& 
\sum_{\kk,e_Y,\eta,s} \sqrt{V(\qq+\GG)} \eta F_{e_Y}(\kk,\qq+\GG) d_{\kk+\qq,-e_Y,\eta,s}^\dagger d_{\kk,e_Y,\eta,s},
\end{align}
where the $M$ matrix $M^{(\eta)}_{e_Y',e_Y}$ and the factors $M_{e_Y}$ and $F_{e_Y}$ are defined in \cref{p5:eq:chiral-MqG1,p5:eq:chiral-MqG2}. 
We assume the initial state as $d_{\kk,e_Y,\eta,s}^\dagger \ket{\Psi_{\nu}^{\nu_+,\nu_-}}$.
Acting the Hamiltonian on the initial state, we obtain two terms: $[H_I-\mu N,d_{\kk,e_Y,\eta,s}^\dagger]\ket{\Psi_{\nu}^{\nu_+,\nu_-}}$ and $d_{\kk,e_Y,\eta,s}^\dagger (H_I-\mu N) \ket{\Psi_{\nu}^{\nu_+,\nu_-}}$.
In the (first) chiral-flat limit, the second term is simply $E_0 d_{\kk,e_Y,\eta,s}^\dagger  \ket{\Psi_{\nu}^{\nu_+,\nu_-}}$ with $E_0$ being the ground state energy.
However, the second term also involve excited states when the (first) chiral symmetry is broken.
To be specific, we have
\begin{align}
d_{\kk,e_Y,\eta,s}^\dagger H_I \ket{\Psi_{\nu}^{\nu_+,\nu_-}}
    =& \frac{1}{2\Omega_{\rm tot}}\sum_{\qq,\GG}
    d_{\kk,e_Y,\eta,s}^\dagger
    (O^0_{-\qq,-\GG} + O^1_{-\qq,-\GG})
    (O^0_{\qq,\GG} + O^1_{\qq,\GG})
    \ket{\Psi_{\nu}^{\nu_+,\nu_-}} \nono\\
    =& \frac{1}{2\Omega_{\rm tot}}\sum_{\qq,\GG}
    d_{\kk,e_Y,\eta,s}^\dagger
    (\delta_{\qq,0} A_{-\GG}A_{\GG} N_M^2 + O^1_{-\qq,-\GG}O^0_{\qq,\GG} + O^0_{-\qq,-\GG} O^1_{\qq,\GG} + O^1_{-\qq,-\GG} O^1_{\qq,\GG})
    \ket{\Psi_{\nu}^{\nu_+,\nu_-}}.
\end{align}
The first term on the right hand side will give the unperturbed ground state energy $E_0$, whereas the other three terms yield excited states.
Now we approximate it by projecting it into the Hilbert space with a single particle excitation. 
Notice that $d_{\kk,e_Y,\eta,s}^\dagger$ generates a particle in an empty Chern band and the $O^1$ operator, by definition, generates particles in the empty Chern bands and holes in occupied Chern bands.
Thus the terms $d_{\kk,e_Y,\eta,s}^\dagger O^1_{-\qq,-\GG}O^0_{\qq,\GG}$ and $d_{\kk,e_Y,\eta,s}^\dagger O^0_{-\qq,-\GG}O^1_{\qq,\GG}$ will at least generate two particles plus one hole. 
Hence they do not contribute to the projected equation. 
Now we consider the term  $d_{\kk,e_Y,\eta,s}^\dagger O^1_{-\qq,-\GG}O^1_{\qq,\GG}$
\begin{align}
d_{\kk,e_Y,\eta,s}^\dagger O^1_{-\qq,-\GG}O^1_{\qq,\GG} 
&= \sum_{\kk_1,\kk_2} \sum_{e_{Y1},e_{Y2}} \sum_{\eta_1 s_1 \eta_2 s_2} 
 V(\qq+\GG) \eta_1 F^{(\eta_1)}_{e_{Y1}}(\kk_1,-\qq-\GG) \eta_2 F^{(\eta_2)}_{e_{Y2}}(\kk_2,\qq+\GG) \nono\\
& \times  d_{\kk,e_Y,\eta,s}^\dagger 
    d_{\kk_1-\qq,-e_{Y1},\eta_1,s_1}^\dagger d_{\kk_1,e_{Y1},\eta_1,s_1}
    d_{\kk_2+\qq,-e_{Y2},\eta_2,s_2}^\dagger d_{\kk_2,e_{Y2},\eta_2,s_2} \ .
\end{align}
According to the Wick's theorem, we have
\begin{align}
& d_{\kk,e_Y,\eta,s}^\dagger 
    d_{\kk_1-\qq,-e_{Y1},\eta_1,s_1}^\dagger d_{\kk_1,e_{Y1},\eta_1,s_1}
    d_{\kk_2+\qq,-e_{Y2},\eta_2,s_2}^\dagger d_{\kk_2,e_{Y2},\eta_2,s_2} \nono\\
=& :d_{\kk,e_Y,\eta,s}^\dagger 
    d_{\kk_1-\qq,-e_{Y1},\eta_1,s_1}^\dagger d_{\kk_1,e_{Y1},\eta_1,s_1}
    d_{\kk_2+\qq,-e_{Y2},\eta_2,s_2}^\dagger d_{\kk_2,e_{Y2},\eta_2,s_2}: + \cdots \nono\\
& + \inn{\Psi_\nu^{\nu_+,\nu_-}| d_{\kk,e_Y,\eta,s}^\dagger d_{\kk_1,e_{Y1},\eta_1,s_1} | \Psi_\nu^{\nu_+,\nu_-} }
    \inn{\Psi_\nu^{\nu_+,\nu_-}| d_{\kk_1-\qq,-e_{Y1},\eta_1,s_1}^\dagger d_{\kk_2,e_{Y2},\eta_2,s_2} | \Psi_\nu^{\nu_+,\nu_-}} d_{\kk_2+\qq,-e_{Y2},\eta_2,s_2}^\dagger \nono\\
& - \inn{\Psi_\nu^{\nu_+,\nu_-}| d_{\kk,e_Y,\eta,s}^\dagger d_{\kk_2,e_{Y2},\eta_2,s_2} | \Psi_\nu^{\nu_+,\nu_-}}
    \inn{\Psi_\nu^{\nu_+,\nu_-}| d_{\kk_1,e_{Y1},\eta_1,s_1} d_{\kk_2+\qq,-e_{Y2},\eta_2,s_2}^\dagger |\Psi_\nu^{\nu_+,\nu_-}} d_{\kk_1-\qq,-e_{Y1},\eta_1,s_1}^\dagger \nono\\
& + \inn{\Psi_\nu^{\nu_+,\nu_-}| d_{\kk_1-\qq,-e_{Y1},\eta_1,s_1}^\dagger d_{\kk_2,e_{Y2},\eta_2,s_2} | \Psi_\nu^{\nu_+,\nu_-}}
    \inn{\Psi_\nu^{\nu_+,\nu_-}| d_{\kk_1,e_{Y1},\eta_1,s_1} d_{\kk_2+\qq,-e_{Y2},\eta_2,s_2}^\dagger |\Psi_\nu^{\nu_+,\nu_-}} d_{\kk,e_Y,\eta,s}^\dagger \ .
\end{align}
Here $:A:=A$ represents the normal ordered form of the operator $A$ with respect to $\ket{\Psi_\nu^{\nu_+,\nu_-}}$, where the operators that annihilate $\ket{\Psi_\nu^{\nu_+,\nu_-}}$ is ordered on the right hand side of the operators that do not.
The second term (``$\cdots$'') represent the normal ordered terms with one contraction.
The first (second) term either annihilate $\ket{\Psi_\nu^{\nu_+,\nu_-}}$ or generate three (two) particles plus two (one) holes.
We hence will omit them. 
The third and fourth terms must vanish since we require the flavor $\{e_Y,\eta,s\}$ to be empty and hence $\bra{\Psi_\nu^{\nu_+,\nu_-}} d_{\kk,e_Y,\eta,s}^\dagger = 0$. 
The last term is the Fock energy correction to the ground state energy.
Therefore, we conclude
\begin{equation}
d_{\kk,e_Y,\eta,s}^\dagger H_I \ket{\Psi_{\nu}^{\nu_+,\nu_-}}
    \approx (E_0 + \Delta E_0) 
    d_{\kk,e_Y,\eta,s}^\dagger  \ket{\Psi_{\nu}^{\nu_+,\nu_-}}, \label{p5eq:charge1-HFenergy}
\end{equation}
where $E_0$ is the unperturbed ground state energy and $\Delta E_0$ is the Fock energy correct.

The excitation energy is hence given by the spectrum of $[H_I-\mu N,d_{\kk,e_Y,\eta,s}^\dagger]$.
Following the calculation in \cref{p5:ChargeCommutationRelationsAppendix}, we obtain
\begin{align}
& [O_{\mathbf{-q,-G}}O_{\mathbf{q,G}}, d_{\mathbf{k},e_Y,\eta,s}^\dag]
= V(\qq+\GG) P^\eta_{e_Y',e_Y}(\kk,\qq+\GG) d_{\mathbf{k},e_Y',\eta,s}^\dag \nono\\
+& \sqrt{V(\mathbf{G}+\mathbf{q})}  (M_{e_Y',e_Y}^{(\eta)}\left(\mathbf{k},\mathbf{q}+\mathbf{G}\right) d_{\mathbf{k+q},e_Y',\eta,s}^\dag O_{-\mathbf{q,-G}} + M_{e_Y',e_Y}^{(\eta)}\left(\mathbf{k},-\mathbf{q}-\mathbf{G}\right) d_{\mathbf{k-q},e_Y',\eta,s}^\dag O_{\mathbf{q,G}}),
\end{align}
where
\begin{equation} \label{p5eq:P-Chern}
P^\eta_{e_Y',e_Y}(\kk,\qq+\GG) = \sum_{e_Y''} M^{(\eta)*}_{e_Y'',e_Y'}(\kk,\qq+\GG) M^{(\eta)}_{e_Y'',e_Y}(\kk,\qq+\GG) \ .
\end{equation}
Acting the commutator of the interaction Hamiltonian and $d^\dagger_{\kk,e_Y,\eta,s}$ on the state $\ket{\Psi_{\nu}^{\nu_+,\nu_-}}$, we have 
\begin{align}
& [H_I -\mu N, d^\dagger_{\kk,e_Y,\eta,s}] \ket{\Psi_{\nu}^{\nu_+,\nu_-}}
=  \frac{1}{2\Omega_{tot}} \sum_{\qq,\GG,e_Y'} \bigg( V(\qq+\GG) P^\eta_{e_Y',e_Y}(\kk,\qq+\GG) d_{\mathbf{k},e_Y',\eta,s}^\dag \nono\\
&\qquad + 2\sqrt{V(\mathbf{G}+\mathbf{q})}  M_{e_Y',e_Y}^{(\eta)} \left(\mathbf{k},\mathbf{q}+\mathbf{G}\right) d_{\mathbf{k+q},e_Y',\eta,s}^\dag O_{-\mathbf{q,-G}} \bigg) \ket{\Psi_{\nu}^{\nu_+,\nu_-}} - \mu  d^\dagger_{\kk,e_Y,\eta,s} \ket{\Psi_{\nu}^{\nu_+,\nu_-}}. \label{p5eq:Rapprox-tmp1}
\end{align}
In the (first) chiral limit, where $O_{-\qq,-\GG} \ket{\Psi_{\nu}^{\nu_+,\nu_-}} = \delta_{\qq,0} A_{-\GG} N_M \ket{\Psi_{\nu}^{\nu_+,\nu_-}}$, the right hand side of the above equation only has one particle excitations.
However, when the (first) chiral symmetry is broken, $d^\dagger_{\kk,e_Y,\eta,s} O_{-\qq,-\GG} \ket{\Psi_{\nu}^{\nu_+,\nu_-}}$ will yield excitations with two particles plus one hole
\begin{align}
&d_{\mathbf{k+q},e_Y',\eta,s}^\dag O_{-\mathbf{q,-G}} \ket{\Psi_{\nu}^{\nu_+,\nu_-}} 
= d_{\mathbf{k+q},e_Y',\eta,s}^\dag (O^0_{-\mathbf{q,-G}} + O^1_{-\mathbf{q,-G}}) \ket{\Psi_{\nu}^{\nu_+,\nu_-}} = \delta_{\qq,0} A_{-\GG} N_M d_{\mathbf{k},e_Y',\eta,s}^\dag \ket{\Psi_{\nu}^{\nu_+,\nu_-}} \nono\\
& +  
\sum_{\kk',e_Y'',\eta',s'} \sqrt{V(\qq+\GG)} \eta' F_{e_Y''}(\kk',-\qq-\GG) d_{\mathbf{k+q},e_Y',\eta,s}^\dag d_{\kk'-\qq,-e_Y'',\eta',s'}^\dagger d_{\kk',e_Y'',\eta',s'} \ket{\Psi_{\nu}^{\nu_+,\nu_-}}
\end{align}
We now approximate the right hand side by projecting it into the one particle Hilbert space:
We only keep the term satisfying $\kk'=\kk+\qq$, $e_Y''=e_Y'$, $\eta'=\eta$, $s'=s$
\begin{align}
& d_{\mathbf{k+q},e_Y',\eta,s}^\dag O_{-\mathbf{q,-G}} \ket{\Psi_{\nu}^{\nu_+,\nu_-}} 
\approx \delta_{\qq,0} A_{-\GG} N_M d_{\mathbf{k},e_Y',\eta,s}^\dag \ket{\Psi_{\nu}^{\nu_+,\nu_-}} \nono\\  
& - \sqrt{V(\qq+\GG)} \eta F_{e_Y'}(\kk+\qq,-\qq-\GG)  n_{e_Y',\eta,s} d_{\kk,-e_Y',\eta,s}^\dagger \ket{\Psi_{\nu}^{\nu_+,\nu_-}},
\end{align}
where $n_{e_Y',\eta,s}$ equals to 1 if the flavor $\{e_Y',\eta,s\}$ is occupied and equals to 0 otherwise. 
If $e_Y'$ in the second term equals to $e_Y$, then there must be $n_{e_Y',\eta,s}=0$ because we require $\{e_Y,\eta,s\}$ to be empty such that $d^\dagger_{\kk,e_Y,\eta,s} \ket{\Psi_{\nu}^{\nu_+,\nu_-}}$ is non-vanishing.
Hence we only need to keep the $e_Y'=-e_Y$ in the second term. 
Then we can rewrite the second term in \cref{p5eq:Rapprox-tmp1} as
\begin{align}
 & \sum_{e_Y'} M_{e_Y',e_Y}^{(\eta)} \left(\mathbf{k},\mathbf{q}+\mathbf{G}\right) d_{\mathbf{k+q},e_Y',\eta,s}^\dag O_{-\mathbf{q,-G}}  \ket{\Psi_{\nu}^{\nu_+,\nu_-}}
 \approx  \sum_{e_Y'} \delta_{\qq,0} A_{-\GG} N_M M_{e_Y',e_Y}^{(\eta)}(\kk,\GG) d_{\mathbf{k},e_Y',\eta,s}^\dag \ket{\Psi_{\nu}^{\nu_+,\nu_-}} 
 \nono\\
& - 
\sqrt{V(\qq+\GG)} M_{-e_Y,e_Y}^{\eta}(\kk,\qq+\GG) \eta F_{-e_Y}(\kk+\qq,-\qq-\GG) n_{-e_Y,\eta,s} d_{\kk,e_Y,\eta,s}^\dagger \ket{\Psi_{\nu}^{\nu_+,\nu_-}} \nono\\
\approx & \sum_{e_Y'} \delta_{\qq,0} A_{-\GG} N_M M_{e_Y',e_Y}^{(\eta)}(\kk,\GG) d_{\mathbf{k},e_Y',\eta,s}^\dag \ket{\Psi_{\nu}^{\nu_+,\nu_-}} 
- n_{-e_Y,\eta,s} \sqrt{V(\qq+\GG)} |F_{e_Y}(\kk,\qq+\GG)|^2  d^\dagger_{e_Y,\eta,s} \ket{\Psi_{\nu}^{\nu_+,\nu_-}}.
\end{align}
Here we have made use of \cref{p5:eq:chiral-MqG2} and \cref{p5:eq:alpha-cond1}. 
With the above approximation, we can write the excitation equation as
\begin{equation}
    [H_I -\mu N, d^\dagger_{\kk,e_Y,\eta,s}] \ket{\Psi_{\nu}^{\nu_+,\nu_-}} \approx \sum_{e_Y'} R^{\eta,s}_{e_Y',e_Y} d^\dagger_{\kk,e_Y',\eta,s} \ket{\Psi_{\nu}^{\nu_+,\nu_-}},
\end{equation}
where
\begin{align}
R^{\eta,s}_{e_Y',e_Y} =& \frac{1}{2\Omega_{\rm tot}} \sum_{\qq,\GG} \bigg( V(\qq+\GG) P^\eta_{e_Y',e_Y}(\kk,\qq+\GG)
 + 2 \delta_{\qq,0} \sqrt{V(\mathbf{G})} A_{-\GG} N_M M_{e_Y',e_Y}^{(\eta)} \left(\mathbf{k},\mathbf{G}\right)  \nono\\
& - 2\delta_{e_Y',e_Y} n_{-e_Y,\eta,s} V(\qq+\GG) |F_{e_Y}(\kk,\qq+\GG)|^2 \bigg) 
- \mu \delta_{e_Y',e_Y}. \label{p5eq:Rapprox}
\end{align}
Notice that both $e_Y$ and $e_Y'$ are limited to empty Chern bands in the valley-spin flavor $\eta,s$.

\begin{figure}
\centering
\includegraphics[width=1\linewidth]{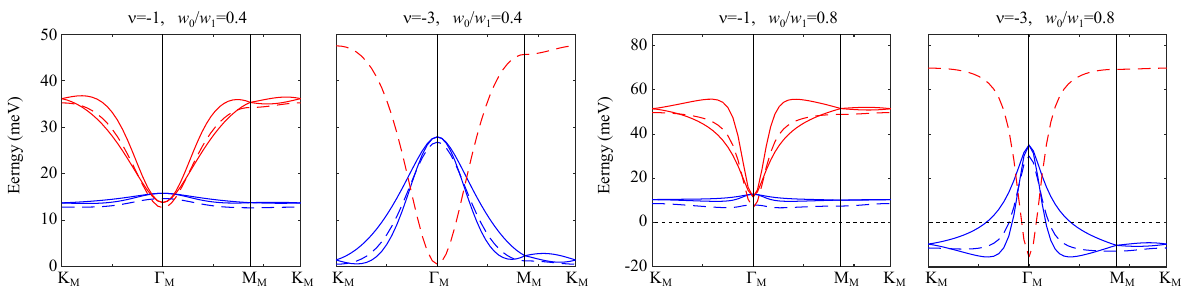}
\caption{Approximate charge $+1$ (blue) and $-1$ (red) excitations at $\theta=1.05^\circ$ at odd fillings in the nonchiral-flat limit.
The dashed bands are the excitations in the half-filled valley-spin sectors.
The solid blue (red) bands are the +1 ($-1$) excitations in the fully empty (occupied) valley-spin sectors. 
The flat metric condition is {\it not} imposed. 
In this plot we set the screening length as $\xi=10$nm and accordingly the interaction strength as $U_\xi=13$meV.
The other parameters are same as in \cref{p5:newapp:Ham}: $v_F=5.944{\rm eV \cdot \mathring{ A} }$, $|K|=1.703\mathring{\rm A}^{-1}$, $w_1=110{\rm meV}$.}
\label{fig:E1approx}
\end{figure}

Let us first consider the charge +1 excitation at $\nu=-3$.
Without loss of generality, we assume the occupied flavor is $\{+1,\up,+1\}$. 
For the excitation in the half-filled valley-spin sector ($\{+1,\up\}$), the $e_Y$ and $e_Y'$ indices in $R^{\eta,s}_{e_Y',e_Y}$ are limited for the empty Chern band ($-1$).
In this case the $R$ matrix is one-by-one with $n_{-e_Y,\eta,s}=1$. 
For the excitation in the other (fully empty) valley-spin sectors, the $e_Y$ and $e_Y'$ indices can be either 1 or $-1$. 
Hence the $R$ matrix is a two-by-two matrix with $n_{-e_Y,\eta,s}=0$.
The calculation for $\nu=-1$ is similar.
Due to Ref. \cite{ourpaper4}, the perturbative ground state in the nonchiral-flat limit at $\nu=-1$ is $\ket{\Psi_{-1}^{2,1}}$ (or its U(4) rotations).
There is a fully occupied valley-spin sector and a half-filled valley-spin sector. 
In the half-filled valley-spin sector $R$ is one-by-one with $n_{-e_Y,\eta,s}=1$ and in the empty valley-spin sectors $R$ is two-by-two with $n_{-e_Y,\eta,s}=0$. 
The approximate charge $+1$ excitations are shown in \cref{fig:E1approx}.

\subsection{Approximate charge -1 excitations at odd fillings in the nonchiral-flat limit}

In the nonchiral-flat limit the states $\ket{\Psi_{\nu}^{\nu_+,\nu_-}}$ at odd fillings $\nu=\pm1,\pm3$ are no longer exact eigenstates of the Hamiltonian.
However, in Ref. \cite{ourpaper4} we have shown that the states $\ket{\Psi_{-3}^{1,0}}$ (or its U(4) rotations) and $\ket{\Psi_{-1}^{2,1}}$ (or its U(4) rotations) are the lowest states at fillings $\nu=-3,-1$ to first order perturbation of the (first) chiral symmetry breaking. 
Using the same method as in \cref{p5app:charge1-approx}, in this subsection, we derive the approximate charge -1 excitations above these perturbative lowest states. 

For the same reason in \cref{p5app:charge1-approx}, the spectrum of excitation is given by the eigenvalues of $[H_I -\mu N, d_{\kk,e_Y,\eta,s}]$. 
Following the calculation in \cref{p5app:charge1-approx}, we have
\begin{align}
& [H_I -\mu N, d_{\kk,e_Y,\eta,s}] \ket{\Psi_{\nu}^{\nu_+,\nu_-}}
=  \frac{1}{2\Omega_{tot}} \sum_{\qq,\GG,e_Y'} \bigg( V(\qq+\GG) P_{e_Y',e_Y}^{\eta*} (\kk,\qq+\GG) d_{\mathbf{k},e_Y',\eta,s} \nono\\
&\qquad - 2\sqrt{V(\mathbf{G}+\mathbf{q})}  M_{e_Y',e_Y}^{(\eta)*} \left(\mathbf{k},\mathbf{q}+\mathbf{G}\right) d_{\mathbf{k+q},e_Y',\eta,s} O_{\mathbf{q,G}} \bigg) \ket{\Psi_{\nu}^{\nu_+,\nu_-}} + \mu  d_{\kk,e_Y,\eta,s} \ket{\Psi_{\nu}^{\nu_+,\nu_-}}, \label{p5eq:Rtapprox-tmp1}
\end{align}
where $P^\eta$ is defined in \cref{p5eq:P-Chern}. 
When the (first) chiral symmetry is broken, the term $d_{\mathbf{k+q},e_Y',\eta,s} O_{\mathbf{q,G}} \ket{\Psi_{\nu}^{\nu_+,\nu_-}}$ yields excitations with two holes plus one particle
\begin{align}
&d_{\mathbf{k+q},e_Y',\eta,s} O_{\mathbf{q,G}} \ket{\Psi_{\nu}^{\nu_+,\nu_-}} 
= d_{\mathbf{k+q},e_Y',\eta,s} (O^0_{\mathbf{q,G}} + O^1_{\mathbf{q,G}}) \ket{\Psi_{\nu}^{\nu_+,\nu_-}} \nono\\
=& \delta_{\qq,0} A_{\GG} N_M d_{\mathbf{k},e_Y',\eta,s} \ket{\Psi_{\nu}^{\nu_+,\nu_-}} +  
\sum_{\kk',e_Y'',\eta',s'} \sqrt{V(\qq+\GG)} \eta' F_{e_Y''}(\kk',\qq+\GG) d_{\mathbf{k+q},e_Y',\eta,s} d_{\kk'+\qq,-e_Y'',\eta',s'}^\dagger d_{\kk',e_Y'',\eta',s'} \ket{\Psi_{\nu}^{\nu_+,\nu_-}}
\end{align}
We now approximate the right hand side by projecting it into the one hole Hilbert space:
We only keep the term satisfying $\kk'=\kk$, $e_Y''=-e_Y'$, $\eta'=\eta$, $s'=s$
\begin{align}
&d_{\mathbf{k+q},e_Y',\eta,s} O_{\mathbf{q,G}} \ket{\Psi_{\nu}^{\nu_+,\nu_-}} 
= d_{\mathbf{k+q},e_Y',\eta,s} (O^0_{\mathbf{q,G}} + O^1_{\mathbf{q,G}}) \ket{\Psi_{\nu}^{\nu_+,\nu_-}} \nono\\
=& \delta_{\qq,0} A_{\GG} N_M d_{\mathbf{k},e_Y',\eta,s} \ket{\Psi_{\nu}^{\nu_+,\nu_-}} +  
 \sqrt{V(\qq+\GG)} \eta F_{-e_Y'}(\kk,\qq+\GG) (1-n_{e_Y',\eta,s}) d_{\kk,-e_Y',\eta,s} \ket{\Psi_{\nu}^{\nu_+,\nu_-}} \ ,
\end{align}
where $n_{e_Y',\eta,s}$ equals to 1 if the flavor $\{e_Y',\eta,s\}$ is occupied and equals to 0 otherwise. 
If $e_Y'$ in the second term equals to $e_Y$, then there must be $n_{e_Y',\eta,s}=1$ because we require $\{e_Y,\eta,s\}$ to be occupied such that $d_{\kk,e_Y,\eta,s} \ket{\Psi_{\nu}^{\nu_+,\nu_-}}$ is non-vanishing.
Hence we only need to keep the $e_Y'=-e_Y$ in the second term. 
Then we can rewrite the second term in \cref{p5eq:Rtapprox-tmp1} as
\begin{align}
 & \sum_{e_Y'}M_{e_Y',e_Y}^{(\eta)*} (\kk,\qq+\GG) d_{\mathbf{k+q},e_Y',\eta,s} O_{\mathbf{q,G}} \bigg) \ket{\Psi_{\nu}^{\nu_+,\nu_-}} \nono\\
=& \sum_{e_Y'} \delta_{\qq,0} A_\GG N_M  M_{e_Y',e_Y}^{(\eta)*} (\kk,\GG) d_{\mathbf{k},e_Y',\eta,s} \ket{\Psi_{\nu}^{\nu_+,\nu_-}} 
+ (1-n_{-e_Y,\eta,s}) \sqrt{V(\qq+\GG)} |F_{e_Y}(\kk,\qq+\GG)|^2 d_{\kk,e_Y,\eta,s} \ket{\Psi_{\nu}^{\nu_+,\nu_-}} \ .
\end{align}
Here we have made use of \cref{p5:eq:chiral-MqG2,p5:eq:alpha-cond1}.
With the above approximation, we can write the excitation equation as
\begin{equation}
    [H_I -\mu N, d_{\kk,e_Y,\eta,s}] \ket{\Psi_{\nu}^{\nu_+,\nu_-}} \approx \sum_{e_Y'} \td{R}^{\eta,s}_{e_Y',e_Y} d_{\kk,e_Y',\eta,s} \ket{\Psi_{\nu}^{\nu_+,\nu_-}},
\end{equation}
where
\begin{align} \label{p5eq:Rtapprox}
\td{R}^{\eta,s}_{e_Y',e_Y} =& \frac{1}{2\Omega_{\rm tot}} \sum_{\qq,\GG} \bigg( V(\qq+\GG) P_{e_Y',e_Y}^{\eta*} (\kk,\qq+\GG)
 - 2 \delta_{\qq,0} \sqrt{V(\mathbf{G})} A_{\GG} N_M M_{e_Y',e_Y}^{(\eta)*} \left(\mathbf{k},\mathbf{G}\right)  \nono\\
& - 2\delta_{e_Y',e_Y} (1-n_{-e_Y,\eta,s}) V(\qq+\GG) |F_{e_Y}(\kk,\qq+\GG)|^2 \bigg) 
+ \mu \delta_{e_Y',e_Y}.
\end{align}
Notice that both $e_Y$ and $e_Y'$ are limited to fully filled Chern bands in the valley-spin flavor $\eta,s$.

Let us first consider the charge -1 excitation at $\nu=-3$.
Without loss of generality, we assume the occupied flavor is $\{+1,\up,+1\}$. 
For the excitation in the half-filled valley-spin sector ($\{+1,\up\}$), the $e_Y$ and $e_Y'$ indices in $\td{R}^{\eta,s}_{e_Y',e_Y}$ are limited for the occupied Chern band ($+1$).
In this case the $\td{R}$ matrix is one-by-one with $n_{-e_Y,\eta,s}=0$. 
And there is no hole excitation in the other (fully empty) valley-spin sectors. 
The calculation for $\nu=-1$ is similar.
Due to Ref. \cite{ourpaper4}, the perturbative ground state in the nonchiral-flat limit at $\nu=-1$ is $\ket{\Psi_{-1}^{2,1}}$ (or its U(4) rotations).
There is a fully occupied valley-spin sector and a half-filled valley-spin sector. 
In the half-filled valley-spin sector $\td{R}$ is one-by-one with $n_{-e_Y,\eta,s}=0$ and in the fully occupied valley-spin sector $\td{R}$ is two-by-two with $n_{-e_Y,\eta,s}=1$. 
The approximate charge $-1$ excitations are shown in \cref{fig:E1approx}.

\subsection{Approximate charge neutral excitations at odd fillings in the nonchiral-flat U(4) limit}

In the nonchiral-flat limit the states $\ket{\Psi_{\nu}^{\nu_+,\nu_-}}$ at odd fillings $\nu=\pm1,\pm3$ are no longer exact eigenstates of the Hamiltonian.
However, in Ref. \cite{ourpaper4} we have shown that the states $\ket{\Psi_{-3}^{1,0}}$ (or its U(4) rotations) and $\ket{\Psi_{-1}^{2,1}}$ (or its U(4) rotations) are the lowest states at fillings $\nu=-3,-1$ to first order perturbation of the (first) chiral symmetry breaking. 
In this subsection, we derive the approximate charge neutral excitations above these perturbative lowest states. 

Since these states have fully occupied Chern bands, it is convenient to work in the Chern band basis. 
Following the calculations in \cref{p5:NeutralExcitationNonChiralappendix}, we have
\begin{align} \label{p5eq:charge-neutral-tmp1}
& [H_I- \mu N, d_{\kk+\pp, e_{Y2}, \eta_2, s_2}^\dagger d_{\kk, e_{Y1}, \eta_1, s_1}] \ket{\Psi_\nu^{\nu_+,\nu_-}} \nonumber \\ 
=& \frac1{2\Omega_{\rm tot}} \sum_{\qq,\GG} \bigg(
\sum_{e_{Y}}  P^{\eta_2}_{e_Y,e_{Y2}}(\kk+\pp,\qq+\GG)
   d_{\mathbf{k}+\pp,e_Y,\eta_2,s_2}^\dag d_{\kk, e_{Y1}, \eta_1, s_1}  
   + \sum_{e_Y'} P_{e_Y',e_{Y1}}^{\eta_1*} (\kk,\qq+\GG)
    d_{\kk+\pp, e_{Y2}, \eta_2, s_2}^\dagger d_{\kk, e_Y', \eta_1, s_1}\nonumber \\ 
& + 2\sqrt{V(\mathbf{G}+\mathbf{q})} \sum_{e_Y}    
    M_{e_Y,e_{Y2}}^{\left(\eta_2\right)}\left(\kk+\pp,\mathbf{q}+\mathbf{G}\right) d_{\kk+\pp+ \qq, e_Y, \eta_2, s_2}^\dagger d_{\kk, e_{Y1}, \eta_1, s_1} O_{-\mathbf{q,-G}}  \nonumber \\ 
&-2\sqrt{V(\mathbf{G}+\mathbf{q})}\sum_{e_Y'} 
     M_{e_Y',e_{Y1}}^{\left(\eta_1\right)*}\left(\mathbf{k},-\mathbf{q}-\mathbf{G}\right) d_{\kk+\pp, e_{Y2}, \eta_2, s_2}^\dagger d_{\kk-\qq, e_Y', \eta_1, s_1} O_{-\mathbf{q,-G}}  \nonumber \\ 
&-2 V(\mathbf{G}+\mathbf{q}) \sum_{e_Y,e_Y'}  
    M_{e_Y,e_{Y2}}^{\left(\eta_2\right)}\left(\mathbf{k}+\pp,\mathbf{q}+\mathbf{G}\right)  M_{e_Y',e_{Y1}}^{\left(\eta_1\right)*}\left(\mathbf{k},\mathbf{q}+ \mathbf{G}\right)  d_{\kk+\pp+\qq, e_Y, \eta_2, s_2}^\dagger d_{\kk+\qq, e_Y', \eta_1, s_1} \bigg) \ket{\Psi_\nu^{\nu_+,\nu_-}} \ .
\end{align}
In the (first) chiral limit, where $O_{-\qq,-\GG} \ket{\Psi_{\nu}^{\nu_+,\nu_-}} = \delta_{\qq,0} A_{-\GG} N_M \ket{\Psi_{\nu}^{\nu_+,\nu_-}}$, the right hand side of the above equation only involve excitations with one pair of particle and hole.
However, when the (first) chiral symmetry is broken, $O_{-\qq,-\GG} \ket{\Psi_{\nu}^{\nu_+,\nu_-}}$ will yield additional particle-hole excitations:
\begin{align}
& d_{\kk+\pp+ \qq, e_Y, \eta_2, s_2}^\dagger d_{\kk, e_{Y1}, \eta_1, s_1} O_{-\mathbf{q,-G}} \ket{\Psi_\nu^{\nu_+,\nu_-}} = d_{\kk+\pp+ \qq, e_Y, \eta_2, s_2}^\dagger d_{\kk, m_1, \eta_1, s_1} ( O^0_{-\qq,-\GG} + O^1_{-\qq,-\GG} ) \ket{\Psi_\nu^{\nu_+,\nu_-}} \nono\\
=& \delta_{\qq,0} A_{-\GG} N_M d_{\kk+\pp, e_Y, \eta_2, s_2}^\dagger d_{\kk, e_{Y1}, \eta_1, s_1} \ket{\Psi_\nu^{\nu_+,\nu_-}} 
    + \sum_{\kk'',e_Y'',\eta'',s''} \sqrt{V(\qq+\GG)} \eta'' F_{e_Y''}(\kk'',-\qq-\GG) \nono\\
&  \qquad \times
    d_{\kk+\pp+ \qq, e_Y, \eta_2, s_2}^\dagger d_{\kk, e_{Y1}, \eta_1, s_1}
    d_{\kk''-\qq,-e_Y'',\eta'',s''}^\dagger d_{\kk'',e_Y'',\eta'',s''} \ket{\Psi_\nu^{\nu_+,\nu_-}} \ . 
\end{align}
We approximate the above equation by projecting it into the Hilbert space of excitations with only one pair of particle and hole. 
According to the Wick's theorem, we have
\begin{align}
&  d_{\kk+\pp+ \qq, e_Y, \eta_2, s_2}^\dagger d_{\kk, e_{Y1}, \eta_1, s_1} d_{\kk''-\qq,-e_Y'',\eta'',s''}^\dagger d_{\kk'',e_Y'',\eta'',s''} 
    = :d_{\kk+\pp+ \qq, e_Y, \eta_2, s_2}^\dagger d_{\kk, e_{Y1}, \eta_1, s_1}
    d_{\kk''-\qq,-e_Y'',\eta'',s''}^\dagger d_{\kk'',e_Y'',\eta'',s''} : \nono\\
& +  \bra{\Psi_\nu^{\nu_+,\nu_-}}d_{\kk+\pp+ \qq, e_Y, \eta_2, s_2}^\dagger d_{\kk, e_{Y1}, \eta_1, s_1} \ket{\Psi_\nu^{\nu_+,\nu_-}} d_{\kk''-\qq,-e_Y'',\eta'',s''}^\dagger d_{\kk'',e_Y'',\eta'',s''}  \nono\\
& + \bra{\Psi_\nu^{\nu_+,\nu_-}} d_{\kk+\pp+ \qq, e_Y, \eta_2, s_2}^\dagger d_{\kk'',e_Y'',\eta'',s''} \ket{\Psi_\nu^{\nu_+,\nu_-}}  d_{\kk, e_{Y1}, \eta_1, s_1} d_{\kk''-\qq,-e_Y'',\eta'',s''}^\dagger \nono\\
& + \bra{\Psi_\nu^{\nu_+,\nu_-}} d_{\kk, e_{Y1}, \eta_1, s_1} d_{\kk''-\qq,-e_Y'',\eta'',s''}^\dagger \ket{\Psi_\nu^{\nu_+,\nu_-} }
    d_{\kk+\pp+ \qq, e_Y, \eta_2, s_2}^\dagger d_{\kk'',e_Y'',\eta'',s''}  \nono\\
& - \bra{\Psi_\nu^{\nu_+,\nu_-}} d_{\kk, e_{Y1}, \eta_1, s_1} d_{\kk''-\qq,-e_Y'',\eta'',s''}^\dagger \ket{\Psi_\nu^{\nu_+,\nu_-} }
    \bra{\Psi_\nu^{\nu_+,\nu_-}} d_{\kk+\pp+ \qq, e_Y, \eta_2, s_2}^\dagger d_{\kk'',e_Y'',\eta'',s''} \ket{\Psi_\nu^{\nu_+,\nu_-}}
\end{align}
Here $:A:$ represents the normal ordered form of the operator $A$ with respect to $\ket{\Psi_\nu^{\nu_+,\nu_-}}$, where all the creation operators of occupied states and annihilation operators of empty states are on the right hand side of the creation operators of empty states and annihilation operators of occupied states.
The first (normal ordered) term is non-vanishing when acted on $\ket{\Psi_\nu^{\nu_+,\nu_-}}$ if the two creation operators are of empty states and the two annihilation operators are of occupied states.
However, we will omit this term because it yields two particle-hole pairs. 
Since we require $d_{\kk,e_{Y1},\eta_1,s_1}$ to be occupied such that the initial state is non-vanishing, there must be $\bra{\Psi_\nu^{\nu_+,\nu_-}}d_{\kk,e_{Y1},\eta_1,s_1}=0$ and hence the last two terms in the above equation vanish. 
Then we obtain
\begin{align}
&  d_{\kk+\pp+ \qq, e_Y, \eta_2, s_2}^\dagger d_{\kk, e_{Y1}, \eta_1, s_1} d_{\kk''-\qq,-e_Y'',\eta'',s''}^\dagger d_{\kk'',e_Y'',\eta'',s''} \nono\\
\approx & \delta_{\pp+\qq,0} \delta_{e_Y,e_{Y1}} \delta_{\eta_2,\eta_1} \delta_{s_2,s_1}
    d_{\kk''-\qq,-e_Y'',\eta'',s''}^\dagger d_{\kk'',e_Y'',\eta'',s''} \nono\\
& -  \delta_{\kk+\pp+\qq,\kk''} \delta_{e_Y,e_Y''} \delta_{\eta_2,\eta''} \delta_{s_2,s''} n_{e_Y,\eta_2,s_2} 
    d_{\kk+\pp,-e_Y,\eta,s}^\dagger d_{\kk, e_{Y1}, \eta_1, s_1} \ ,
\end{align}
where $n_{e_Y,\eta,s}$ equals to 1 if the flavor $\{e_Y,\eta,s\}$ is occupied and equals to 0 otherwise.
Here we have made use of $\{ d_{\kk+\pp,-e_Y,\eta,s}^\dagger, d_{\kk, e_{Y1}, \eta_1, s_1}\}=0$, which is because $\{e_{Y1},\eta_1,s_1\}$ is required to be occupied and $\{-e_Y,\eta,s\}$ is required to be empty and hence  $\{e_{Y1},\eta_1,s_1\} \neq \{-e_Y,\eta,s\}$.
Then we obtain
\begin{align} \label{p5eq:ddO-tmp1}
& d_{\kk+\pp+ \qq, e_Y, \eta_2, s_2}^\dagger d_{\kk, e_{Y1}, \eta_1, s_1} O_{-\mathbf{q,-G}} \ket{\Psi_\nu^{\nu_+,\nu_-}} 
\approx \delta_{\qq,0} A_{-\GG} N_M d_{\kk+\pp, e_Y, \eta_2, s_2}^\dagger d_{\kk, e_{Y1}, \eta_1, s_1} \ket{\Psi_\nu^{\nu_+,\nu_-}} \nono\\
& - \sqrt{V(\qq+\GG)} \eta_2 F_{e_Y}(\kk+\pp+\qq,-\qq-\GG) n_{e_Y,\eta_2,s_2}  
    d_{\kk+\pp,-e_Y,\eta,s}^\dagger d_{\kk, e_{Y1}, \eta_1, s_1} \nono\\
& + \delta_{\pp+\qq,0} \delta_{e_Y,e_{Y1}} \delta_{\eta_2,\eta_1} \delta_{s_2,s_1}
    \sum_{\kk'',e_Y'',\eta'',s''} \sqrt{V(\qq+\GG)}\eta'' F_{e_Y''}(\kk'',-\qq-\GG) 
    d_{\kk''-\qq,-e_Y'',\eta'',s''}^\dagger d_{\kk'',e_Y'',\eta'',s''} \ .
\end{align}
Similarly, we have
\begin{align}
&  d_{\kk+\pp, e_{Y2}, \eta_2, s_2}^\dagger d_{\kk-\qq, e_{Y}', \eta_1, s_1} d_{\kk''-\qq,-e_Y'',\eta'',s''}^\dagger d_{\kk'',e_Y'',\eta'',s''} \nono\\
\approx & 
    \bra{\Psi_\nu^{\nu_+,\nu_-}} d_{\kk+\pp, e_{Y2}, \eta_2, s_2}^\dagger  d_{\kk-\qq, e_{Y}', \eta_1, s_1} \ket{\Psi_\nu^{\nu_+,\nu_-}}
    d_{\kk''-\qq,-e_Y'',\eta'',s''}^\dagger d_{\kk'',e_Y'',\eta'',s''} \nono\\
& + \bra{\Psi_\nu^{\nu_+,\nu_-}} d_{\kk+\pp, e_{Y2}, \eta_2, s_2}^\dagger d_{\kk'',e_Y'',\eta'',s''}  \ket{\Psi_\nu^{\nu_+,\nu_-}}
    d_{\kk-\qq, e_{Y}', \eta_1, s_1} d_{\kk''-\qq,-e_Y'',\eta'',s''}^\dagger \nono\\
& + \bra{\Psi_\nu^{\nu_+,\nu_-}}  d_{\kk-\qq, e_{Y}', \eta_1, s_1} d_{\kk''-\qq,-e_Y'',\eta'',s''}^\dagger \ket{\Psi_\nu^{\nu_+,\nu_-}}
    d_{\kk+\pp, e_{Y2}, \eta_2, s_2}^\dagger d_{\kk'',e_Y'',\eta'',s''} \nono\\
& - \bra{\Psi_\nu^{\nu_+,\nu_-}}  d_{\kk-\qq, e_{Y}', \eta_1, s_1} d_{\kk''-\qq,-e_Y'',\eta'',s''}^\dagger \ket{\Psi_\nu^{\nu_+,\nu_-}}
    \bra{\Psi_\nu^{\nu_+,\nu_-}} d_{\kk+\pp, e_{Y2}, \eta_2, s_2}^\dagger d_{\kk'',e_Y'',\eta'',s''} \ket{\Psi_\nu^{\nu_+,\nu_-}}. \nono\\
\approx & \delta_{\kk,\kk''} \delta_{e_Y',-e_Y''} 
    \delta_{\eta_1,\eta''} \delta_{s_1,s''}
    (1- n_{e_Y',\eta_1,s_1} )
    d_{\kk+\pp, e_{Y2}, \eta_2, s_2}^\dagger d_{\kk'',e_Y'',\eta'',s''},
\end{align}
where we have made use of $\bra{\Psi_\nu^{\nu_+,\nu_-}} d_{\kk+\pp, e_{Y2}, \eta_2, s_2}^\dagger=0$, and hence 
\begin{align} \label{p5eq:ddO-tmp2}
& d_{\kk+\pp, e_{Y2}, \eta_2, s_2}^\dagger d_{\kk-\qq, e_{Y}', \eta_1, s_1} O_{-\mathbf{q,-G}} \ket{\Psi_\nu^{\nu_+,\nu_-}}  
    \approx \delta_{\qq,0} A_{-\GG} N_M d_{\kk+\pp, e_Y, \eta_2, s_2}^\dagger d_{\kk, e_{Y1}, \eta_1, s_1} \ket{\Psi_\nu^{\nu_+,\nu_-}} \nono\\
& + \sqrt{V(\qq+\GG)} \eta_1 F_{-e_{Y1}}(\kk,-\qq-\GG) (1-n_{e_{Y}',\eta_1,s_1})  d_{\kk+\pp,e_{Y2},\eta,s}^\dagger d_{\kk,-e_{Y}', \eta_1, s_1} \ .
\end{align}
Substituting \cref{p5eq:ddO-tmp1,p5eq:ddO-tmp2} into \cref{p5eq:charge-neutral-tmp1}, we obtain
{\small
\begin{align}
& [H_I- \mu N, d_{\kk+\pp, e_{Y2}, \eta_2, s_2}^\dagger d_{\kk, e_{Y1}, \eta_1, s_1}] \ket{\Psi_\nu^{\nu_+,\nu_-}} \nonumber \\ 
\approx & \bigg(
    \sum_{e_{Y}}  R^{\eta_2,s_2}_{e_Y,e_{Y2}}(\kk+\pp)
   d_{\mathbf{k}+\pp,e_Y,\eta_2,s_2}^\dag d_{\kk, e_{Y1}, \eta_1, s_1}  
   + \sum_{e_Y'} \td{R}_{e_Y',e_{Y1}}^{\eta_1,s_1*} (\kk)
    d_{\kk+\pp, e_{Y2}, \eta_2, s_2}^\dagger d_{\kk, e_Y', \eta_1, s_1} 
    \bigg) \ket{\Psi_\nu^{\nu_+,\nu_-}}   \nonumber \\ 
&- \frac{1}{\Omega_{\rm tot}}\sum_{\qq,\GG} V(\mathbf{G}+\mathbf{q}) \sum_{e_Y,e_Y'}  
    M_{e_Y,e_{Y2}}^{\left(\eta_2\right)}\left(\mathbf{k}+\pp,\mathbf{q}+\mathbf{G}\right)  M_{e_Y',e_{Y1}}^{\left(\eta_1\right)*}\left(\mathbf{k},\mathbf{q}+ \mathbf{G}\right)  d_{\kk+\pp+\qq, e_Y, \eta_2, s_2}^\dagger d_{\kk+\qq, e_Y', \eta_1, s_1}  \ket{\Psi_\nu^{\nu_+,\nu_-}} \nono\\
&+\frac{\delta_{\eta_2 \eta_1} \delta_{s_2,s_1}}{\Omega_{\rm tot}} \sum_\GG \sum_{\kk''e_Y'' \eta'' s''} V(-\pp+\GG) 
    M^{(\eta_2)}_{e_{Y1},e_{Y2}}(\kk+\pp,-\pp+\GG) \eta'' F_{e_Y''}(\kk'',\pp-\GG) 
    d_{\kk''+\pp,-e_Y'',\eta'',s''}^\dagger d_{\kk'',e_Y'',\eta'',s''} 
    \ket{\Psi_\nu^{\nu_+,\nu_-}} \ ,
\end{align} }%
where $R^{\eta_2,s_2}$ and $\td{R}^{\eta_1,s_1}$ are given by \cref{p5eq:Rapprox,p5eq:Rtapprox}, respectively. 
The last term is nonzero only if $\{e_{Y1},\eta_1,s_1\}$ is occupied and $\{e_{Y2},\eta_2,s_2\}$ is empty, which, provided $\eta_2=\eta_1$ and $s_2=s_1$, also implies $e_{Y1}$ must equal to $-e_{Y2}$. 
Thus we can rewrite the last term as 
{\small
\begin{align}
& \frac{\delta_{e_{Y2},-e_{Y1}} \delta_{\eta_2 \eta_1} \delta_{s_2,s_1}}{\Omega_{\rm tot}}  
    \sum_\GG \sum_{\kk'' e_Y  \eta s} {V(-\pp+\GG)} 
    \eta_2 F_{e_{Y2}}(\kk+\pp,-\pp+\GG) \eta F_{-e_Y}(\kk'',\pp-\GG) 
    d_{\kk''+\pp,e_Y,\eta,s}^\dagger d_{\kk'',-e_Y,\eta,s} 
    \ket{\Psi_\nu^{\nu_+,\nu_-}} \nono\\
=& \frac{\delta_{e_{Y2},-e_{Y1}} \delta_{\eta_2 \eta_1} \delta_{s_2,s_1}}{\Omega_{\rm tot}} 
    \sum_\GG \sum_{\kk'' e_Y  \eta s} {V(-\pp+\GG)} 
    \eta_2 F_{e_{Y2}}(\kk,\pp-\GG) \eta F_{e_Y}^*(\kk'',\pp-\GG) 
    d_{\kk''+\pp,e_Y,\eta,s}^\dagger d_{\kk'',-e_Y,\eta,s} 
    \ket{\Psi_\nu^{\nu_+,\nu_-}} \ ,
\end{align} }%
where we have made use of \cref{p5:eq:chiral-MqG2,p5:eq:alpha-cond1}.
Therefore, we can write the scattering equation as 
\begin{align}
& [H_I- \mu N, d_{\kk+\pp, e_{Y2}, \eta_2, s_2}^\dagger d_{\kk, e_{Y1}, \eta_1, s_1}] \ket{\Psi_\nu^{\nu_+,\nu_-}} \nono\\
\approx & \sum_{\eta,s, \eta',s'} \sum_{e_Y,e_Y'} \sum_{\qq} S^{\eta,s,\eta',s'; \eta_2,s_2,\eta_1,s_1}_{e_Y,e_Y'; e_{Y2}, e_{Y1} }(\kk+\qq,\kk;\pp) 
     d_{\kk+\qq+\pp, e_{Y}, \eta, s}^\dagger d_{\kk+\qq, e_{Y}', \eta', s'} 
     \ket{\Psi_\nu^{\nu_+,\nu_-}} \ ,
\end{align}
where the scattering matrix is 
\begin{align} \label{p5eq:Smatrix-approx}
& S^{\eta,s,\eta',s' ; \eta_2,s_2,\eta_1,s_1}_{e_Y,e_Y'; e_{Y2}, e_{Y1} }(\kk+\qq,\kk;\pp) = \delta_{\eta,\eta_2} \delta_{s,s_2} \delta_{\eta',\eta_1} \delta_{s',s_1}
    \bigg( \delta_{\qq,0} \Big( R^{\eta_2,s_2}_{e_Y,e_{Y2}}(\kk+\pp) \delta_{e_Y',e_{Y1}} +  \delta_{e_Y,e_{Y2}}  \td{R}^{\eta_1,s_1}_{e_Y',e_{Y1}}(\kk)  \Big) \nono\\
& \qquad - \frac{1}{\Omega_{\rm tot}}\sum_{\GG} V(\mathbf{G}+\mathbf{q}) 
    M_{e_Y,e_{Y2}}^{\left(\eta_2\right)}\left(\mathbf{k}+\pp,\mathbf{q}+\mathbf{G}\right)  M_{e_Y',e_{Y1}}^{\left(\eta_1\right)*}\left(\mathbf{k},\mathbf{q}+ \mathbf{G}\right)  \bigg) \nono\\
& \qquad + \delta_{e_{Y2},-e_{Y1}} \delta_{\eta_2 \eta_1} \delta_{s_2,s_1}
    \delta_{e_Y,-e_Y'} \delta_{\eta\eta'} \delta_{ss'}
    \frac1{\Omega_{\rm tot}}
    \sum_\GG {V(-\pp+\GG)} 
    \eta F_{e_Y}^*(\kk+\qq,\pp-\GG) \eta_2 F_{e_{Y2}}(\kk,\pp-\GG) \ .
\end{align}
The last term is nonzero if both the $\{\eta_2, s_2\}$ and $\{\eta,s\}$ valley-spin flavors are halfly filled. 
It couples all the halfly filled valley-spin flavors, which are independent in the (first) chiral-flat limit, to each other. 

\begin{figure}
\centering
\includegraphics[width=1\linewidth]{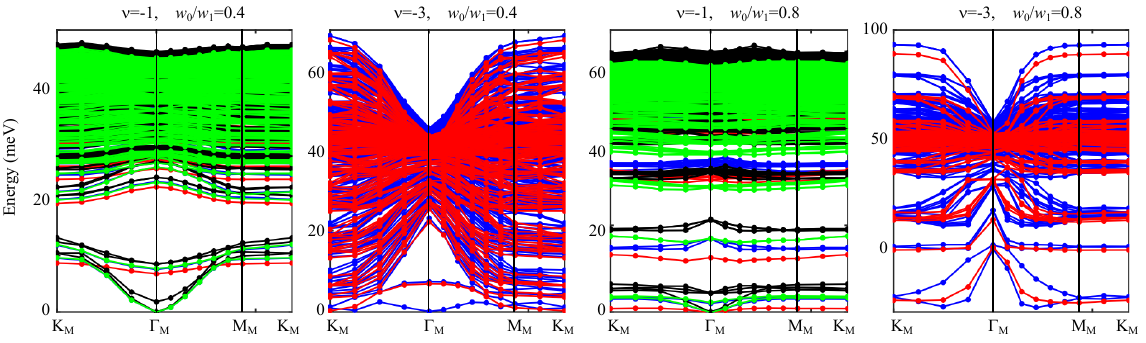}
\caption{Approximate charge neutral excitations at $\theta=1.05^\circ$ at odd fillings in the nonchiral-flat limit.
The flat metric condition is {\it not} imposed. 
Blue, red, black, and green bands are in the empty-half, half-half, empty-filled, half-filled sectors, respectively. 
(See the text for the definition of sectors.)
In this plot we set the screening length as $\xi=10$nm and accordingly the interaction strength as $U_\xi=13$meV.
The other parameters are same as in \cref{p5:newapp:Ham}: $v_F=5.944{\rm eV \cdot \mathring{ A} }$, $|K|=1.703\mathring{\rm A}^{-1}$, $w_1=110{\rm meV}$.}
\label{fig:E0approx}
\end{figure}

According to \cite{ourpaper4}, in the nonchiral-flat limit the state $\Psi_{-3}^{1,0}$ (or its U(4) rotations) is still the perturbative ground state at $\nu=-3$. 
Without loss of generality, we assume the occupied flavor is $\{+1,\up,+1\}$. 
Then we can divide the neutral excitations into the following sectors:
\begin{enumerate}
\item The half-half sector, where $\eta_2=\eta_1=+1$, $s_2=s_1=\up$. The delta functions in the first term of \cref{p5eq:Smatrix-approx} require $\eta=\eta'=+1$, $s=s'=\up$. The delta functions in the second term requires $\eta=\eta'$ and $s=s'$. Since $d_{\eta,s,e_Y}^\dagger$ and $d_{\eta',s',e_Y'}$ must belong to empty and occupied bands, there must also be $\eta=\eta'=+1$, $s=s'=\up$ in the second term.  Then it follows that $e_Y=e_{Y2}=-1$, $e_Y'=e_{Y1}=+1$. At given $\kk,\pp,\qq$, the $S$ matrix is a one-by-one matrix.
\item The empty-half sector, where $\{\eta_2,s_2\}$ is an empty valley-spin sector and $\{\eta_1,s_1\}$ is a half-filled valley-spin sector. The second term of \cref{p5eq:Smatrix-approx} vanish due to the delta function $\delta_{\eta_2\eta_1} \delta_{s_2s_1}$. 
The delta functions in the first term requires $\eta=\eta_2$, $s=s_2$, $\eta'=\eta_1$, $s'=s_1$. It follows that $e_Y'=e_{Y1}=+1$ and $e_Y,e_{Y2}$ take values in $\pm1$. At given $\kk,\pp,\qq$, the $S$ matrix is a two-by-two matrix.
\end{enumerate}
According to \cite{ourpaper4}, in the nonchiral-flat limit the state $\Psi_{-1}^{2,1}$ (or its U(4) rotations) is still the perturbative ground state at $\nu=-1$. 
There is a fully occupied valley-spin sector and a half filled valley-spin sector. 
Without loss of generality, we assume the occupied flavors as $\{+1,\up,+1\}$, $\{+1,\up,-1\}$, $\{+1,\down,+1\}$.
Then we can divide the neutral excitations into the following sectors:
\begin{enumerate}
\item The half-half sector, where $\eta_2=\eta_1=+1$, $s_2=s_1=\down$. The delta functions in the first term of \cref{p5eq:Smatrix-approx} require $\eta=\eta'=+1$, $s=s'=\down$. The delta functions in the second term requires $\eta=\eta'$ and $s=s'$. Since $d_{\eta,s,e_Y}^\dagger$ and $d_{\eta',s',e_Y'}$ must belong to empty and occupied bands, there must also be $\eta=\eta'=+1$, $s=s'=\down$ (the half filled valley-spin sector) in the second term. Then it follows that $e_Y=e_{Y2}=-1$, $e_Y'=e_{Y1}=+1$. At given $\kk,\pp,\qq$, the $S$ matrix is a one-by-one matrix.
\item The empty-half sector, where $\{\eta_2,s_2\}$ is an empty valley-spin sector and $\{\eta_1,s_1\}$ is a half-filled valley-spin sector. The second term of \cref{p5eq:Smatrix-approx} vanish due to the delta function $\delta_{\eta_2\eta_1} \delta_{s_2s_1}$. 
The delta functions in the first term requires $\eta=\eta_2$, $s=s_2$, $\eta'=\eta_1$, $s'=s_1$. It follows that $e_Y'=e_{Y1}=+1$ and $e_Y,e_{Y2}$ take values in $\pm1$. At given $\kk,\pp,\qq$, the $S$ matrix is a two-by-two matrix.
\item The half-occupied sector, where $\{\eta_2,s_2\}$ is the half filled valley-spin sector $\{+1,\down\}$ and $\{\eta_1,s_1\}$ is the fully occupied valley-spin sector $\{+1,\up\}$. The second term of \cref{p5eq:Smatrix-approx} vanish due to the delta function $\delta_{\eta_2\eta_1} \delta_{s_2s_1}$. 
The delta functions in the first term requires $\eta=\eta_2$, $s=s_2$, $\eta'=\eta_1$, $s'=s_1$. It follows that $e_Y',e_{Y1}$ take values in $\pm1$ and $e_Y=e_{Y2}=-1$. At given $\kk,\pp,\qq$, the $S$ matrix is a two-by-two matrix.
\item The empty-occupied sector, where $\{\eta_2,s_2\}$ is an empty valley-spin sector  and $\{\eta_1,s_1\}$ is the fully occupied valley-spin sector $\{+1,\up\}$. The second term of \cref{p5eq:Smatrix-approx} vanish due to the delta function $\delta_{\eta_2\eta_1} \delta_{s_2s_1}$. 
The delta functions in the first term requires $\eta=\eta_2$, $s=s_2$, $\eta'=\eta_1$, $s'=s_1$. It follows that $e_Y',e_{Y1},e_Y,e_{Y2}$ all take values in $\pm1$. At given $\kk,\pp,\qq$, the $S$ matrix is a four-by-four matrix.
\end{enumerate}
The numerical results are shown in \cref{fig:E0approx}.

\end{document}